\documentclass[11pt,a4paper]{article}
\pdfoutput=1
\usepackage{amsfonts}
\usepackage{jheppub}
\usepackage{amsmath}
\usepackage{bm}
\usepackage{times}
\usepackage{braket}
\usepackage{color,graphicx}
\usepackage{multirow}
\usepackage[T1]{fontenc}
\usepackage[utf8]{inputenc}
\usepackage[table]{xcolor}
\usepackage{hyperref}
\usepackage{lscape}
\usepackage[normalem]{ulem}
\usepackage{slashed}
\usepackage{bbold}

\usepackage{arydshln}

\usepackage{colortbl}
\definecolor{Gray}{gray}{0.95}
\definecolor{RGray}{gray}{0.85}
\definecolor{CGray}{gray}{0.92}

\makeatletter
\g@addto@macro\bfseries{\boldmath}
\makeatother

\definecolor{violet}{cmyk}{0,1,0,0.2}

\usepackage{xspace}

\def\eq#1{{Eq.~(\ref{#1})}}
\def\eqs#1#2{{Eqs.~(\ref{#1})--(\ref{#2})}}
\def\fig#1{{Fig.~\ref{#1}}}

\def\Table#1{{Table~\ref{#1}}}

\def\sect#1{{Sec.~\ref{#1}}}

\def\app#1{{App.~\ref{#1}}}
\def\apps#1#2{{Apps.~\ref{#1}--\ref{#2}}}
\def\vev#1{\left\langle #1\right\rangle}
\def\abs#1{\left| #1\right|}

\def\Im{\mbox{Im}\,}
\def\Re{\mbox{Re}\,}
\def\trace{\mbox{Tr}\,}
\def\Tr{\mbox{Tr}\,}

\renewcommand{\bar}{\overline}

\definecolor{LightCyan}{rgb}{0.88,1,1}
\definecolor{piggypink}{rgb}{0.99, 0.87, 0.9}
\definecolor{applegreen}{rgb}{0.55, 0.71, 0.0}
\definecolor{darkpastelgreen}{rgb}{0.01, 0.75, 0.24}
\definecolor{green-yellow}{rgb}{0.68, 1.0, 0.18}

\newcommand{\beq}{\begin{equation}}
\newcommand{\eeq}{\end{equation}}
\newcommand{\bea}{\begin{eqnarray}}
\newcommand{\eea}{\end{eqnarray}}

\title{Maximal Flavour Violation: \\a Cabibbo mechanism for leptoquarks}

\author[a]{Luca Di Luzio,} 
\author[b]{Javier Fuentes-Martin,}
\author[c, d]{Admir Greljo,}
\author[e, f]{Marco Nardecchia,}
\author[c]{Sophie Renner}

\affiliation[a]{Institute for Particle Physics Phenomenology, Department of Physics, Durham University, DH1 3LE, Durham, United Kingdom}
\affiliation[b]{Physik-Institut, Universit\"{a}t Z\"{u}rich, CH-8057 Z\"{u}rich, Switzerland}
\affiliation[c]{PRISMA Cluster of Excellence and Mainz Institute for Theoretical Physics, Johannes Gutenberg-Universit\"at Mainz, 55099 Mainz, Germany}
\affiliation[d]{Faculty of Science, University of Sarajevo, Zmaja od Bosne 33-35, 71000 Sarajevo, Bosnia and Herzegovina}
\affiliation[e]{Theoretical Physics Department, CERN, Geneva, Switzerland}
\affiliation[f]{INFN, Sezione di Trieste, Italy}

\emailAdd{luca.di-luzio@durham.ac.uk}
\emailAdd{fuentes@physik.uzh.ch} 
\emailAdd{admgrelj@uni-mainz.de}
\emailAdd{marco.nardecchia@cern.ch} 
\emailAdd{sorenner@uni-mainz.de}
  
\abstract{We propose a mechanism that allows for sizeable flavour violation in quark-lepton currents, 
while suppressing flavour changing neutral currents in quark-quark and lepton-lepton sectors. 
The mechanism is applied to the recently proposed ``$4321$'' renormalizable model, 
which can accommodate the current experimental anomalies in $B$-meson decays, both in charged and neutral currents,    
while remaining consistent with all other indirect flavour and electroweak precision measurements and direct searches at high-$p_T$.  To support this claim, we present an exhaustive phenomenological survey of this fully calculable UV complete model and highlight the rich complementarity between indirect and direct searches. 
}

\arxivnumber{1808.00942}

\begin{document}

\maketitle


\section{Introduction}

In the recent years the central question of flavour physics beyond the Standard Model (SM) has been the following: 
\emph{``How is it possible to reconcile TeV-scale new physics (NP) (as suggested e.g.~by naturalness) with the absence of indirect signals in flavour changing neutral currents (FCNC)?''.} 
One possible answer was given by the principle of Minimal Flavour Violation~\cite{DAmbrosio:2002vsn}, which allowed 
for exciting NP at ATLAS and CMS while predicting less room for serendipity at LHCb. 
Somewhat unexpectedly, we are faced with the fact that experimental 
data seem to rather suggest the opposite situation. 
In fact, a coherent pattern of SM deviations in semileptonic $B$-decays, 
which goes under the widely accepted name of ``flavour anomalies",
keeps building up 
since $2012$~\cite{Lees:2012xj,Lees:2013uzd,Aaij:2013qta,Aaij:2014ora,Aaij:2015yra,Aaij:2015oid,Huschle:2015rga,Sato:2016svk,Hirose:2016wfn,Hirose:2017dxl,Aaij:2017vbb,Aaij:2017deq}.
Were these anomalies due to NP, 
they would certainly imply a shift of paradigm in flavour physics. 

A unified explanation of the whole set of anomalous data 
minimally requires:
$i)$ a NP contribution in $b \to s \mu \mu$ neutral currents 
that interferes destructively with the SM and $ii)$ a NP contribution in charged currents 
that enhances the decay rates of $b \to c \tau \nu$ transitions.
Despite many models being proposed so far 
for the combined explanation of the anomalies (see 
\cite{Bhattacharya:2014wla,Alonso:2015sja,Greljo:2015mma,Calibbi:2015kma,Bauer:2015knc,Fajfer:2015ycq,
Barbieri:2015yvd,Das:2016vkr,Boucenna:2016wpr,Boucenna:2016qad,Becirevic:2016yqi,
Hiller:2016kry,Bhattacharya:2016mcc,Barbieri:2016las,Becirevic:2016oho,Bordone:2017anc,Megias:2017ove, 
Crivellin:2017zlb,Cai:2017wry,Altmannshofer:2017poe,Dorsner:2017ufx,Buttazzo:2017ixm,Assad:2017iib,
DiLuzio:2017vat,Calibbi:2017qbu,Bordone:2017bld,Choudhury:2017ijp,Barbieri:2017tuq,Sannino:2017utc,
Blanke:2018sro,Greljo:2018tuh,Marzocca:2018wcf,Asadi:2018wea,Greljo:2018ogz,Robinson:2018gza,Azatov:2018knx,Bordone:2018nbg,
Becirevic:2018afm,Kumar:2018kmr,Trifinopoulos:2018rna,Azatov:2018kzb} for an incomplete list), 
it is fair to say that the majority of these works suffer from various issues: 
neglect of key observables (both at low energy and high-$p_T$), 
missing UV completion, breakdown of the perturbative expansion, unnatural and tuned values of the parameters, 
etc.
The difficulties in constructing a viable and coherent NP interpretation of the flavour anomalies 
(both in charged and neutral currents) are due to the simultaneous presence of the following aspects of the phenomenological situation: 
\begin{enumerate}
\item  
the NP contribution in $b \to c \tau \nu$ needs to be very large, 
since it must compete with a SM tree-level process;  
\item there is an absence of NP signals in direct searches at the LHC;
\item there are very severe constraints from flavour observables in pure hadronic channels, 
most notably in $\Delta F = 2$ transitions;
\item there are very severe constraints from flavour observables in pure leptonic channels, 
most notably in processes violating lepton universality and lepton flavour. 
\end{enumerate}
Since the first point clearly contrasts with the remaining ones,  
finding a coherent NP framework to explain all these facts remains a non-trivial challenge.  
However, the points above are also suggesting in a (qualitative) way their own solutions. 
Indeed a viable NP scenario should:
\begin{enumerate}
\item contain a leptoquark 
with \textit{large} flavour violating couplings 
in order to trigger the anomalous semileptonic decays in charged currents;
\item only introduce new states that are \emph{heavy} enough to escape direct detection;
\item have a protecting flavour symmetry in the purely quark sector, such as a $U(2)$ acting on the first two families of quarks;
\item have a protecting flavour symmetry in the purely lepton sector, such as $U(1)_e \times U(1)_{\mu} \times U(1)_{\tau}$.
\end{enumerate}
Does a model with such properties exist?
In this paper we are going to present a phenomenological attempt to answer this question,  
by exploring a specific limit of the ``4321 model'' introduced in Ref.~\cite{DiLuzio:2017vat}. 
Here, 4321 stands for the gauge structure of the model, 
which is invariant under the local group $SU(4) \times SU(3)' \times SU(2)_L \times U(1)'$. 
The symmetry breaking down to the SM delivers a TeV-scale vector leptoquark, 
$U_\mu \sim (\mathbf{3},\mathbf{1},1/3)$, 
with the most favourable quantum numbers in order to mediate the flavour anomalies, 
as inferred from recent simplified-model analyses \cite{Buttazzo:2017ixm,Kumar:2018kmr}.

While the aspects that we are going to discuss will be exemplified 
in the context of the 4321 model (the detailed phenomenological analysis of this model 
is in fact one of the main goals of this paper), 
we believe that the mechanism presented here should 
be a welcome ingredient for any extension aiming at a consistent 
description of the whole set of anomalies. 
This ingredient is nothing but a generalisation of the well-known Cabibbo mixing 
\cite{Cabibbo:1963yz} to the leptoquark sector. 
The up- and down-quark sectors in the SM, when taken in isolation, 
preserve their own $U(1)^3$ family symmetry. It is only 
the simultaneous presence of up and down Yukawa matrices 
that provides a flavour violating misalignment of the size of the Cabibbo angle. 
Our proposal follows in close analogy: quarks and leptons in isolation 
preserve their own original symmetries, 
while flavour violation is a product of the \emph{collective breaking} coming from the two sectors. The misalignment between the second and third family of quark and lepton doublets, $\theta_{LQ}$, 
is the generalisation of the Cabibbo angle, $\theta_C$. 
As a consequence, tree-level neutral currents are (practically) absent 
and all the relevant flavour violating interactions only involve the exchange of the leptoquark. 
The individual (assumed) larger symmetries in the quark and lepton sectors 
guarantee enough flavour protection from low-energy indirect probes, 
while a sizeable $\theta_{LQ}$ allows for large effects in the desired $b \to c \tau \nu$ transitions at tree level. 
Crucially, a large 3-2 leptoquark transition 
allows the scale of NP to be raised and relaxes in turn the bounds from 
LHC direct searches. This approach differs from those scenarios in which the NP is 
aligned along the third generation and the 3-2 transitions are obtained via an $\mathcal{O}(V_{cb})$ 
rotation. In the latter case,  
the flavour suppression in the NP amplitude has to be compensated either by a lower value of the NP scale 
or by large couplings arising from non-perturbative dynamics. 
In both cases, one is faced with very serious challenges 
both from precisely measured $Z$-pole observables and $\tau$ decays \cite{Feruglio:2016gvd,Feruglio:2017rjo} 
and direct searches (see e.g.~\cite{Faroughy:2016osc,Greljo:2017vvb}). 

The connection between low- and high-energy phenomenology in the 4321 model 
goes even further. 
In fact, the large flavour breaking between second and third generation in the leptoquark sector 
is responsible for $\Delta F=2$ quark transitions at the one-loop level, 
for which a GIM-like mechanism is at work: 
a sufficient suppression of $B_s$ and $D$ mixing 
is guaranteed by the lightness of the lepton partners present in the radiative amplitudes.
We hence obtain upper bounds on heavy lepton partners from indirect searches 
and lower bounds from direct searches at the LHC. Remarkably, in a large part of the parameter space 
this mass window is very narrow: low-energy probes are suggesting a clear target for direct searches 
at high-$p_T$. The role of the heavy lepton partners in the 4321 model recall in a sense the charm 
prediction from kaon meson mixing in the SM \cite{Glashow:1970gm,Gaillard:1974hs}. 


The paper is structured as follows: 
in \sect{sec:4321} we introduce the main elements of the 4321 model and in \sect{sec:cabibbo} 
we discuss the leptoquark Cabibbo mechanism 
making use of symmetry arguments and analogies with the SM. 
In \sect{sec:lowen} we collect the main observables relevant for the low-energy phenomenology, 
including the flavour anomalies and the relevant constraints from indirect searches.  
In \sect{sec:highpT} we present the status of direct searches, and show that a large breaking in the 3-2 sector 
is needed to lift the NP scale in order to escape direct detection. 
In \sect{sec:concl} we summarize our main predictions and conclude. 
A thorough discussion of several theoretical aspects 
of the 4321 model is deferred to \app{details4321}. 


\section{The $4321$ model}
\label{sec:4321}

In this section we summarise the main features of the 4321 model presented in \cite{DiLuzio:2017vat} (see also \cite{Diaz:2017lit}). 
Further details are provided in~\app{details4321}. 
The goal of the model's construction is to generate a coupling of the vector leptoquark $U \sim (\mathbf{3},\mathbf{1},2/3)$ 
mainly to \emph{left-handed} SM fermions. 
This allows $i)$ to match with the model-independent fits to $B$-anomalies \cite{Buttazzo:2017ixm,Kumar:2018kmr} 
and $ii)$ to tame strong constraints from chirality-enhanced meson decays into lepton pairs 
(for an updated analysis see Ref.~\cite{Smirnov:2018ske}). 
To this end we consider the gauge group $\mathcal{G}_{\rm 4321} \equiv SU(4) \times SU(3)' \times SU(2)_L \times U(1)'$, 
which extends the SM group $\mathcal{G}_{\rm 321} \equiv SU(3)_c \times SU(2)_L \times U(1)_Y$ 
by means of an extra $SU(4)$ factor. 
The embedding of colour 
and hypercharge into $\mathcal{G}_{\rm 4321}$ is defined as $SU(3)_c = \left( SU(3)_4 \times SU(3)' \right)_{\rm diag}$ and 
$Y = \sqrt{2/3} \, T^{15} + Y'$, with $SU(3)_4\subset SU(4)$ and $T^{15}$ being one of the generators of $SU(4)$.\footnote{For a complete list of $SU(4)$ generators see \app{SU4gt}.} 
Apart from the SM gauge fields, the gauge boson spectrum comprises three new massive vectors belonging 
to $\mathcal{G}_{\rm 4321} / \mathcal{G}_{\rm 321}$ 
and transforming under $\mathcal{G}_{\rm 321}$ as $U \sim (\mathbf{3},\mathbf{1},2/3)$, 
$g' \sim (\mathbf{8},\mathbf{1},0)$ and $Z' \sim (\mathbf{1},\mathbf{1},0)$. 
Their definition in terms of the $\mathcal{G}_{\rm 4321}$ gauge fields, 
as well as their masses, 
are given in \app{gbspec}. 

An important point to be stressed is that the three massive vectors are  
connected by gauge symmetry breaking and it is not possible to parametrically 
decouple the $g'$ (hereafter called ``coloron'') 
and the $Z'$ from the leptoquark mass 
scale. In \app{app:gaugescpectgt} we show that this feature persists also in non-minimal scalar 
sectors responsible for $\mathcal{G}_{\rm 4321}$ breaking. 
Moreover, the peculiar embedding of the SM into $\mathcal{G}_{\rm 4321}$ 
allows for suppressed coupling of the $Z'$ and coloron to light quarks (cf.~\sect{intmb}). 
That is not the case in more standard Pati Salam \cite{Pati:1974yy} 
embeddings such as in \cite{Calibbi:2017qbu}, where the $Z'$ has unsuppressed 
$\mathcal{O}(g_s)$ couplings to valence quarks. 

The matter content of the model is summarised in \Table{fieldcontent}, 
where we have emphasised with a grey background the states added on top of the SM-like fields.
The new gauge bosons receive a TeV-scale mass 
induced by the vacuum expectation value (VEV) of three scalar multiplets:
$\Omega_1 \sim \left( \mathbf{\bar 4}, \mathbf{1},\mathbf{1}, -1/2 \right)$, 
$\Omega_3 \sim \left( \mathbf{\bar 4}, \mathbf{3}, \mathbf{1}, 1/6 \right)$ and 
$\Omega_{15} \sim \left( \mathbf{15}, \mathbf{1}, \mathbf{1}, 0 \right)$,
responsible for the breaking of $\mathcal{G}_{\rm 4321} \to \mathcal{G}_{\rm 321}$.
While only~$\Omega_3$ would suffice for the breaking, the role of the other fields is of phenomenological 
nature as discussed below. 
By means of a suitable scalar potential (analysed in \app{scalpot}) 
it is possible to achieve a VEV configuration 
ensuring the proper $\mathcal{G}_{\rm 4321} \to \mathcal{G}_{\rm 321}$ breaking. 
After removing the linear combinations corresponding to the would-be Goldstone bosons (GB), 
the massive scalar spectrum featuring the radial modes is detailed in \app{scalspect}.  
The final breaking of $\mathcal{G}_{\rm 321}$ is obtained via the Higgs doublet field transforming as 
$H \sim (\mathbf{1},\mathbf{1},\mathbf{2},1/2)$.  

\begin{table}[htp]
\begin{center}
\begin{tabular}{|c|c|c|c|c||c|c|}
\hline
Field & $SU(4)$ & $SU(3)'$ & $SU(2)_L$ & $U(1)'$ & $U(1)_{B'}$ & $U(1)_{L'}$ \\
\hline
\hline
$q'^i_L$ & $\mathbf{1}$ & $\mathbf{3}$ & $\mathbf{2}$ & $1/6$ & $1/3$ & 0 \\
$u'^i_R$ & $\mathbf{1}$ & $\mathbf{3}$ & $\mathbf{1}$ & $2/3$ & $1/3$ & 0 \\
$d'^i_R$ & $\mathbf{1}$ & $\mathbf{3}$ & $\mathbf{1}$ & $-1/3$ & $1/3$ & 0 \\
$\ell'^i_L$ & $\mathbf{1}$ & $\mathbf{1}$ & $\mathbf{2}$ & $-1/2$ & 0 & $1$ \\
$e'^i_R$ & $\mathbf{1}$ & $\mathbf{1}$ & $\mathbf{1}$ & $-1$ & 0 & $1$ \\ \rowcolor{CGray}
$\Psi^i_L$ & $\mathbf{4}$ & $\mathbf{1}$ & $\mathbf{2}$ & 0 & $1/4$ & $1/4$ \\ \rowcolor{CGray}
$\Psi^i_R$ & $\mathbf{4}$ & $\mathbf{1}$ & $\mathbf{2}$ & 0 & $1/4$ & $1/4$ \\
\hline
\hline
$H$ & $\mathbf{1}$ & $\mathbf{1}$ & $\mathbf{2}$ & 1/2 & 0 & 0 \\ \rowcolor{CGray}
$\Omega_1$ & $\mathbf{\bar 4}$ & $\mathbf{1}$ & $\mathbf{1}$ & $-1/2$ & $-1/4$ & $3/4$ \\ \rowcolor{CGray}
$\Omega_3$ & $\mathbf{\bar 4}$ & $\mathbf{3}$ & $\mathbf{1}$ & $1/6$ & $1/12$ & $-1/4$ \\ \rowcolor{CGray}
$\Omega_{15}$ & $\mathbf{15}$ & $\mathbf{1}$ & $\mathbf{1}$ & 0 & $0$ & $0$ \\ 
\hline
\end{tabular}
\end{center}
\caption{\sf Field content of the 4321 model. The index $i=1,2,3$ runs over generations, 
while $U(1)_{B'}$ and $U(1)_{L'}$ are accidental global symmetries 
(see text for further clarifications). Particles added to the SM matter content are shown
on a grey background.   
}
\label{fieldcontent}
\end{table}

The would-be SM fermion fields, denoted with a prime, are singlets of $SU(4)$ and are charged under the 
$SU(3)' \times SU(2)_L \times U(1)'$ subgroup with SM-like charges. 
Like in the SM,  they come in three copies of flavour. 
Being $SU(4)$ singlets, they do not couple to the vector leptoquark directly.
In order to induce the required leptoquark interactions to SM fermions, we introduce
three vector-like heavy fermions that mix with the SM-like fermions once $\Omega_{1,3}$ acquire a 
VEV (cf.~also \fig{fig:mixing}). 
The vector-like fermions transform under $\mathcal{G}_{\rm 4321}$ 
as $\Psi_{L,R} = (Q'_{L,R}, L'_{L,R})^T \sim (\mathbf{4},\mathbf{1},\mathbf{2},0)$, with $Q'_{L,R} \sim (\mathbf{3}, \mathbf{2},1/6)$ 
and $L'_{L,R}\sim(\mathbf{1}, \mathbf{2},-1/2)$ when decomposed under $\mathcal{G}_{\rm 321}$. 
The vector-like masses of $Q'$ and $L'$ are split by the VEV of $\Omega_{15}$. 
The mixing among the left-handed SM-like and vector-like fermions is described by 
the Yukawa Lagrangian 
$\mathcal{L}_Y = \mathcal{L}_{\rm SM-like} + \mathcal{L}_{\rm mix}$, with 
\begin{align}
\label{LYUK1}
\mathcal{L}_{\rm SM-like} &=  - \bar{q}'_L \,Y_d \, H d'_R - \bar{q}'_L \,Y_u \, \tilde H u'_R - \bar{\ell}'_L \, Y_e \, H e'_R + \text{h.c.} \, ,  \\
\label{LYUK2}
\mathcal{L}_{\rm mix} &= - \bar {q}'_L \, \lambda_q \, \Omega_3^T \Psi_R - \bar {\ell}'_L \, \lambda_\ell \, \Omega_1^T \Psi_R 
- \bar \Psi_L \left( M + \lambda_{15}\, \Omega_{15} \right) \Psi_R + \text{h.c.} \, . 
\end{align}
Here, $\tilde H = i \sigma_2 H^*$ and $Y_{u,d,e}$, $\lambda_{q,\ell,15}$, $M$ are $3 \times 3$ flavour matrices. 
The flavour structure of the 4321 model will be discussed in detail in \sect{sec:cabibbo}. 

The full Lagrangian (including also the scalar potential in \eq{eqscalpot}) 
is invariant under the \textit{accidental} global symmetries $U(1)_{B'}$ and $U(1)_{L'}$,   
whose action on the matter fields is displayed in the last two 
columns of \Table{fieldcontent}.\footnote{Note that these global symmetries are anomalous under $SU(2)_L \times U(1)'$.} 
The VEVs of $\Omega_3$ and $\Omega_1$ break spontaneously 
both the gauge and the global symmetries,  
leaving unbroken two new global $U(1)$s: 
$B = B'+\frac{1}{\sqrt{6}}\,T^{15}$ and $L = L'-\sqrt{\frac{3}{2}}\,T^{15}$, 
which for the SM eigenstates correspond respectively to ordinary baryon and lepton number. 
These symmetries protect proton stability, make neutrinos massless
and prevent the appearance of massless state related to the spontaneous breaking of 
$U(1)_{B'}$ and $U(1)_{L'}$. 
Non-zero neutrino masses can be achieved by introducing an
explicit breaking of $U(1)_{L'}$, e.g.~via a $d=5$ effective operator 
$\ell' \ell' H H / \Lambda_{\slashed{L}}$, where the effective scale of lepton number violation, 
$\Lambda_{\slashed{L}}$, is well above the TeV scale.  
In contrast, recent proposals which address the anomalies based on a non-minimal Pati-Salam extension 
with gauged $B-L$ broken at the TeV, such as e.g.~\cite{Calibbi:2017qbu,Bordone:2018nbg}, generically 
predict too large neutrino masses. The latter either require a strong fine-tuning in the Yukawa structure or a very specific (untuned) realisation of the neutrino mass matrix by the inverse seesaw mechanism~\cite{Perez:2013osa,Greljo:2018tuh}.



\section{Cabibbo mechanism for leptoquarks}
\label{sec:cabibbo}
 
Our goal is to introduce the flavour structure required by the anomalies in the quark-lepton transitions, 
while simultaneously suppressing the most dangerous quark-quark and lepton-lepton flavour violating 
operators.\footnote{For a partially related discussion in the context of the neutral current anomalies, see \cite{Guadagnoli:2018ojc}.}
This step can be neatly understood in terms of the global symmetries of the Yukawa Lagrangian. 

Let us first consider the $\mathcal{L}_{\rm mix} \to 0$ limit.  
The surviving term in Eq.~(\ref{LYUK1}) corresponds to the SM Yukawa Lagrangian.  
Exploiting the $U(3)^5$ invariance of the kinetic term of the SM-like fields we choose,
without loss of generality, a basis where $Y_d= \hat{Y}_d$, 
$Y_u= V^{\dagger}\, \hat{Y}_u$ and $Y_e = \hat{Y}_e$ 
(a hat denotes a diagonal matrix with positive eigenvalues and $V$ is the CKM matrix).
For later convenience, we recall some well-known features of the SM quark Yukawa sector. 
In the $Y_u \to 0$ limit, the term
$\overline{q}'_L \hat{Y}_d \tilde{H} d_R$ leaves invariant the subgroup $U(1)_d \times U(1)_s \times U(1)_b$, 
thus implying the absence of flavour violation in the down sector. 
Similarly, for $Y_d \to 0$ we are left with $\overline{q}'_L V^{\dagger}\, \hat{Y}_u \tilde{H} u_R$ in the up sector. 
Reabsorbing $V$ into $q'$ 
bears no physical effects and the subgroup $U(1)_u \times U(1)_c \times U(1)_t$ is left unbroken. 
If both $Y_u$ and $Y_d$ are present, the two $U(1)^3$ are not independent any more due to the
$SU(2)_L$ gauge symmetry that forces the transformations of the left-handed down and up fields to be the same. 
The intersection of the two subgroups yields\footnote{Here $U(1)_{d+u}$ stands for the simultaneous transformation
$d \to e^{i\theta} d$ and $u \to e^{i\theta} u$, where $e^{i\theta}$ is an element of $U(1)_{d+u}$. 
The generalisation to non-abelian factors, which is employed later on, follows in analogy.} 
\begin{align}
&(U(1)_d \times U(1)_s \times U(1)_b ) \cap (U(1)_u \times U(1)_c \times U(1)_t) \supseteq \nonumber \\
&U(1)_{d+u} \times U(1)_{s+c} \times U(1)_{b+t} \xrightarrow[]{V \neq \mathbb{1}}
U(1)_B \, ,
\end{align}
where the last step of breaking is due to the CKM mixing and $U(1)_B$ is the baryon number. 
The consequences of this \emph{collective breaking} are: 
$i)$ No tree-level FCNC are generated. These are forbidden by the two $U(1)^3$ 
symmetries in isolation, either in the up or in the down sector. 
$ii)$ Flavour changing charged currents are generated by the misalignment between the up and down sectors, 
which is parametrised by the CKM matrix $V$. 
In the unitary gauge, the physical effects of flavour violation are fully encoded in the coupling of the $W$ boson 
to the up and down quark fields. 

Let us consider now the pattern of global symmetries when $\mathcal{L}_{\rm mix} \neq 0$. 
The role of the scalar representations $\Omega_{i}$ in $\mathcal{L}_{\rm mix}$ is the following:
\begin{itemize}
\item  $\langle \Omega_{3} \rangle$ mixes the would-be SM state $q'_L$ with 
$Q'_L \subset \Psi_L$. In this way the SM quark doublet enters into the $SU(4)$ 
representation $\Psi_L$ and feels the leptoquark interaction.
\item  $\langle \Omega_{1} \rangle$ mixes the would-be SM state $\ell'_L$ with $L'_L \subset \Psi_L$. 
In this way the SM lepton doublet enters into the $SU(4)$ representation $\Psi_L$ and feels the leptoquark interaction.
\item $\langle \Omega_{15} \rangle$ splits the bare masses of quark and lepton partners. 
We can hence effectively trade $M$ and $\lambda_{15} \langle \Omega_{15} \rangle$ for $M_Q$ and $M_L$.
\end{itemize}
Without loss of generality, we use the $U(3)^7$ symmetry of the fermionic kinetic term 
to pick up the following basis: 
\begin{align}\label{eq:Yukawas}
\mathcal{L}_{\rm SM-like} &=- \overline{q}'_L V^{\dagger} \hat{Y}_u u'_R \, \tilde{H}   - \overline{q}'_L \hat{Y}_d d'_R \, H - \overline{\ell}'_L \hat{Y}_e e'_R \, H + \text{h.c.} \, , \\
\mathcal{L}_{\rm mix} &= -\overline{q}'_L \lambda_q \Psi_R \, \Omega_3 - \overline{\ell}'_L \lambda_{\ell} \Psi_R \, \Omega_1  - \overline{\Psi}_L (\hat{M} + \lambda_{15} \Omega_{15} ) \Psi_R + \text{h.c.} \, ,  
\end{align}
where $\lambda_q$, $\lambda_{\ell}$ and $\lambda_{15}$ are matrices in flavour space. 
If the latter were generic, we would expect large flavour violating effects both in quark and lepton processes. 
We are going to argue that, \emph{assuming} the following flavour structure: 
\begin{align}\label{lamqfs}
\begin{aligned}
\lambda_q &= \hat{\lambda}_q \equiv \text{diag} \left( \lambda^q_{12}, \lambda^q_{12}, \lambda^q_{3} \right)
\, , \\
\lambda_{\ell} &= \hat{\lambda}_{\ell}\,  W^\dagger \equiv
\text{diag} \left( \lambda^{\ell}_{1}, \lambda^{\ell}_{2}, \lambda^{\ell}_{3} \right)
\left(
\begin{array}{ccc}
1 & 0 & 0 \\
0 & \cos \theta_{LQ} & -\sin \theta_{LQ} \\
0 & \sin \theta_{LQ} & \cos \theta_{LQ}
\end{array}
\right) \, , \\ 
\lambda_{15} & \propto \hat{M} \propto \mathbb{1} \, , 
\end{aligned}
\end{align}
provides a good starting point to 
comply with flavour constraints. 
Later on we will comment about the plausibility of our assumptions, 
but for the moment let us inspect the physical consequences of \eq{lamqfs}.

Mimicking the pure SM discussion, we examine the surviving global symmetries 
of $\mathcal{L}_{\rm mix}$ in either of the limits $\lambda_{\ell} \to 0$ or $\lambda_q \to 0$.
In the former case $\mathcal{L}_{\rm mix}$ is invariant under the action of the global 
symmetry group 
$\mathcal{G}_{Q} \equiv U(2)_{q' + \Psi} \times U(1)_{q'_3 + \Psi_3}$, 
with the non-abelian factor acting on the first and second generation. 
Basically, we are promoting the approximate $U(2)_{q'}$ of the SM 
(emerging in the limit where only $(Y_{u,d})_{33} \neq 0$) 
to be also a symmetry of the NP. This guarantees in turn:
\begin{itemize}
\item the absence of tree-level FCNC for down quarks (note that $Y_d$ and $\lambda_q$ are diagonal in the same basis). 
Such a \emph{down alignment} mechanism was already introduced in Ref.~\cite{DiLuzio:2017vat}. 
\item a strong suppression of tree-level FCNC for up quarks. This suppression is guaranteed by the underlying $U(2)$ symmetry 
and the physical effects are proportional to the small breaking induced by the SM-like Yukawa $Y_u$ via the CKM. 
We will show in \sect{sec:lowen} that this protection is crucial in order to pass the bounds from $D$-$\bar D$ mixing.
\end{itemize}
We continue with the discussion of the lepton sector when $\lambda_q \to 0$.
In this limit $\mathcal{L}_{\rm mix}$ has a $U(1)^3$ symmetry which is just the generalisation of the accidental 
symmetries of the SM in the lepton sector. To show this let us reabsorb $W$ in a redefinition of the field $\Psi$, 
via $\tilde{\Psi} \equiv W^\dagger \Psi$. With such a redefinition $\mathcal{L}_{\rm mix}$ reads
\begin{equation}
\mathcal{L}_{\rm mix}(\lambda_q \to 0) = - \overline{\ell}'_L \hat{\lambda}_{\ell} \tilde{\Psi}_R \, \Omega_1  - \overline{\tilde{\Psi}}_L (\hat{M} + \hat{\lambda}_{15} \Omega_{15} ) \tilde{\Psi}_R + \text{h.c.} \, . 
\end{equation}
Since everything is diagonal, the global symmetry is identified as 
$\mathcal{G}_{L} = U(1)_{\ell'_1+\tilde{\Psi}_{1}} \times U(1)_{\ell'_2+\tilde{\Psi}_{2}} \times U(1)_{\ell'_3+\tilde{\Psi}_{3}} $. 
The limit $\lambda_q \to 0$ thus implies: 
\begin{itemize}
\item the absence of tree-level FCNC for (charged) leptons. Note indeed that there exists a basis where $Y_e$ and $\lambda_{\ell}$ 
are simultaneously diagonal. 
\item that the $W$ matrix is unphysical.  
\end{itemize}
Let us consider now the case where both $\lambda_q$ and $\lambda_{\ell}$ are simultaneously present in $\mathcal{L}_{\rm mix}$. 
The symmetries in the quark ($\mathcal{G}_Q$) and lepton ($\mathcal{G}_L$) sectors are not independent 
due to the presence of the underlying $SU(4)$ gauge symmetry which locks together the transformations 
of the $Q$ and $L$ fields. The intersection of the two groups yields
\begin{equation}
\mathcal{G}_{Q} \cap \mathcal{G}_{L} 
\supseteq U(1)_{q'_1+\ell'_1+\Psi_1}  \times U(1)_{q'_2+\ell'_2+\Psi_2} \times U(1)_{q'_3+\ell'_3+\Psi_3} 
\xrightarrow[]{W \neq \mathbb{1}} U(1)_{q'_1+\ell'_1+\Psi_1} \times U(1)_{q'+\ell'+\Psi}  \, ,
\end{equation}
where the last step of breaking is a consequence 
of the specific structure of the $W$ matrix in \eq{eq:Yukawas} featuring only 3-2 mixing. 
The unbroken groups correspond to the quantum number of the first family of quarks and leptons, $U(1)_{q'_1+\ell'_1+\Psi_1}$, and to the total fermion number $U(1)_{q'+\ell'+\Psi}$, namely 
the simultaneous re-phasing of all the fermion fields in $\mathcal{L}_{\rm mix}$. 
The latter is nothing but $3 B' + L'$ (cf.~\Table{fieldcontent}), which in combination with 
with $T^{15}$ yields ordinary baryon and lepton number after $\mathcal{G}_{4321}$ breaking. 

To simplify our analysis even more we can set the coupling $\lambda^{\ell}_1$ to zero, 
thus implying a further enhancement of the symmetry: 
$U(1)_{q'_1+\ell'_1+\Psi_1} \to U(1)_{q'_1+\Psi_1} \times U(1)_{\ell'_1+\Psi_1}$
which forbids flavour violating transitions involving either down quark or electron fields.
On the other hand, we can still have a large mixing between the second and third family of quarks and leptons, 
whose misalignment is parametrised by the matrix $W$. 
Such an effect appears in the coupling of $U_\mu$ with quarks and leptons, 
in complete analogy with the flavour violation involving the $W^{\pm}$ boson and the quark doublet in the SM. 
Working e.g.~in the basis $\Psi_L = (Q'_L, L'_L)^T = (Q_L, W L_L)^T$,  
the interaction of $U_\mu$ with quarks and leptons can be readily extracted from the 
covariant derivative: 
\begin{equation}
i \overline{\Psi}_L \gamma^{\mu} D_{\mu} \Psi_L 
\supset  \frac{g_4}{\sqrt{2}} U_{\mu} \, \overline{Q}_L \gamma^{\mu}  
\left(
\begin{array}{ccc}
1 & 0 & 0 \\
0 & \cos \theta_{LQ} & \sin \theta_{LQ} \\
0 & -\sin \theta_{LQ} & \cos \theta_{LQ}
\end{array}
\right)
L_L 
\, .
\end{equation}
In the same way that the Cabibbo angle $\theta_C$ represents the misalignment between the up and down quarks 
of the first two families within an $SU(2)_L$ doublet, 
here $\theta_{LQ}$ represent the misalignment between the quark and lepton fields of the second and third generation 
within an $SU(4)$ quadruplet. 
Note, however, that the states $Q_L$ and $L_L$ have to be projected along the light SM mass eigenstates, 
since the breaking induced by $\langle \Omega_3 \rangle$ and $\langle \Omega_1 \rangle$ 
redirects part of the SM quark and lepton doublets into $\Psi_L$. The net effect is given by (cf.~\app{intmb})
\beq
\frac{g_4}{\sqrt{2}} \beta_{ij} U_{\mu} \, \overline{q}^i_L \gamma^{\mu} \ell^j_L \, ,
\end{equation}
where $\beta$ is a $3 \times 3$ matrix describing the flavour structure of the leptoquark interactions 
with the light SM mass eigenstates:
\beq
\beta = \textrm{diag}(s_{q_{12}}, s_{q_{12}}, s_{q_3}) 
\, W \, 
\textrm{diag}(0, s_{\ell_2}, s_{\ell_3}) 
= 
\left(
\begin{array}{ccc}
0 & 0 & 0 \\
0 & c_{\theta_{LQ}} s_{q_{12}}s_{\ell_2} & s_{\theta_{LQ}} s_{q_{12}}s_{\ell_3} \\
0 & -s_{\theta_{LQ}} s_{q_3} s_{\ell_2} & c_{\theta_{LQ}} s_{q_3}s_{\ell_3}
\end{array}
\right) \, .
\eeq
The definitions of the mixing angles in terms of the fundamental parameters of the Yukawa Lagrangian 
are given in \app{app:rot_massbasis}. 

A crucial aspect that breaks the analogy with the SM is however the following: 
while the global symmetries in the Yukawa sector of the SM are accidental, in our phenomenological limit the symmetry groups 
$\mathcal{G}_Q$, $\mathcal{G}_L$ and their relative orientation parametrised by 
$W$ have been assumed. 
This clearly calls for a UV understanding in terms of some flavour dynamics above the scale of $\mathcal{G}_{4321}$ breaking.  
On the other hand, since the symmetries that we imposed for phenomenological reasons 
are nothing but a generalisation of the accidental and approximate symmetries already present in the SM, 
the possibility to create a link between the flavour structure of the SM and $\mathcal{G}_{Q,L}$ is 
well motivated, and proposals such as those in Refs.~\cite{Bordone:2017bld,Bordone:2018nbg} might play a role in achieving this goal. 
It appears instead more difficult to provide flavour dynamics responsible for the misalignment induced by $W$, 
since a large 3-2 misalignment points to flavour-breaking spurions beyond those of the SM Yukawas.  
This notwithstanding, our phenomenological limit turns out to be robust against higher-order effects and is not tuned. 
It also allows us to identify the most important observables and understand suppressions or enhancements directly in terms of the 
symmetries of the fundamental Lagrangian. 
Another difference with respect to the SM is the presence of radial modes contained 
in the scalar fields $\Omega_i$ which can mediate flavour violation beyond that induced by the massive vectors. 
It can be shown, however, (see \sect{sec:lowen}) that flavour violating effects mediated by the radial modes are phenomenologically under control. 

{\footnotesize
\begin{table}[htp]
\begin{center}
\begin{tabular}
{|p{6cm}|p{6cm}|}
\hline
$\mathcal{G}_{321}$ & $\mathcal{G}_{4321}$ \\
\hline
$\theta_{C}$ & $\theta_{LQ}$ \\
$V$ & $W$ \\
$W_{\mu}$ & $U_{\mu}$ \\
$q_L = \left(
\begin{array}{c}
u_L \\
V d_L 
\end{array}
\right)$ &
$\Psi_L = \left(
\begin{array}{c}
Q_L \\
W L_L 
\end{array}
\right)$ \\
$Y_u$, $Y_d$ & $\lambda_q$, $\lambda_{\ell}$ \\
$SU(2)_L$ & $SU(4)$ \\
$U(1)_u \times U(1)_c \times U(1)_t$ & $U(2)_{q'+\Psi} \times U(1)_{q'_3+\Psi_3}$ \\
$U(1)_d \times U(1)_s \times U(1)_b$ & $U(1)_{\ell'_1+\tilde{\Psi}_{1}} \times U(1)_{\ell'_2+\tilde{\Psi}_{2}} \times U(1)_{\ell'_3+\tilde{\Psi}_{3}}$ \\
$U(1)_B$ & $U(1)_{q'_1+\ell'_1+\Psi_1} \times U(1)_{q'+\ell'+\Psi}$ \\
$u \to d \, \textrm{ tree level}$ & $Q \to L \,  \textrm{ tree level}$ \\
$u_i \to u_j \, \textrm{ loop level}$ & $Q_i \to Q_j \, \textrm{ loop level}$ \\
$d_i \to d_j \, \textrm{ loop level}$ & $L_i \to L_j \, \textrm{ loop level}$ \\
\hline
\end{tabular}
\end{center}
\caption{\sf 
Analogies between the SM and the 4321 model.
}
\label{SManalogy}
\end{table}
}

We conclude this section by summarizing the main features of the \emph{Cabibbo mechanism for leptoquarks} 
advocated above (cf.~also \Table{SManalogy} for a SM analogy): 
\begin{itemize}
\item We have found a mechanism that allows for large flavour violation in semi-leptonic decays in the $3$-$2$ sector, 
as required by the flavour anomalies. 
\item Tree-level FCNC involving down quarks and charged leptons are absent. 
\item Tree-level FCNC in the up sector are protected by the small $U(2)_{q'}$ breaking of the SM Yukawas.
\item FCNC not protected by the $U(2)_{q'}$ symmetry (both in up and down sectors) are induced at one loop. 
While flavour changing processes involving electrons and down quarks are forbidden,  
the leptoquark contributes at one loop to $B_s$ and $D$ mixing, 
as well as lepton flavour violating (LFV) processes such as $\tau \to \mu \gamma$ and other EW observables. 
In \sect{sec:lowen} we show that these bounds can be satisfied, also thanks to an 
extra dynamical GIM-like suppression provided by the lepton partners running in the loop. 
One-loop effects due to the exchange of the coloron, $Z'$ and scalar radial modes are also under control.  
\item 
We can now match the UV-complete 4321 model with the simplified-model analysis performed in \cite{Buttazzo:2017ixm}. 
Most importantly, since the theory is fully calculable, we are also able to provide precise predictions in $\Delta F=2$ and LFV observables. 
\end{itemize}
All these aspects will be addressed in a quantitative way in the next section.


\section{Low-energy phenomenology}
\label{sec:lowen}

\begin{figure}[t]
\centering
\includegraphics[width=\textwidth]{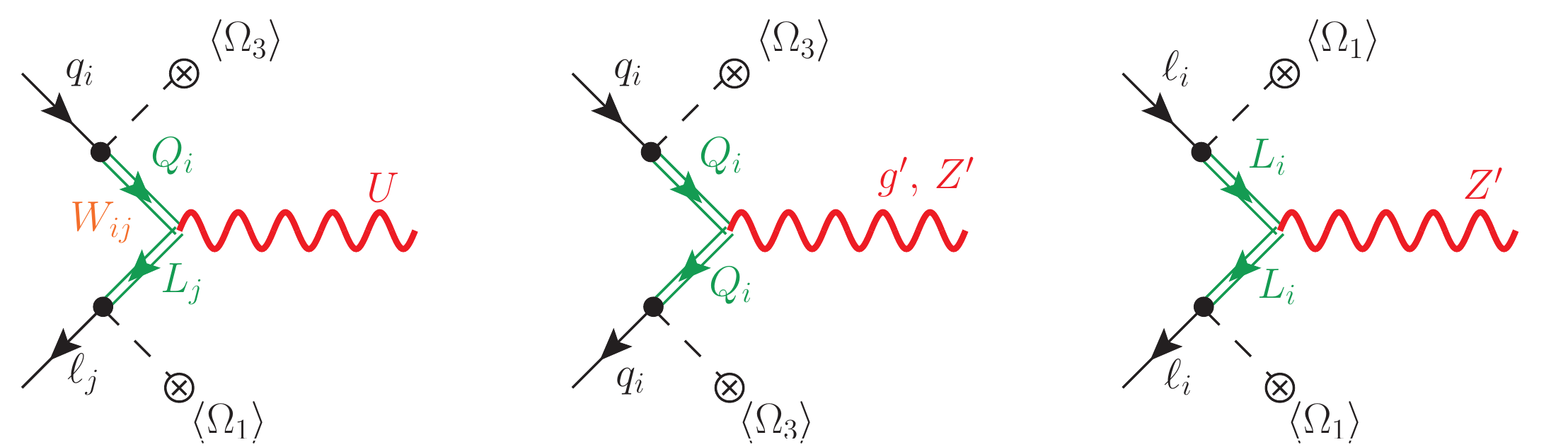}
\caption{\sf Interactions of the SM fermions with the heavy vectors induced by the fermion mixing. \label{fig:mixing}}
\end{figure}

The scope of this section is to discuss the main low-energy observables of the 4321 model, 
together with the relevant constraints coming from electroweak precision tests and FCNC.
Let us start by outlining the main interactions of the new vectors with the SM fermions, 
described in terms of mixing angles between the would-be SM fermions and their vector-like partners. 
The flavour structure of our model, defined by our assumptions in Eq.~\eqref{lamqfs}, is such that (up to CKM rotations) each SM family mixes with \textit{only one} fermion partner, see Fig.~\ref{fig:mixing} for illustration. 
The only non-trivial source of flavour breaking is found in the $W$ matrix, introduced in the previous section, which 
is responsible for a misalignment between quarks and leptons in the leptoquark interactions. The resulting vector leptoquark interactions with SM fermions closely follow those introduced in~\cite{Buttazzo:2017ixm}, 
which were shown to provide a successful explanation of the $b\to s\ell\ell$ and $R(D^{(*)})$ anomalies. We write these interactions in the mass basis in a similar fashion\footnote{In this section we show only the interactions of the new gauge bosons with the SM fermions for illustration. Full expressions, including also the couplings to vector-like fermions, can be found in \app{intmb}.}
\begin{align}\label{eq:Ucouplings}
\mathcal{L}_U&\supset\frac{g_4}{\sqrt{2}}\,U_\mu\left[\beta_{ij}\,\bar q^i\gamma^\mu \ell^j+h.c.\right]\,,
\end{align}
with
\begin{align}\label{eq:beta}
\beta&= 
\begin{pmatrix}
s_{q_1}\,s_{\ell_1} & 0 & 0\\
0 & c_{\theta_{LQ}}\,s_{q_2}\,s_{\ell_2} & s_{\theta_{LQ}}\,s_{q_2}\,s_{\ell_3}\\
0 & -s_{\theta_{LQ}}\,s_{q_3}\,s_{\ell_2} & c_{\theta_{LQ}}\,s_{q_3}\,s_{\ell_3}
\end{pmatrix}\,,
&
q^i&=
\begin{pmatrix}
V^*_{ji}\,u_L^j\\
d_L^i
\end{pmatrix}\,,
\,
&
\ell^i&=
\begin{pmatrix}
\nu_L^i\\
e_L^i
\end{pmatrix}\,.
\end{align}
and $V$ the CKM matrix. The interactions of these new gauge bosons with SM fermions read
\begin{align}
\begin{aligned}
\mathcal{L}_{g^\prime}&\supset g_s\,\frac{g_4}{g_3}\,g^{\prime a}_\mu\left[\kappa_q^{ij}\,\bar q^i\gamma^\mu T^a q^j+\kappa_u^{ij}\,\bar u_R^i\gamma^\mu T^a u_R^j+\kappa_d^{ij}\,\bar d_R^i\gamma^\mu T^a d_R^j\right]\,,\\
\mathcal{L}_{Z^\prime}&\supset \frac{g_Y}{2\sqrt{6}}\,\frac{g_4}{g_1}\,Z^\prime_\mu\left[\xi_q^{ij}\,\bar q^i\gamma^\mu q^j+\xi_u^{ij}\,\bar u_R^i\gamma^\mu u_R^j+\xi_d^{ij}\,\bar d_R^i\gamma^\mu d_R^j-3\,\xi_\ell^{ij}\,\bar \ell^i\gamma^\mu \ell^j-3\,\xi_e^{ij}\,\bar e_R^i\gamma^\mu e_R^j\right]\,,
\end{aligned}
\end{align}
with
\begin{align}\label{eq:kappaxi}
\begin{aligned}
\kappa_q&\approx
\begin{pmatrix}
s_{q_1}^2 & 0 & 0\\
0 & s_{q_2}^2 & 0\\
0 & 0 & s_{q_3}^2\\
\end{pmatrix}
-\frac{g_3^2}{g_4^2}\,\mathbb{1}\,,&\qquad&& \kappa_u\approx\kappa_d&\approx-\frac{g_3^2}{g_4^2}\,\mathbb{1}\,,\\
\xi_q&\approx
\begin{pmatrix}
s_{q_1}^2 & 0 & 0\\
0 & s_{q_2}^2 & 0\\
0 & 0 & s_{q_3}^2\\
\end{pmatrix}
-\frac{2\,g_1^2}{3\,g_4^2}\,\mathbb{1}\,,&&& \xi_u\approx\xi_d&\approx-\frac{2\,g_1^2}{3\,g_4^2}\,\mathbb{1}\,,\\
\xi_\ell&\approx
\begin{pmatrix}
s_{\ell_1}^2 & 0 & 0\\
0 & s_{\ell_2}^2 & 0\\
0 & 0 & s_{\ell_3}^2\\
\end{pmatrix}
-\frac{2\,g_1^2}{3\,g_4^2}\,\mathbb{1}\,,&&& \xi_e&\approx-\frac{2\,g_1^2}{3\,g_4^2}\,\mathbb{1}\,.
\end{aligned}
\end{align}
Note that the $W$ mixing matrix cancels by unitary in the neutral current sector and hence it does not enter in the $Z^\prime$ and $g^\prime$ interactions. This situation is completely analogous to the SM, in which the CKM cancels in the $\gamma$ and $Z$ interactions. Also note that the assumed \textit{down-aligned} flavour structure implies no tree-level FCNC in the down-quark and charged-lepton sectors mediated by these extra gauge bosons. In the case where $s_{q_1}\neq s_{q_2}\neq s_{q_3}$, FCNC in the up sector proportional to the CKM matrix elements are induced. These transitions yield potentially dangerous contributions in $\Delta C=2$ observables. Assuming $\theta_{q_1}=\theta_{q_2}\equiv\theta_{q_{12}}$ ensures an additional $U(2)$-like protection of the FCNC in the up sector. As we show in Sec.~\ref{sec:Dmix}, this extra protection plays a crucial role in keeping the effects in $D-\bar D$ mixing under control. An even larger protection against FCNCs can be achieved when $\theta_{q_{12}}=\theta_{q_3}$, which we denote as \textit{full-alignment limit}. In this limit the flavour matrices in Eq.~\eqref{eq:kappaxi} become proportional to the identity, yielding, as with the $W$ matrix, a unitarity cancellation of the CKM matrix in the up sector and thus resulting in a complete absence of tree-level FCNC mediated by the $g^\prime$ and the $Z^\prime$. As we show in Secs.~and \ref{sec:Dmix} and \ref{colliderconstr}, this latter limit is disfavoured by low-energy and high-$p_T$ data.

The relevant low-energy phenomenology of the model is described in terms of the fermion mixing angles: $\theta_{q_i}$ and  $\theta_{\ell_i}$, the $W$ matrix, the ratios of fermion masses to the leptoquark mass, and the following combinations of gauge couplings and vector masses
\begin{align}\label{eq:Cs}
C_U&=\frac{g_4^2v^2}{4M_U^2}\,, & C_{Z^\prime}&=\frac{g_Y^2}{24\,g_1^2}\,\frac{g_4^2v^2}{4\,M_{Z^\prime}^2}\,, & C_{g^\prime}&=\frac{g_s^2}{g_3^2}\,\frac{g_4^2v^2}{4\,M_{G^\prime}^2}\,,
\end{align}
which measure the strength of the new gauge boson interactions relative to the weak interactions.  
In the limit $g_4\gg g_{1,3}$, in which we are working, we have $g_Y\approx g_1$ and $g_s\approx g_3$. Moreover, in the phenomenological limit $v_3\gg v_1\gg v_{15}$, the following approximate relation among vector masses holds (see \app{gbspec}):
\begin{align}
M_{g'} : M_U : M_{Z'} \approx \sqrt{2} : 1 : \frac{1}{\sqrt{2}}~,
\end{align}
while for the NP scale constants we find:
\begin{align}
C_{g'} : C_U : C_{Z'} \approx \frac{1}{2} : 1 : \frac{1}{12}~.
\end{align}
In what follows, we describe the main low-energy constraints on these model parameters.

\subsection{Constraints on fermion mixing}
\label{sec:FermConst}

The fermion mass mixing induced by Eqs.~\eqref{LYUK1} and \eqref{LYUK2} is the essential ingredient in our construction. While the full fermion mass diagonalization is discussed in~\app{app:rot_massbasis}, here we give a simplified discussion and comment on the main constraints on the mixing angles. To a good approximation, this mixing is such that each family of the SM fermions mixes with a single vector-like family. We introduce the following notation,
\begin{align}\label{eq:epsx}
\epsilon^i_{u(d)} = \frac{Y_{u(d)}^i\,v}{\sqrt{2} \hat M_Q}~, \qquad x_{q_i} = ~\frac{\lambda_q^i\,v_3}{\sqrt{2} \hat M_Q}~,
\end{align}
where $i = 1,2,3$ is the family index, and analogously for the lepton sector. The quark and lepton mixing angles, expanded in small $\epsilon^i_x$, are given by
\begin{align}\label{eq:angles}
\tan \theta_{q_i,\ell_i} \approx x_{q_i,\ell_i}~, \qquad \tan \theta_{u^i_R (d^i_R)} \approx \frac{m_{u^i(d^i)}}{m_{U^i(D^i)}}\,\tan \theta_{q_i}\,,  \qquad \tan \theta_{e^i_R} \approx \frac{m_{e^i}}{m_{E^i}}\,\tan \theta_{\ell_i}~.
\end{align}
The physical masses are instead given by
\begin{align}\label{eq:maseteskih}
\begin{aligned}
m_{u_i(d_i)} &\approx \frac{Y_{u(d)}^i\,v}{\sqrt{2}} \cos \theta_{q_i}~, &\qquad&& M_{U_i(D_i)} &\approx \frac{\hat M_Q}{\cos \theta_{q_i}} ~,\\
m_{e_i} &\approx \frac{Y_e^i\,v}{\sqrt{2}} \cos \theta_{\ell_i}~, &&& M_{E_i(N_i)} &\approx \frac{\hat M_L}{\cos \theta_{\ell_i}} ~.
\end{aligned}
\end{align}
Note that large left-handed mixing angles of the third generation quarks and leptons are required by the $R(D^{(*)})$ 
anomaly (cf.~Eq.~\eqref{eq:RDscaling}). There are a few subtleties regarding the top quark mixing due to its large mass. After electroweak symmetry breaking, contributions to electroweak precision tests are generated, setting important limits on the right-handed top mixing. In particular, $Z \to b_L \bar b_L$ decay and the $\rho$ parameter, both induced at one-loop, set upper limits of $\tan \theta_{u^3_R} \lesssim 0.4$ and $\tan \theta_{u^3_R} \lesssim 0.15$, respectively (for more details see Ref.~\cite{Fajfer:2013wca}). As a consequence, the two charged components of the doublet are almost degenerate ($M_T \approx M_B$) since the relative mass difference, $M_{U_i}/M_{D_i} - 1 \sim \frac{1}{2}( \tan^2 \theta_{u^i_R}- \tan^2 \theta_{d^i_R})$.  In addition, setting $\sin \theta_{q_3} = 0.8$, the second relation in Eq.~\eqref{eq:angles} implies a lower limit $M_T \gtrsim 1.7$~TeV. 
The maximal size of the mixing angles is also limited by the perturbativity of the Yukawa couplings (cf.~\sect{sect:pert}). 
For example, setting $\sin \theta_{q_3}=0.8$ implies $y_t \approx 1.7$ (see \eq{eq:maseteskih}).
Similarly, large values for $\lambda^3_q$ and $\lambda^3_\ell$ are also required to keep these angles maximal.

\subsection{Semileptonic processes}

A key element of the Cabibbo mechanism introduced in \sect{sec:cabibbo} is that NP effects in flavour-violating semileptonic transitions are expected to be maximal. In particular, the relative misalignment in flavour space between quark and leptons, parametrised by the $W$ matrix, is responsible for sizeable 3-2 transitions mediated by the leptoquark. In what follows, we describe the main NP effects in this sector, paying particular attention to the anomalies in $b \to c \tau \nu$ and $b \to s \mu \mu$ transitions. 

\subsubsection{Charged currents}
Current measurements of the $R(D^{(*)})=\mathcal{B}(B\to D^{(*)}\tau\nu)/\mathcal{B}(B\to D^{(*)}\ell\nu)$ $(\ell=e,\mu)$ ratios performed by BaBar~\cite{Lees:2012xj,Lees:2013uzd}, Belle~\cite{Huschle:2015rga,Sato:2016svk,Hirose:2016wfn,Hirose:2017dxl} and LHCb~\cite{Aaij:2015yra,Aaij:2017uff,Aaij:2017deq} point to a large deviation away from
lepton flavour universality (LFU). We define possible NP contributions to these LFU ratios as 
\begin{align}
\Delta R_{D^{(*)}}=\frac{R(D^{(*)})_{\rm exp}}{R(D^{(*)})_{\rm SM}}-1\,.
\end{align}
Effects in these observables are induced in our model by the tree-level exchange of $U_\mu$. Since $U_\mu$ only couples to SM fermions of left-handed chirality (see \eq{eq:Ucouplings}), the NP effect has the same structure as the SM one mediated by the $W_\mu^{\pm}$. As a result, our model predicts the same NP contributions to $R(D)$ and $R(D^*)$, compatible with current experimental data. Using the HFLAV experimental average for the $R(D^{(*)})$ ratios~\cite{Amhis:2016xyh} (summer 2018), taking the arithmetic average of latest SM predictions for these observables~\cite{Bigi:2016mdz,Bernlochner:2017jka,Bigi:2017jbd,Jaiswal:2017rve}, and assuming $\Delta R_D=\Delta R_{D^*}$ (as predicted by the model) we find: $\Delta R_{D^{(*)}}=0.217\pm0.053$. The model contribution 
to this observable (taking only the leading interference contribution) reads 
\begin{align}
\Delta R_{D^{(*)}}\approx2\,C_U\,\beta_{b\tau}\left(\beta_{b\tau}+\beta_{s\tau}\,\frac{V_{cs}}{V_{cb}}\right)\,.
\end{align}
In the limit $\beta_{b\tau}|V_{ts}|\ll\beta_{s\tau}$, we can neglect the first term in the equation above. This allows us to derive the following approximate 
expression 
\begin{align}\label{eq:RDscaling}
\Delta R_{D^{(*)}} \approx 0.2 \left (\frac{2~\textrm{TeV}}{M_U} \right)^{2} \left(\frac{g_4}{3.5} \right)^{2} \sin (2 \theta_{LQ}) \left(\frac{s_{\ell_3}}{0.8}\right)^2 \left(\frac{s_{q_3}}{0.8}\right) \left(\frac{s_{q_2}}{0.3}\right)~, 
\end{align}
which is helpful in order to understand the parametric dependence:  
a successful explanation of the $R(D^{(*)})$ anomaly requires large mixing angles with third-generation fermions. Moreover, setting $\theta_{LQ}=\pi/4$ and the third family mixing and $g_4$ nearly 
to the maximum value compatible with perturbativity, the NP contribution to $R(D^{(*)})$ is fixed in terms of $s_{q_2}$ and the NP scale, see blue contours in Fig.~\ref{fig:bsmix}. Interesting constraints on the value of $s_{q_2}$ arise from $\Delta F=2$ observables and high-$p_T$ searches for vector-like partners, which are addressed 
respectively in Secs.~\ref{sec:Dmix} and~\ref{colliderconstr}.

An interesting remark is that the $W$ matrix introduces an additional source of $U(2)_{q}$ breaking other than that 
discussed in~\cite{Buttazzo:2017ixm}. As a result, 
we predict a NP enhancement that is different for $b\to c$ and $b\to u$ transitions. In particular, we have that 
\begin{align}
\Delta_{b\to c}: \Delta_{b\to u}=\left(\beta_{b\tau}+\beta_{s\tau}\frac{V_{cs}}{V_{cb}}\right):\left(\beta_{b\tau}+\beta_{s\tau}\frac{V_{us}}{V_{ub}}\right)\,,
\end{align}
where $\Delta_{b\to c (u)}$ is the ratio of the NP amplitude over the SM one for the $b \to c (u) \tau \nu$ transitions.
The previous formula, in the phenomenological limit of the parameters that we are considering ($\beta_{b\tau}\,|V_{cb}|\ll\beta_{s\tau}$), reduces to
\begin{align}
\label{Deltas}
\Delta_{b\to c}: \Delta_{b\to u}\approx 1: \frac{V_{us}V_{cb}}{V_{ub}V_{cs}}\approx 1: 1+2.5\,i\,. 
\end{align}
While the real part of the NP contribution shows the same universal enhancement, 
relatively large non-interfering effects in $b\to u$ transitions are predicted in our model, which could allow differentiation of this solution from the one in~\cite{Buttazzo:2017ixm}. 
So far, the only experimental measurement of $b\to u\tau\nu$ transitions is $\mathcal{B}(B\to\tau\nu)$. The modification of $\mathcal{B}(B\to\tau\nu)$ compared to the SM is dictated by $\Delta_{b\to u}$, while $\Delta_{b\to c}$ is fixed by $\Delta{R_{D^{(*)}}}$. Using Eq.(\ref{Deltas}) we can derive the following prediction:
\begin{equation}
\Delta \mathcal{B} (B \to \tau \nu) = \frac{\mathcal{B} (B \to \tau \nu)_{\rm exp}}{\mathcal{B} (B \to \tau \nu)_{\rm SM}}-1 = \Delta R_{D^{(*)}} + 1.8 \, \left(\Delta R_{D^{(*)}} \right)^2 \, , 
\end{equation}
which for $\Delta R_{D^{(*)}} = 0.22$ yields $\Delta \mathcal{B} (B \to \tau \nu) = 0.31$. 
Remarkably, using the PDG value~\cite{Patrignani:2016xqp} for the experimental input and the UTFit 
value \cite{UTfit2016WEB} for the SM prediction, we have $\Delta \mathcal{B}(B\to\tau\nu)=0.35\pm0.31$, 
which supports the model prediction, 
but still has a very large error. 
Future improvements of the sensitivity could be used to test this prediction 
and possibly discriminate among different sources of $U(2)_{q}$ breaking.

\subsubsection{Neutral currents} 

\begin{figure}[ht]
\begin{minipage}{6in}
  \centering
  $\vcenter{\hbox{\includegraphics[width=0.35\textwidth]{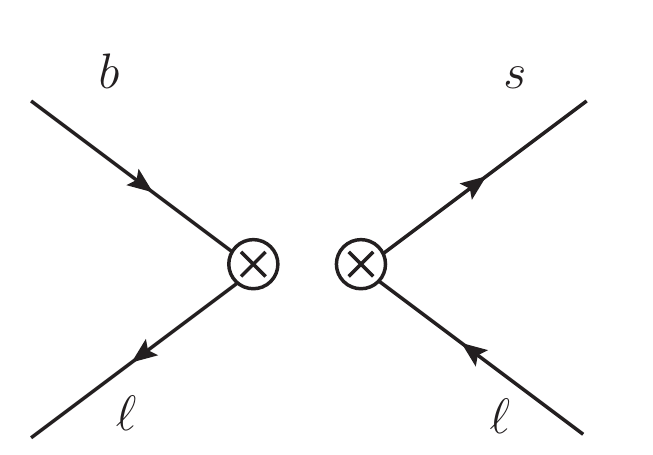}}}$
  \hspace*{40pt}
  $\vcenter{\hbox{\includegraphics[width=0.3\textwidth]{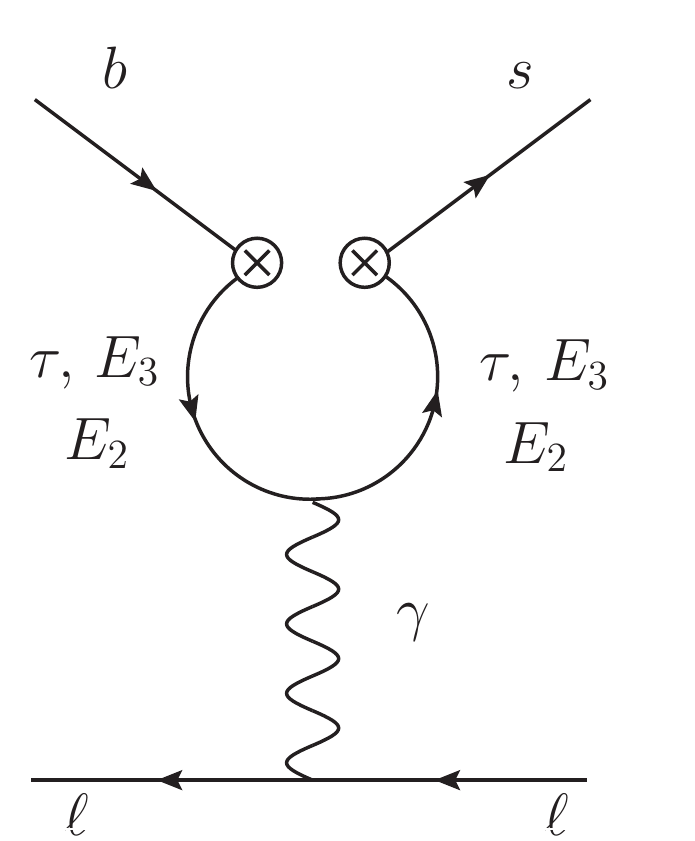}}}$
\end{minipage}
  \vspace{10pt}
\caption{\sf EFT diagrams contributing to $b\to s\ell\ell$ transitions after $U_\mu$ has been integrated out. \label{fig:bsll}}
\end{figure}

The effective Hamiltonian describing $b \rightarrow s \ell \ell$ transitions reads
\begin{align}
\mathcal{H}_{\mbox{\scriptsize eff}}=   - \frac{4G_F}{\sqrt{2}}\,\frac{\alpha}{4\pi}\,  V_{tb}V_{ts}^*  
\sum_{i}   \left( C_i^{\ell} \mathcal{O}^{\ell}_i  +  C_i^{\prime \ell} \mathcal{O}^{\prime \ell}_i 
\right),
\end{align}
with
\begin{align}\label{eq:WCWET}
\mathcal{O}^{\ell}_9&=\left(\overline{s}\gamma_\mu P_Lb\right)\left(\overline{\ell}\gamma^\mu
 \ell\right)\,,    & \mathcal{O}^{\ell}_{10}&= \left(\overline{s}\gamma_\mu P_Lb\right)\left(\overline{\ell}\gamma^\mu\gamma_5 \ell\right)\,,
\end{align}
where we ignore the scalar and chirality-flipped (primed) operators which receive negligible contributions in our model and are thus irrelevant for the present discussion. Due to the assumed down-aligned flavoured structure, only the leptoquark mediates tree-level contributions to $b\to s\ell\ell$ transitions. Since the leptoquark only couples to left-handed SM fields, NP contributions to these transitions are of the form (we define $C_i=C_i^{\rm SM}+\Delta C_i$)
\begin{align}\label{eq:C9tree}
\left.\Delta C_9^{\mu\mu}\right|_{\rm tree}=-\left.\Delta C_{10}^{\mu\mu}\right|_{\rm tree}=-\frac{2\pi}{\alpha\,V_{tb}V_{ts}^*}\,C_U\,\beta_{s\mu}\,\beta_{b\mu}^*\,.
\end{align}
As recently put forward in~\cite{Crivellin:2018yvo}, given the large values of the $\beta_{s\tau}$ leptoquark coupling required in our setup to explain the $R(D^{(*)})$ anomaly, 
one-loop log-enhanced contributions to these Wilson coefficients at the scale of the bottom mass can be sizeable. The most relevant of such contributions is given by a photon penguin with a $\tau$ in the loop, see \fig{fig:bsll}.

This yields a contribution only to $C_9$ that is universal for all leptons. We find ($\ell=e,\mu,\tau$)
\begin{align}\label{eq:C9LD}
\begin{aligned}
\left.\Delta C_9^{\ell\ell}\right|_{\rm loop}(m_b^2)&\approx-\frac{1}{V_{tb}V_{ts}^*}\,\frac{2}{3}\,C_U\left(\beta_{s\tau}\,\beta_{b\tau}^*\,\log x_b+\beta_{sE_2}\,\beta_{bE_2}^*\,\log x_{E_2}+\beta_{sE_3}\,\beta_{bE_3}^*\,\log x_{E_3}\right)\\
&\approx-\frac{1}{V_{tb}V_{ts}^*}\,\frac{2}{3}\,C_U\,\frac{1}{2}\sin2\theta_{LQ}\,s_{q_2}\,s_{q_3}\left(s_\tau^2\,\log x_b+c_\tau^2\log x_{E_3}-\log x_{E_2}\right)\,,
\end{aligned}
\end{align}
with $x_\alpha=m_\alpha^2/M_U^2$. In the computation we neglected fermion mixing in the muon sector (i.e.~we took $s_{\ell_2}\approx0$), which amounts to a very small correction. Our result is in agreement with the one in~\cite{Crivellin:2018yvo}, however in our setup we also have computed the contributions involving the vector-like leptons $E_{2,3}$.

The violations of LFU measured in the $R_K$~\cite{Aaij:2014ora} and $R_{K^*}$~\cite{Aaij:2017vbb} ratios (as well as $Q_{4,5} = P^{\mu\prime}_{4,5} - P^{e\prime}_{4,5}$~\cite{Wehle:2016yoi}) fix the (non-universal) tree-level contribution. Combining the experimental measurements of LFU observables in $b\to s\ell\ell$ transitions yields the following preferred region for the non-universal NP effect~\cite{Capdevila:2017bsm} (see also \cite{Ciuchini:2017mik,Altmannshofer:2017yso,Geng:2017svp,DAmico:2017mtc,Alok:2017sui,Hiller:2017bzc})
\begin{align}
\left.\Delta C_9^{\mu\mu}\right|_{\rm tree}=-\left.\Delta C_{10}^{\mu\mu}\right|_{\rm tree}=-0.66\pm0.18\,.
\end{align}
This  value can be perfectly accommodated by fixing $s_{\ell_2}$ in terms of the remainaing parameters. 
Taking typical values for the other model parameters, we find that $s_{\ell_2}\sim\mathcal{O}(0.1)$ is required in order to fit the $R_{K^{(*)}}$ anomaly.

Concerning the one-loop contribution, we can connect its value to the NP shift in $\Delta R_{D^{(*)}}$ in the $\beta_{b\tau}|V_{ts}|\ll\beta_{s\tau}$ limit. In this limit the following approximate relation holds
\begin{align}
\left.\Delta C_9^{\ell\ell}\right|_{\rm loop}(m_b^2)\approx\frac{1}{3}\,\Delta R_{D^{(*)}}\left(\log x_b-\frac{1}{s_\tau^2}\,\log x_{E_2}\right)\,.
\end{align}
For $s_\tau=0.8$, $M_U=2.5$~TeV, $M_E=850$~GeV and $\Delta R_{D^{(*)}}=0.1$, we find $\left.\Delta C_9^{\ell\ell}\right|_{\rm loop}(m_b^2)\approx-0.3$. The presence of this universal contribution predicts a further enhancement of 
$P_5^\prime$ \cite{DescotesGenon:2012zf}, 
beyond the one given by the tree-level effect. As shown in~\cite{Crivellin:2018yvo}, this prediction is in good agreement with current data. 

The assumed flavour structure also implies large NP effects mediated by the leptoquark in $b\to s\tau\tau$ transitions, 
\begin{align}
\Delta C_9^{\tau\tau}=-\Delta C_{10}^{\tau\tau}=-\frac{2\pi}{\alpha\,V_{tb}V_{ts}^*}\,C_U\,\beta_{s\tau}\,\beta_{b\tau}^*\,.
\end{align}
For typical values of the model parameters, this contribution is $\mathcal{O}(100)$ larger than the corresponding NP effect in the $\mu$ channel. Such large NP effects are compatible with current experimental data and provide an interesting smoking-gun signature that can be tested by future experiments such as Belle II, see e.g.~\cite{Alonso:2015sja,Crivellin:2017zlb,Buttazzo:2017ixm,Capdevila:2017iqn}.

Concerning NP contributions to $b\to s\nu\nu$, again here the flavour structure of the model forbids tree-level contributions mediated by the $Z^\prime$. Moreover, being an $SU(2)_L$ singlet, the leptoquark does not contribute to these transitions at 
tree level. The leading effects to these observables thus arise at one loop from $U$ and $W$ boxes and $U$ penguins with a tree-level $Z$ or $Z^\prime$. In contrast to the $b\to s\ell\ell$ case, the contributing penguin diagrams do not have large log-enhancements and/or are mass-suppressed, thanks to the additional suppression from the $Z$ mass. As a result, we find the model contributions to $b\to s\nu\nu$ to be well below the current experimental limits.

\subsubsection{Lepton Flavour Violating transitions}
The protection from our flavour structure (aligned to the charged-lepton sector), forbids tree-level lepton flavour violation (LFV) mediated by the $Z^\prime$. As a result, the dominant LFV effects are mediated by the leptoquark and hence they necessarily involve semileptonic processes for the tree-level effects (see \sect{sec:leptonic} for a discussion on one-loop induced LFV transitions). Further assuming $s_{\ell_1}=0$ implies no NP effects in the electron sector, and offers an additional protection from dangerously large LFV effects in the $\mu-e$ sector such as in $K_L\to\mu e$. Small departures from charged-lepton alignment and/or $s_{\ell_1}=0$ are possible. However, for simplicity, in the following discussion we only consider this limit (the possible departures are not connected to the anomalies and hence they are more model dependent). In this case, the leptoquark contributes to the following LFV transitions at tree-level:

\begin{itemize}
\item[i)] $\boldsymbol{\tau\to \mu ss.}$ This is the most promising LFV channel, since it is enhanced in the large $\beta_{s\tau}$ limit, of phenomenological interest for the $R(D^{(*)})$ anomaly. The most relevant observable involving this transition is $\tau\to\mu\,\phi$, for which we find the following expression for the branching fraction
\begin{align}
\mathcal{B}(\tau\to\mu\,\phi) &=\frac{1}{\Gamma_\tau}\frac{f_\phi^2\,m_\tau^3}{32\pi v^4}\,C_U^2\,|\beta_{s\mu}\,\beta_{s\tau}^*|^2\left(1-\eta_\phi\right)^2\left(1+2\eta_\phi\right)\,,
\end{align}
with $f_\phi\approx225$~MeV and $\eta_\phi\equiv m_\phi^2/m_\tau^2\approx0.33$. In the large $\beta_{s\tau}$ limit (or equivalently $|V_{cb}|\ll s_{q_2}$ and $\theta_{\rm LQ}\approx\pi/4$) , the following approximate expression holds
\begin{align}
\mathcal{B}(\tau\to\mu\,\phi) &\approx4.7\times10^{-9}\,\left(\frac{0.8}{s_{q_3}}\right)^2\left(\frac{s_{q_2}}{0.3}\right)^2\left(\frac{\Delta R_{K^{(*)}}}{0.3}\right)\left(\frac{\Delta R_{D^{(*)}}}{0.2}\right)\,.
\end{align}
This is to be compared with the current $90\%$~CL experimental limit by the Belle Collaboration~\cite{Miyazaki:2011xe}: $\mathcal{B}(\tau\to\mu\,\phi)<8.4\times10^{-8}$. Our model prediction is found to lie well below the current experimental sensitivity for the range of model parameters considered here. As emphasised in~\cite{Kumar:2018kmr},  bounds from this observable can arise in the very large $\beta_{s\tau}$ limit (i.e. $\beta_{b\tau}<\beta_{s\tau}$), which in our model yields to the following upper bound: $s_{q_2}\lesssim1.6\,s_{q_3}$. However such extreme values of $s_{q_2}$ are largely incompatible, in our model, with other low-energy observables as well as with direct searches (see discussion in Secs.~and \ref{sec:Dmix} and \ref{colliderconstr}) and hence are not considered.

\item[ii)] $\boldsymbol{b \to s\tau\mu.}$ These transitions are parametrised by (see Eq.~\eqref{eq:WCWET} for the definition of the Wilson coefficients)
\begin{align}
\begin{aligned}
C_9^{\tau\mu}=-C_{10}^{\tau\mu}=-\frac{2\pi}{\alpha\,V_{tb}V_{ts}^*}\,C_U\,\beta_{s\mu}\,\beta_{b\tau}^*\,,\qquad
C_9^{\mu\tau}=-C_{10}^{\mu\tau}=-\frac{2\pi}{\alpha\,V_{tb}V_{ts}^*}\,C_U\,\beta_{s\tau}\,\beta_{b\mu}^*\,.
\end{aligned}
\end{align}
Taking the explicit expression for $\beta$ in~\eq{eq:beta}, we find $C_i^{\tau\mu}=-\,C_i^{\mu\tau}~(i=9,\,10)$ when $\theta_{LQ}=\pi/4$. Using the expressions in~\cite{Crivellin:2015era} (see also~\cite{Becirevic:2016zri}), we derive the following limits in the large $\beta_{s\tau}$ limit, 
\begin{align}
\begin{aligned}
\mathcal{B}(B\to K\tau^\pm\mu^\mp) &\approx 2.0\,\times10^{-6}\left(\frac{\Delta R_{K^{(*)}}}{0.3}\right)\left(\frac{\Delta R_{D^(*)}}{0.2}\right)\,,\\[2pt]
\mathcal{B}(B\to K^*\tau^\pm\mu^\mp) &\approx 3.9\,\times10^{-6}\left(\frac{\Delta R_{K^{(*)}}}{0.3}\right)\left(\frac{\Delta R_{D^(*)}}{0.2}\right)\,.
\end{aligned}
\end{align} 
Experimental results are only available for the $K^+$ channel. 
The experimental limit at $90\%$~CL from the BaBar Collaboration reads~\cite{Lees:2012zz}: $\mathcal{B}(B^+\to K^+\tau^\pm\mu^\mp) < 4.8\times10^{-5}$.

Another interesting observable in this category is $\mathcal{B}(B_s\to\tau^\pm\mu^\mp)$, whose expression in terms of the Wilson coefficients reads
\begin{align}
\mathcal{B}(B_s\to \tau^\pm\mu^\mp) &=\frac{\tau_{B_s}\,m_\tau^2\,m_{B_s} f_{B_s}^2}{32\pi^3}\,\alpha^2G_F^2\left|V_{tb}V_{ts}^*\right|^2\left(1-\frac{m_\tau^2}{m_{B_s}^2}\right)^2\left(\left|\mathcal{C}_9^{\tau\mu}\right|^2 + \left|\mathcal{C}_{10}
^{\tau\mu}\right|^2 \right)\,.
\end{align}
Again, in the large $\beta_{s\tau}$ limit we can write the model prediction in terms of the NP effect in $R_{K^{(*)}}$ and $R(D^{(*)})$;
\begin{align}
\begin{aligned}
\mathcal{B}(B_s\to \tau^\pm\mu^\mp) &\approx 2\,\times10^{-6}\left(\frac{\Delta R_{K^{(*)}}}{0.3}\right)\left(\frac{\Delta R_{D^{(*)}}}{0.2}\right)\,.
\end{aligned}
\end{align}
However, no experimental measurement of this observable is currently available.

\item[iii)] $\boldsymbol{b\bar b\to \tau\mu.}$ In contrast to the case of $\tau\to\mu ss$ transitions, the transitions in this category are suppressed in the large $\beta_{s\tau}$  limit and are therefore less interesting. The only measured observables in this category are $\mathcal{B}(\Upsilon(nS)\to\tau\mu)$ ($n=1,2,3$)~\cite{Love:2008ys,Lees:2010jk}. The model predictions for these observables are found to lie far below the current experimental sensitivity.
\end{itemize}

\subsection{Hadronic processes}\label{sec:PhenoHad}
The most important constraints in this category arise from $\Delta F=2$ transitions. As anticipated in Sec.~\ref{sec:cabibbo}, the assumed flavour structure offers a protection from the stringent limits set on these transitions. In particular, the \textit{down-alignment} hypothesis implies no tree-level contributions to meson mixing observables in the down-quark sector mediated by the $Z^\prime$ and $g^\prime$.\footnote{As shown in~\cite{Bordone:2018nbg}, deviations from this hypothesis are possible, and could even be welcome, if we allow for CP violating couplings (see also~\cite{DiLuzio:2017fdq}). For simplicity we restrict ourselves here to the down-aligned scenario.} Furthermore, the $U(2)_q$ symmetry arising from setting $\theta_{q_1}=\theta_{q_2}\equiv\theta_{q_{12}}$ is enough to keep the tree-level contributions to $D-\bar D$ mixing under control. As a result we find that the dominant NP contribution to these observables arises from loops mediated by the 
leptoquark, and is proportional to the $W$ matrix. This has two important implications:
\begin{itemize}
\item[i)] The assumption that $W$ rotates only second- and third-generation fermion partners, 
required to maximise the NP contribution to $R(D^{(*)})$, 
implies no NP contributions to $B_d-\bar B_d$ or $K-\bar K$ mixing at one loop. 
\item[ii)] Unitarity of the $W$ matrix provides a GIM-like protection similar to that in the SM arising from CKM unitarity.
\end{itemize}
In what follows we detail the model contributions to $B_s$ and $D$ mixing.

\begin{figure}[t]
\centering
\includegraphics[width=\textwidth]{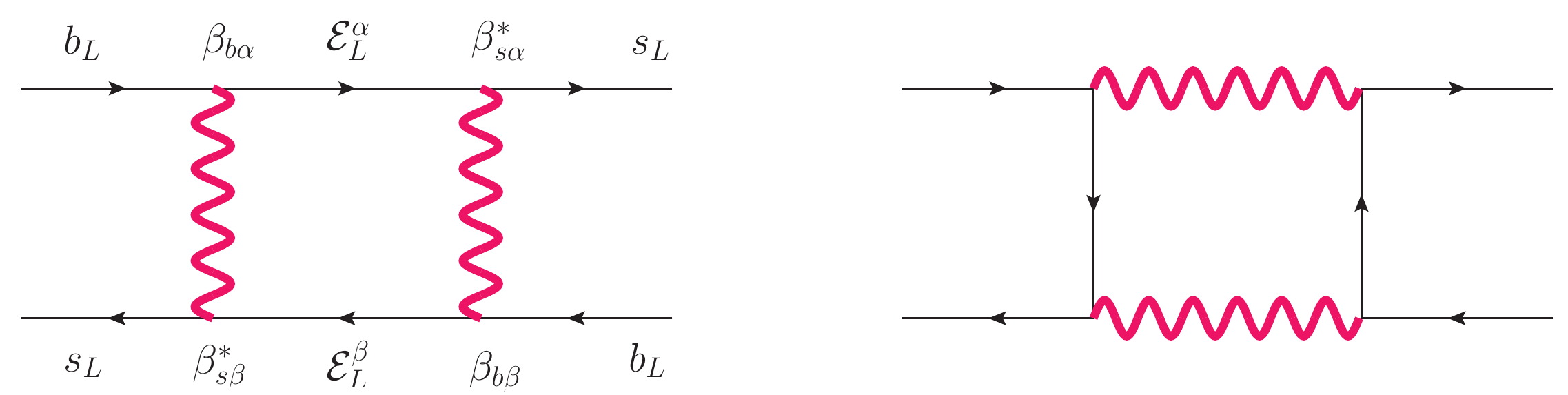}
\caption{\sf Leptoquark mediated one-loop diagrams contributing to $B_s-\bar B_s$ mixing. The symbol $\mathcal{E}$ denotes a six-dimensional vector containing SM charged-leptons and their partners.\label{fig:bsmix_dia}}
\end{figure}

\subsubsection{$B_s-\bar B_s$ mixing}
\begin{figure}[ht]
\centering
\includegraphics[width=0.5\textwidth]{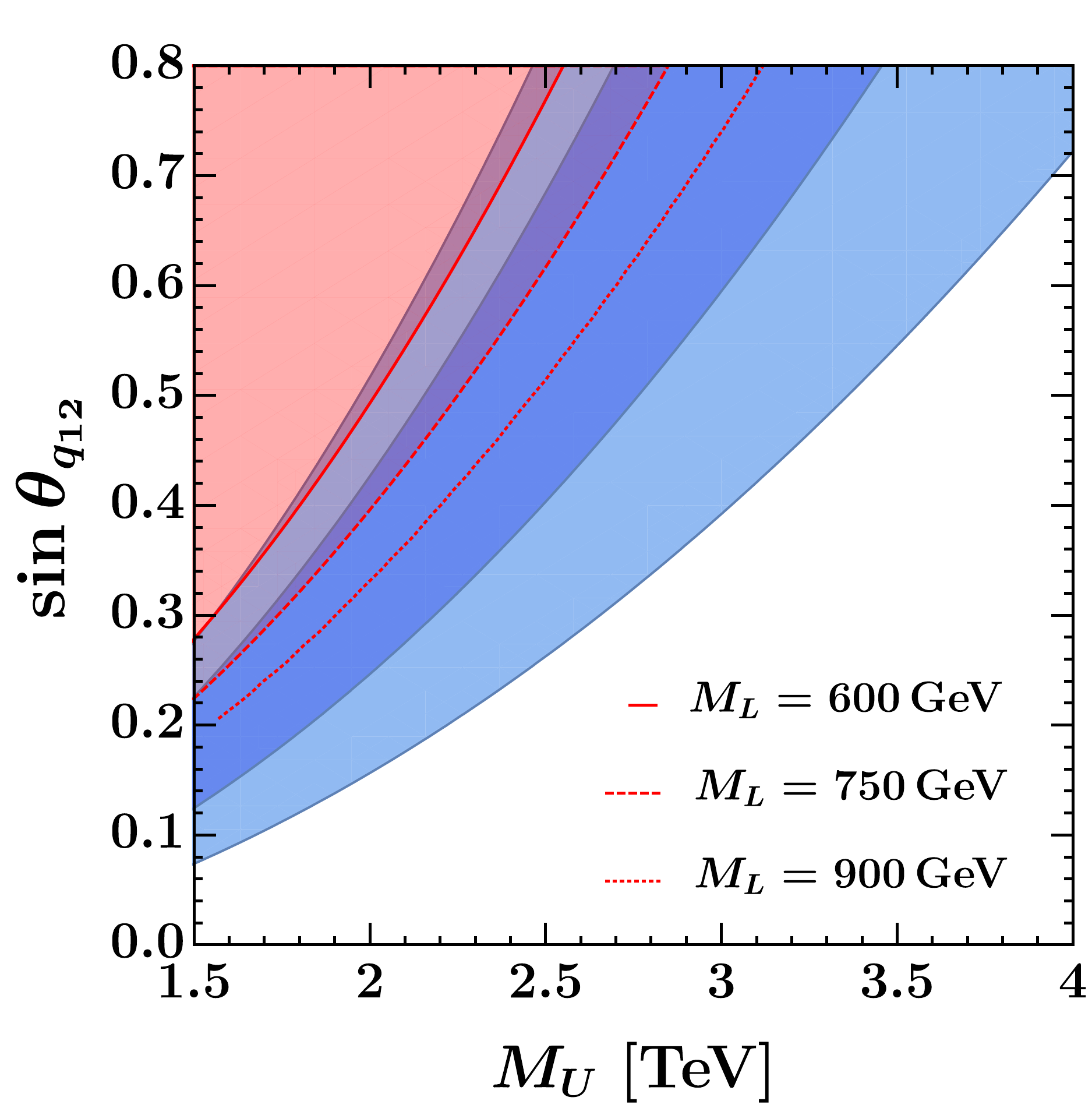}
\caption{\sf Constraints from $\Delta M_s$ at $95\%$~CL using the 2015 SM prediction \cite{Artuso:2015swg}, 
for different values of the vector-like lepton mass parameter $M_L$. The $1\sigma$ and $2\sigma$ preferred regions by the 
$R(D^{(*)})$ anomaly are shown in dark and light blue, respectively. We use as input for the model parameters: $g_4=3.5$, $s_{q_3}=0.8$, $s_{\ell_3}=0.8$, $v_3 = 1.75\, v_1$ and $\lambda_{15}=2.5$. \label{fig:bsmix}}
\end{figure}

The leading NP contribution to the mixing amplitude is given by the leptoquark box diagrams shown in Fig.~\ref{fig:bsmix_dia}. The resulting 
leptoquark contribution follows a very similar structure as that of the SM with a $W_\mu^{\pm}$ boson 
(see e.g.~\cite{Branco:1999fs}). Defining NP contributions to the $B_s$ meson-anti-meson mass difference, 
$\Delta M_s$, as $C_{bs}^{LL}\equiv\Delta M_s/\Delta M_s^{\rm SM}-1$, we find
\begin{align}\label{eq:Cbs}
C_{bs}^{LL}=-\frac{g_4^2}{64\pi^2}\,C_U\,\frac{1}{(V_{tb}V_{ts}^*)^2\,R_{\rm SM}^{\rm loop}}\sum_{\alpha,\beta}\lambda^B_\alpha\lambda^B_\beta\,F(x_\alpha,x_\beta)\,,
\end{align}
with $\alpha$ and $\beta$ running over all the leptons, including the vector-like partners, and where $R^{\rm loop}_{\rm SM} = \sqrt{2}\,G_F\,m^2_W\,\hat{\eta}_B\,S_0(x_t)/16 \pi^2 = 1.34 \times 10^{-3}$, with $S_0(x_t)\approx2.37$ being the Inami-Lim function~\cite{Inami:1980fz}. 
In this expression $F(x_\alpha, x_\beta)$ is a loop function defined as 
\begin{align}\label{eq:loop_func}
F(x_\alpha, x_\beta) &=  \frac{1}{(1-x_\alpha)(1-x_\beta)} \left( \frac{7 x_\alpha x_\beta}{4} - 1 \right) \nonumber \\ 
&+ \frac{x^2_\alpha \log x_\alpha}{(x_\beta - x_\alpha) (1-x_\alpha)^2} \left( 1 - 2 x_\beta + \frac{x_\alpha x_\beta}{4} \right) \nonumber \\ 
& + \frac{x^2_\beta \log x_\beta}{(x_\alpha - x_\beta) (1-x_\beta)^2} \left( 1 - 2 x_\alpha + \frac{x_\alpha x_\beta}{4} \right) 
\,,
\end{align}
with $x_\alpha=m_\alpha^2/M_U^2$ and $\lambda_\alpha^B=\beta_{b\alpha}\,\beta_{s\alpha}^*$, where $\beta$ denote 
the leptoquark couplings to left-handed fermions given in~\eq{eq:Beta}. The explicit form of $\lambda_\alpha^B$ in terms of fermion mixing angles reads
\begin{align}
\lambda_\alpha^B&=\frac{1}{2}\,\sin\,2\theta_{LQ}\sin\theta_{q_3}\sin\theta_{q_{12}}\left(\sin^2\theta_{\ell_3}\,\delta_{\alpha3}+\cos^2\theta_{\ell_3}\,\delta_{\alpha6}-\sin^2\theta_{\ell_2}\,\delta_{\alpha2}-\cos^2\theta_{\ell_2}\,\delta_{\alpha5}\right)\,.
\end{align}
Note that, analogously to the SM case, the flavour parameter $\lambda_\alpha^B$ has the key property $\sum_\alpha \lambda_\alpha^B=0$, related to the unitarity of the flavour rotation matrices 
(and to the assumed down-aligned flavour structure).  
This property, similarly to the GIM-mechanism in the SM, 
is essential to render the loop finite and is required to derive the expression in~\eq{eq:Cbs}. 
As a result of this GIM-like protection, we find that the leptoquark contribution to $C_{bs}^{LL}$ receives an additional mass suppression proportional to $M_L^2/M_U^2$ with respect to the naive dimensional analysis expectation with generic leptoquark couplings and no vector-like fermions.\footnote{This GIM-like behaviour has been qualitatively noticed also in 
a different model presented in Ref.~\cite{Calibbi:2017qbu}. 
On the other hand, models that address the $R(D^{(*)})$ anomaly with scalar leptoquarks 
do not exhibit this suppression, see Eq.~(5.18) in \cite{Marzocca:2018wcf}.} 
In particular, we find that the NP contribution to $\Delta M_s$ follows the approximate scaling
\begin{align}
C_{bs}^{LL}\sim\Delta R_{D^{(*)}}^2\,M_L^2\,,
\end{align}
and therefore it is completely controlled by $M_L$, for fixed $R(D^{(*)})$ anomaly and leptoquark gauge coupling. This scaling is made manifest in Fig.~\ref{fig:bsmix} where we show the constraints arising from the leptoquark 
contribution to $C_{bs}^{LL}$ in the $M_U - s_{q_{12}}$ plane, together with the preferred region for $R(D^{(*)})$,
and for different values of $M_L$. The experimental limit on $C_{bs}^{LL}$ is obtained using the SM determination 
in~\cite{Artuso:2015swg,Lenz:2011ti,Lenz:2006hd}\footnote{A recent lattice QCD simulation  
from the Fermilab/MILC collaboration \cite{Bazavov:2016nty} finds a larger central value (and a smaller error) 
for the non-perturbative parameter $f_{B_s} \sqrt{\hat{B}}$ entering the determination of $\Delta M_s$. 
That would imply a 1.8$\,\sigma$ tension with respect to the SM and translates into 
very stringent limits for purely left-handed NP contributions featuring real couplings \cite{DiLuzio:2017fdq}. 
Given the fact that the new lattice result has not been confirmed yet by other collaborations, 
we conservatively use the pre-2016 determination in \cite{Artuso:2015swg}. 
} 
and the experimental measurement from~\cite{Amhis:2016xyh}. We have 
\begin{align}
C_{bs}^{LL}=1.03\pm0.15\,.
\end{align}

The radial excitation arising from the linear combination of $\Omega_1$ and $\Omega_3$ (see \apps{scalpot}{scalspect}) could also potentially yield dangerous NP contributions not protected by the $U(2)_q$ symmetry. These contributions depend on other parameters 
(masses and couplings) that are not directly connected to the anomalies 
and are therefore more model dependent. Moreover, in the phenomenological limit $v_3\gg v_1$ we find the coupling of the radial mode to be suppressed by  $\cot \beta_T=v_1/v_3$ (see \app{app:radmode}).\footnote{Note that in this phenomenological limit purely leptonic transitions mediated by the radial excitations would receive additional $\tan \beta_T$ enhancements. However, we find the bounds from this sector to be significantly smaller and thus they do not pose any relevant constraint on these effects (see \sect{sec:leptonic}).} As an estimate of the size of such contributions, we compute the box diagrams with two radial modes (similar to the ones in Fig.~\ref{fig:bsmix_dia} but with the leptoquark replaced by the radial excitations). Recasting the result in~\cite{Bertolini:1990if} for the up squark box we find in our model 
{\small
\begin{align}
C_{bs}^{LL}&=\frac{1}{(V_{tb}V_{ts}^*)^2\,R_{SM}^{\rm loop}}\,\frac{G_F^2\,C_U^2\,M_3^4}{128\,\pi^2}\,t_{\beta_T}^{-4}\,s_{q_2}^2s_{q_3}^2\,\left\{\frac{1}{m_{L_2}^2}\, \left[G^\prime(x_{T_R\,L_2},x_{T_R\,L_2},1)\right.\right.\\
&\quad\left.\left.-G^\prime(x_{T_R\,L_2},x_{T_R\,L_2},x_{L_3\,L_2})\right]+\frac{1}{m_{L_3}^2}\, \left[G^\prime(x_{T_R\,L_3},x_{T_R\,L_3},1)-G^\prime(x_{T_R\,L_3},x_{T_R\,L_3},x_{L_2\,L_3})\right]\right\}\,, \nonumber
\end{align}}
with $x_{ab}=m_a^2/m_b^2$ and the loop function $G^\prime$ defined in \cite{Bertolini:1990if}. 
Assuming typical values for the model parameters, we estimate that values as small as $\tan \beta_T\gtrsim 1.75$ are enough to keep this radial-mode contribution to $C_{bs}^{LL}$ to be below $1\%$ and therefore small enough to be ignored. Mixed contributions involving both the leptoquark and the radial mode are present as well. Assuming similar sizes for the loop functions and including the $\cot \beta_T= 1/1.75$ suppression in the radial-mode coupling, we find such contribution to be also sufficiently suppressed to be neglected.

\subsubsection{$D-\bar D$ mixing}\label{sec:Dmix}
Following the analysis from UTfit~\cite{UTfit2018,Carrasco:2014uya}, the constraint obtained from $D-\bar D$ transitions can be expressed in terms of bounds on the Wilson coefficients of the four-fermion effective Hamiltonian
\begin{align}
\mathcal{H}_{\rm eff}^{\Delta C=2}\supset C_1^D\left(\bar c_L\gamma_\mu u_L\right)^2\,.
\end{align}
The latest constraints on $C_1^D$ from UTFit read \cite{UTfit2018}
\begin{align}
\begin{aligned}
\mathrm{Re}\,(C_1^D)&=\left(0.3\pm1.4\right)\times10^{-7}~\mathrm{TeV}^{-2}\,,\\[5pt]
\mathrm{Im}\left(C_1^D\right)&=(-0.03\pm0.46)\times10^{-8}\;\textrm{TeV}^{-2}\,.
\end{aligned}
\end{align}
In our model, NP effects are induced in both the real and the imaginary parts of $C_1^D$. Also, in contrast to the $B_s$ mixing case, 
the model yields contributions both at tree level and at one loop. In what follows we describe both contributions. 

\bigskip
\noindent
\textbf{Tree level.} The $Z^\prime$ and $g^\prime$ mediate tree-level contributions to the $D-\bar D$ amplitude proportional to the CKM matrix elements. These are given by
\begin{align}
\begin{aligned}
\left.C_1^D\right|_{\rm tree}&=\frac{4 G_F}{ \sqrt{2} }\left(C_{Z^\prime}+\frac{C_{g^\prime}}{3}\right)\, (V_{ub}^*\,V_{cb})^2\,\left(\sin^2\theta_{q_3}+\sin^2\theta_{q_2} \frac{V_{us}^*\,V_{cs}}{V_{ub}^*\,V_{cb}}+\sin^2\theta_{q_1}\frac{V_{ud}^*\,V_{cd}}{V_{ub}^*\,V_{cb}}\right)^2\,,
\end{aligned}
\end{align}
with $C_{Z^\prime}$ and $C_{g^\prime}$ defined in~\eq{eq:Cs}. Setting $\theta_{q_1}=\theta_{q_2}\equiv\theta_{q_{12}}$ and using CKM unitarity, the expression above simplifies into
\begin{align}\label{eq:C1D_tree}
\begin{aligned}
\left.C_1^D\right|_{\rm tree}&=\frac{4 G_F}{ \sqrt{2} }\left(C_{Z^\prime}+\frac{C_{g^\prime}}{3}\right)\,(V_{ub}^*\,V_{cb})^2\,\left(\sin^2\theta_{q_3}-\sin^2\theta_{q_{12}}\right)^2\,.
\end{aligned}
\end{align}
This assumption on the mixing angles ensures a $U(2)$-like protection, rendering the tree-level contribution to $C_1^D$ sufficiently small to pass the stringent constraints from $D-\bar D$ mixing. In particular, we find that for values of the NP scale compatible with an explanation of the $R(D^{(*)})$ anomaly, the tree-level contributions to both the real and the imaginary parts of $C_1^D$ are $\approx10^{-9}~\mathrm{TeV}^{-2}$, and are thus compatible with the present bounds. It is also interesting to note from~\eq{eq:C1D_tree} that in the \textit{full-alignment limit}, corresponding to $\theta_{q_{12}}=\theta_{q_3}$, the tree-level contribution to $C_1^D$ would completely vanish by unitarity, 
as expected from the discussion at the beginning of this section.

\begin{figure}[t]
\centering
\includegraphics[width=\textwidth]{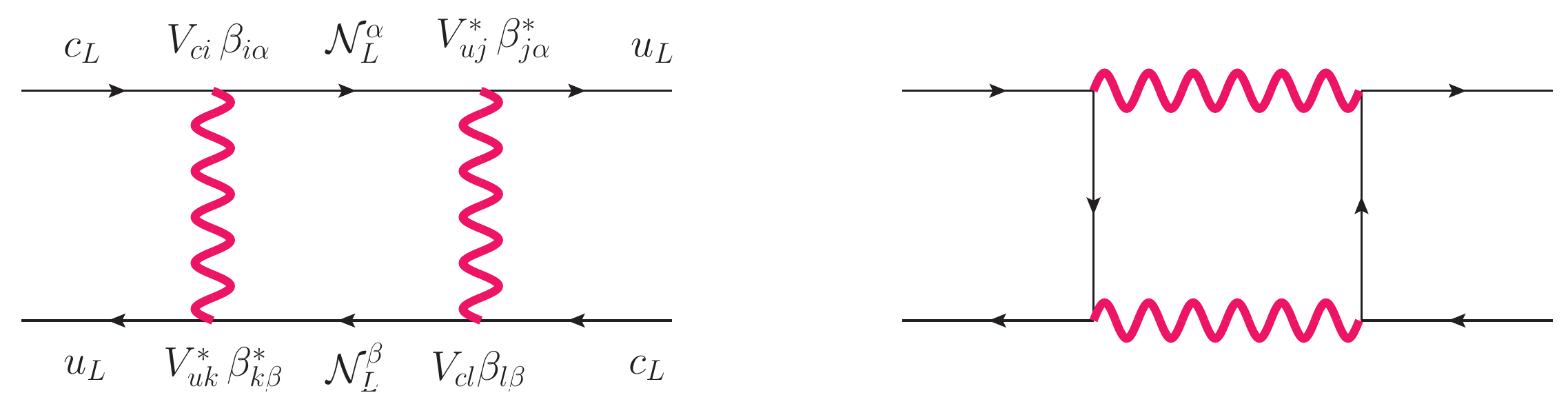}
\caption{\sf Leptoquark mediated one-loop diagrams contributing to $D-\bar D$ mixing. The symbol $\mathcal{N}$ denotes a six-dimensional vector containing both SM neutrinos and their partners.\label{fig:Dmix_dia}}
\end{figure}

\begin{figure}[t]
\centering
\includegraphics[width=0.42\textwidth]{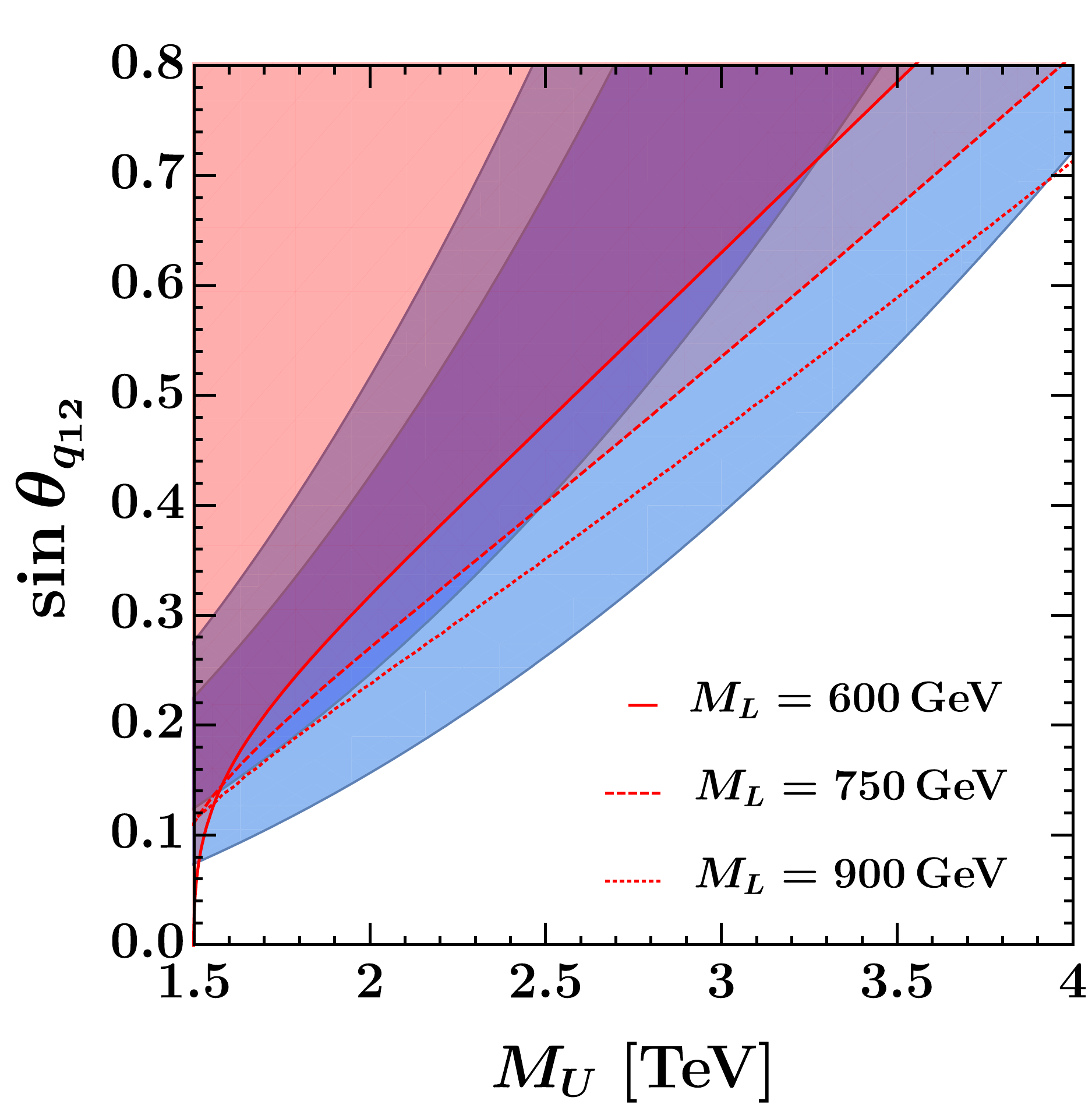}\qquad\quad\includegraphics[width=0.42\textwidth]{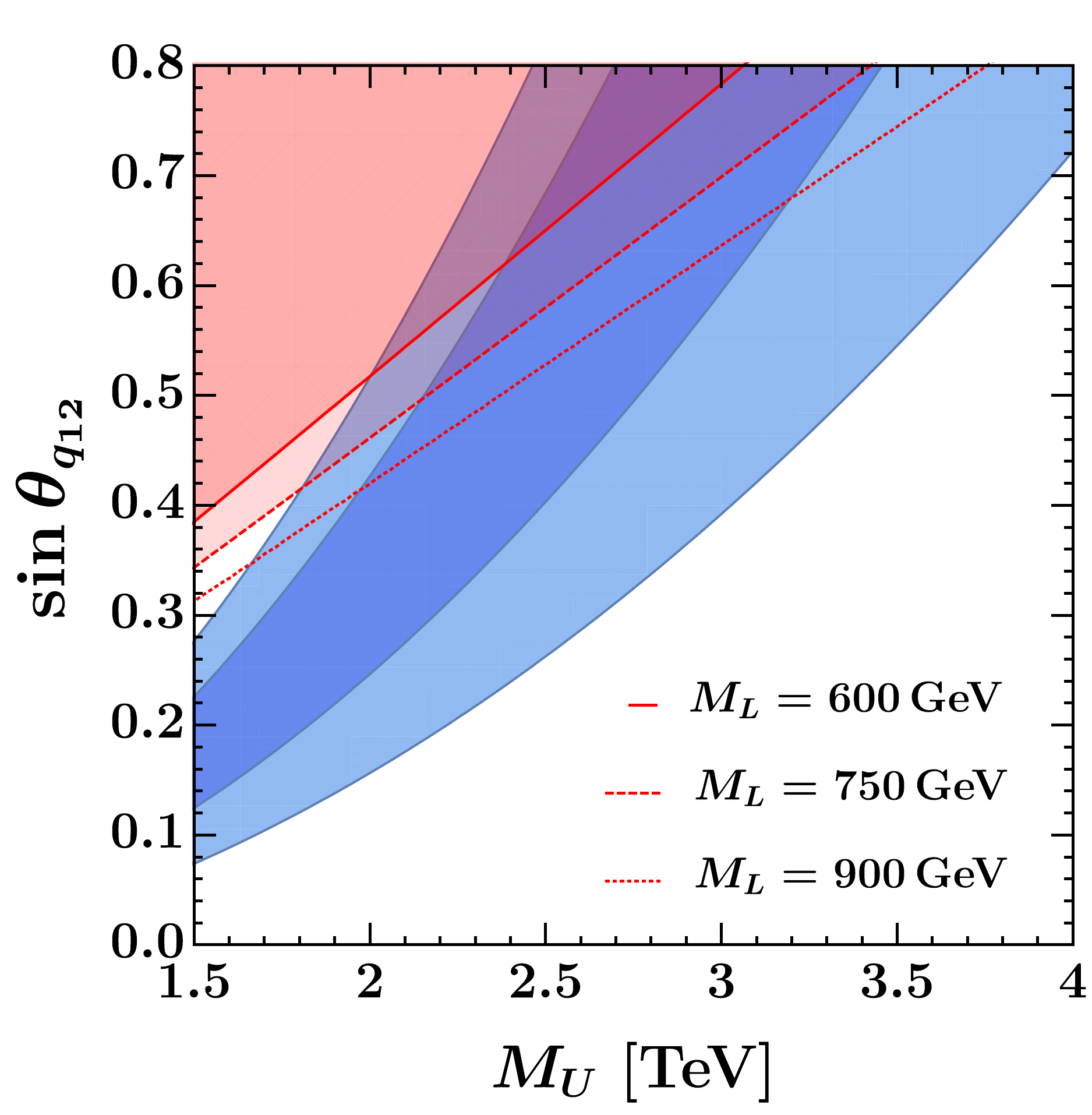}
\caption{\sf Constraints from $D-\bar D$ mixing, for different values of the bare vector-like lepton mass parameter $M_L$. In the left bounds arising from $\Im(C^D_1)$ and in the right those arising from $\Re(C^D_1)$ at $95\%$~CL (see text for more details). The $1\sigma$ and $2\sigma$ preferred regions by the $R(D^{(*)}$ anomaly are shown in dark and light blue, respectively. We use as input for the model parameters: $g_4=3.5$, $s_{q_3}=0.8$, $s_{\ell_3}=0.8$, $v_3 = 1.75\, v_1$ and $\lambda_{15}=2.5$\label{fig:dmix}.}
\end{figure}

\bigskip
\noindent
\textbf{One loop.}  In this case, the computation of the loop effects is technically more challenging than in the previous section, since now also the $g^\prime$ and $Z^\prime$ mediate NP contributions at one loop. However, it is important to note that thanks to the flavour structure of the model all these additional contributions are protected by the same $U(2)_q$ symmetry that protected the tree-level contribution (i.e.~they are proportional to $(V_{ub}^*V_{cb})^2$), and therefore they are much smaller than the (already small) tree-level effect.  As in the $B_s$-mixing case, we find that the dominant contributions to $C_1^D$ arise from loop diagrams involving the leptoquark (see Fig.~\ref{fig:Dmix_dia}), which are not protected by the $U(2)_q$ symmetry, and that these effects are proportional to the $W$ matrix. Neglecting corrections of $\mathcal{O}((V_{ub}^*V_{cb})^2)$, we find
\begin{align}
\left.C_1^D\right|_{\rm loop}\approx-\frac{4G_F}{\sqrt{2}}\frac{g_4^2}{64\pi^2}\,C_U\sum_{\alpha,\beta}\lambda_\alpha^D\lambda_\beta^D\,F(x_\alpha,x_\beta)\,,
\end{align}
where the loop function is defined as in~\eq{eq:loop_func}, and $\lambda_\alpha^D=V_{ci}V_{uj}^*\, \beta_{i\alpha}\,\beta_{j\alpha}^*$ with $\beta$ the leptoquark coupling to fermions. Keeping only the $U(2)_q$-violating contributions, $\lambda_\alpha^D$ can be written in terms of CKM matrix elements and fermion mixing angles as
\begin{align}
\begin{aligned}
\lambda_\alpha^D&\approx V_{cs}V_{us}^*\;s_{q_{12}}^2\left[c_{LQ}^2\,s_{\ell_2}^2\,\delta_{\alpha2}+c_{LQ}^2\,c_{\ell_2}^2\,\delta_{\alpha5}+s_{LQ}^2\,s_{\ell_3}^2\,\delta_{\alpha3}+s_{LQ}^2\,c_{\ell_3}^2\,\delta_{\alpha6}-\delta_{\alpha4}\right]\\
&\quad-(V_{cb}V_{us}^*+V_{cs}V_{ub}^*)\;s_{q_{12}}s_{q_3}\,c_{LQ}\,s_{LQ}\left[\,s_{\ell_2}^2\,\delta_{\alpha2}+c_{\ell_2}^2\,\delta_{\alpha5}-s_{\ell_3}^2\,\delta_{\alpha3}-c_{\ell_3}^2\,\delta_{\alpha6}\right]\\
&\quad+V_{cb}V_{ub}^*\;s_{q_3}^2\left[c_{LQ}^2\,s_{\ell_2}^2\,\delta_{\alpha2}+c_{LQ}^2\,c_{\ell_2}^2\,\delta_{\alpha5}+s_{LQ}^2\,s_{\ell_3}^2\,\delta_{\alpha3}+s_{LQ}^2\,c_{\ell_3}^2\,\delta_{\alpha6}-\delta_{\alpha4}\right]\,.
\end{aligned}
\end{align}
Also in this case, the GIM-like protection encoded in $\sum_\alpha \lambda_\alpha^D\approx 0$ ensures an additional suppression of the box contributions. More precisely, we find the following approximate scaling connecting the NP effect in $R(D^{(*)})$  with the one in $D-\bar D$ mixing
\begin{align}\label{eq:DDbar_scaling}
\left.C_1^D\right|_{\rm loop}&\sim s_{q_{12}}^2\, \Delta R_{D^{(*)}}^2\,M_L^2\,.
\end{align}
Interestingly, this different scaling results in an upper limit on the maximum allowed value for $s_{q_{12}}$. This is shown in Fig.~\ref{fig:dmix}, where we plot the constraints from $D-\bar D$ mixing (both for the real and imaginary contributions) together with the preferred region by $R(D^{(*)})$. In the low-$s_{q_{12}}$ region of the left figure there is a small violation of the scaling in~\eq{eq:DDbar_scaling}. This violation is due to the tree-level contribution in~\eq{eq:C1D_tree}, which for the real part plays a marginal role.

Finally, concerning the contribution from the scalar radial modes, similarly to the case of $B_s$ mixing we find that these receive $\cot\beta_T$ suppressions in the phenomenological limit $v_3>v_1$, making their effect sufficiently small to be neglected.

\subsection{Leptonic processes}
\label{sec:leptonic}

The fully leptonic transitions play a less important role in the low-energy phenomenology than hadronic processes. As already mentioned, the assumption of flavour alignment in the charged-lepton sector forbids tree-level LFV transitions mediated by the $Z^\prime$. The leading effects are therefore those mediated by the leptoquark at one loop, and are completely controlled by the $W$ matrix. The assumed structure for this matrix, i.e.~$\theta_{LQ} = \pi/4$, chosen to maximise the NP contribution in $R(D^{(*)})$, implies no NP contributions to fully leptonic LFV transitions involving electrons (even for $s_{\ell_1}\neq0$). Furthermore, the loop suppression, together with the additional suppression coming from the mixing angle of the muon, $s_{\ell_2}\approx0.1$, are sufficient to render the model contributions to $\tau\to 3\mu$ and $\tau\to\mu\gamma$ well below the current experimental sensitivity. Purely-leptonic and electroweak operators
generated by the renormalisation-group running of the semi-leptonic operators from the mass scale of the leptoquark 
down to the electroweak scale \cite{Feruglio:2016gvd,Feruglio:2017rjo}, are already 
taken into account in the global fits of \cite{Buttazzo:2017ixm} and in the limit of large 3-2 mixing 
studied in this paper they are even less important.

\subsection{Perturbativity}
\label{sect:pert}

\begin{figure}[t]
  \centering
 \includegraphics[width=0.45\textwidth]{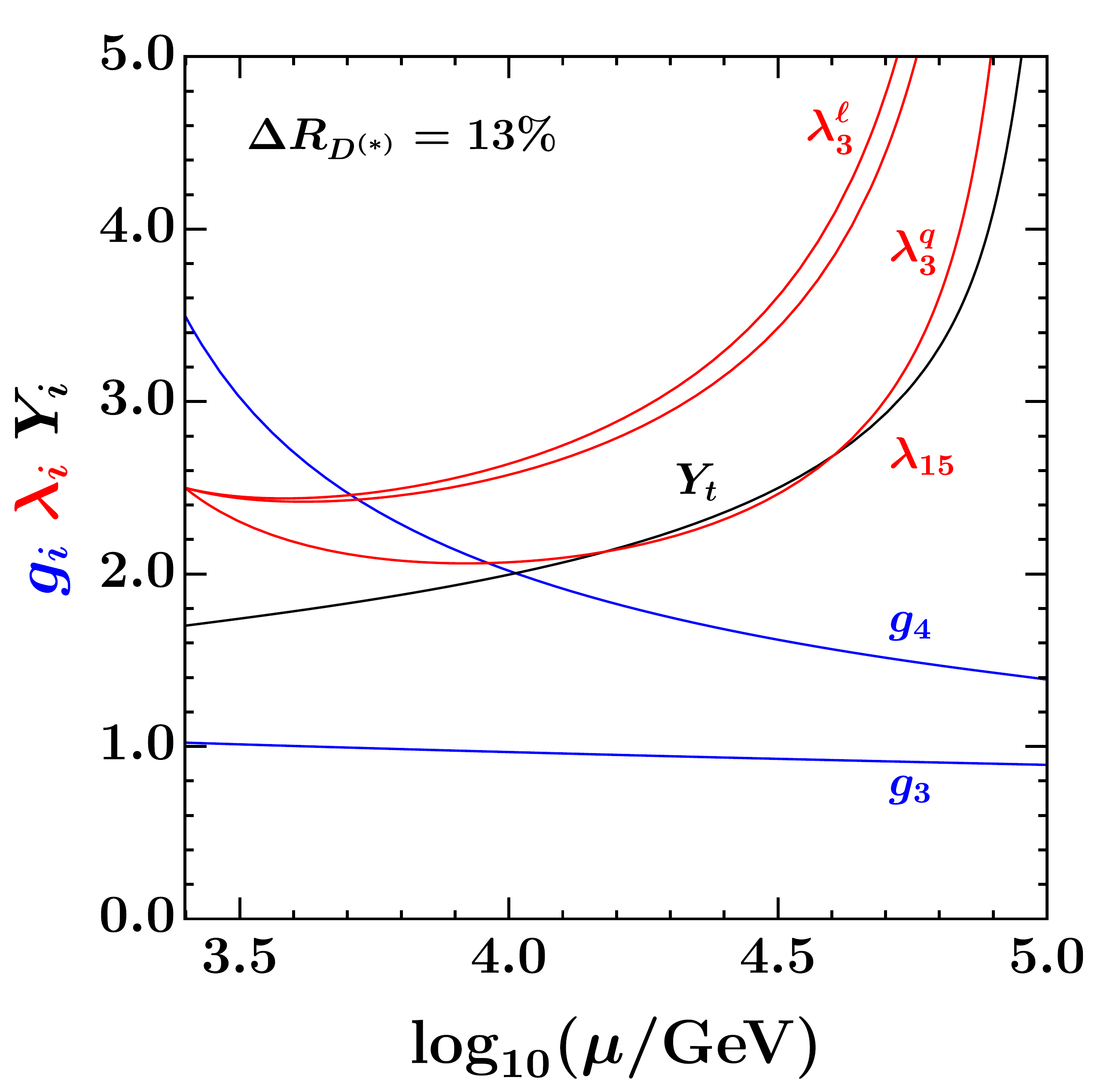}~~~~~~~~~~~~
 \includegraphics[width=0.45\textwidth]{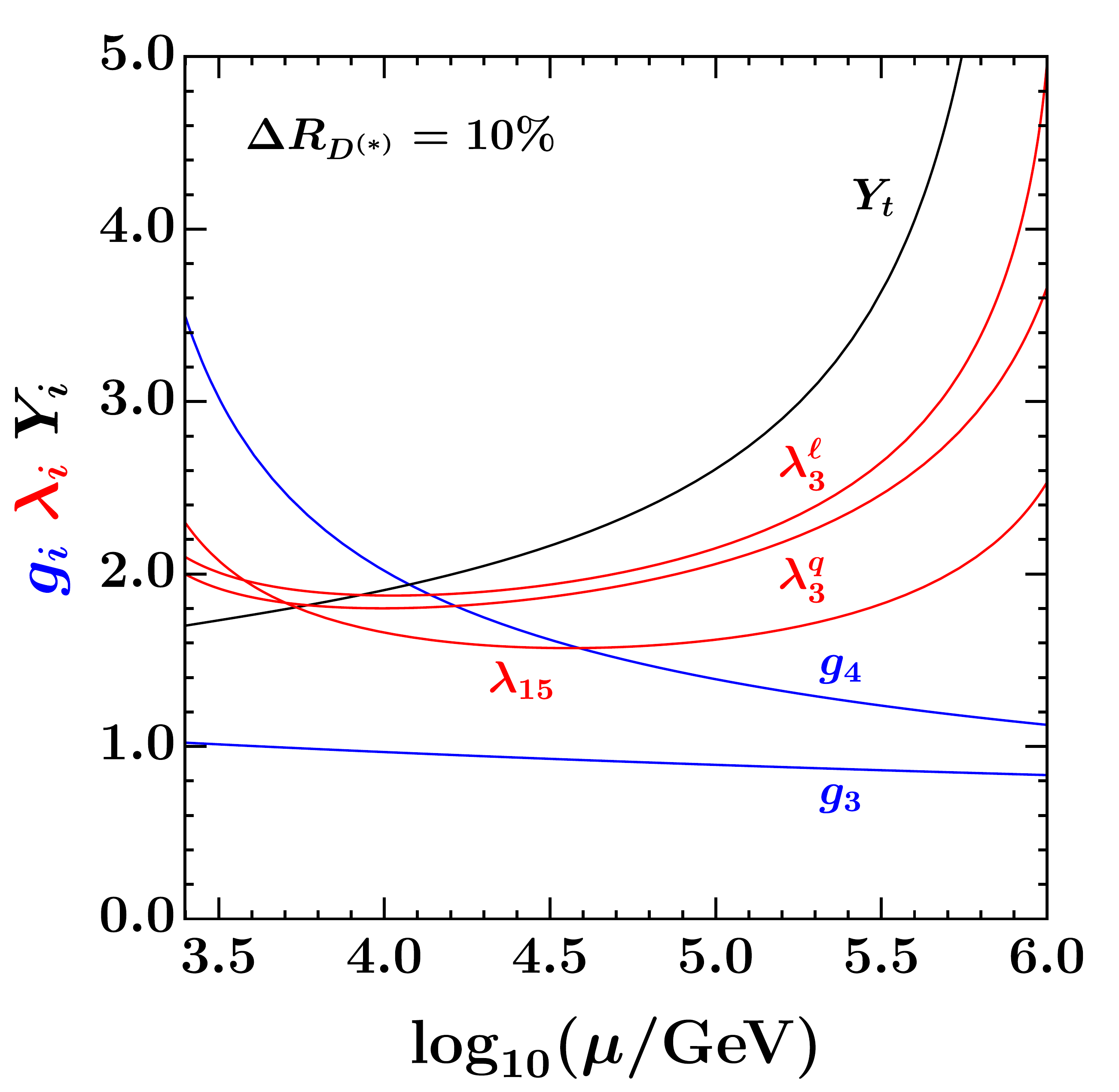} 
  \caption{\sf Left: Sample one-loop RGE evolution for the benchmark point: 
  $g_4 = 3.5$ and $\lambda_{q} = \lambda_{\ell} = \lambda_{15} = 2.5$ and $Y_t=1.7$ 
  at the matching scale $\mu = 2.5$ TeV. Other (subleading) couplings are not shown in the plot. 
  Right: same for $g_4 = 3.5$, $\lambda_{q} =2.1$, $\lambda_{\ell} = 2.0$, $\lambda_{15} = 2.3$ and $Y_t=1.7$. 
  } 
  \label{fig:runningBP}
\end{figure}

The fit of the $R(D^{(*)})$ anomaly (cf.~\eq{eq:RDscaling}) requires simultaneously large 
$g_4$ and mixing angles $s_{q_3}$ and $s_{\ell_3}$, which translate to sizeable 
third generation Yukawa couplings $\lambda_q$ and $\lambda_\ell$, 
thus pushing the model close to the boundary of the perturbative domain. 
When assessing the issue of perturbativity, there are two conceptually different questions that one could 
address: the first (more conservative) is to which extent low-energy observables are 
calculable in perturbation theory and the second (more ambitious) 
is up to which energy the model can be extrapolated in the UV before entering the strongly coupled regime.  Regarding the convergence of the perturbative expansion at low energy, the most important coupling is 
$g_4$, which for typical benchmarks is $\sim 3$. This is still within the limits imposed by standard perturbativity criteria: 
e.g.~the beta function criterium of \cite{Goertz:2015nkp} $\beta_{g_4} / g_4 < 1$  yields $g_4 < 4 \pi / \sqrt{10} \approx 4$, 
while perturbative unitarity of leptoquark-mediated $2 \to 2$ fermion scattering amplitudes requires 
$g_4 < \sqrt{8 \pi} \approx 5$ \cite{DiLuzio:2017chi,DiLuzio:2016sur}.
Remarkably, the phenomenological requirement of a large $g_4$ coupling in the IR 
does not prevent extrapolation of the theory in the UV,  
thanks to the (one-loop) asymptotic freedom of the $SU(4)$ gauge factor. 
Following the $g_4$ evolution from the UV to the IR the theory flows towards the confining phase, 
until the running is frozen by the spontaneous breaking of $SU(4)$ via the Higgs mechanism. 

From the point of view of the UV extrapolation, the problematic couplings are actually the Yukawas, 
which are required to be large in order to generate sizeable mixings between the third generation SM 
fields and their vector-like partners. To investigate their effects we have computed the one-loop
renormalisation group equations (RGEs) of the 4321 model 
(which are reported for completeness in \app{RGE}). 
In \fig{fig:runningBP} 
we show the RGE evolution for two typical benchmark points which are compatible 
with low-energy and high-$p_T$ observables and which  
yield a $13\%$ (left panel) and $10\%$ (right panel) contribution to 
$\Delta R_{D^{(*)}}$. 
Depending on the initial values of the 33  components of the $\lambda_{q,\ell,15}$ and $Y_u$ matrices, 
the theory can be extrapolated in the UV for several decades of energy before hitting a Landau pole. 
These figures also clearly give an idea of the tension between the need to give a sizeable contribution to 
$\Delta R_{D^{(*)}}$ and that of extrapolating the 4321 model in the UV.


\section{High-$p_T$ signatures}
\label{sec:highpT}

In this section we survey the main high-$p_T$ signatures of the $4321$ model in $p p$ collisions at the LHC. After reviewing the main features of the resonances spectrum in Sec.~\ref{sec:resspec}, we describe the leading decay channels in Sec.~\ref{sec:DecCha}. In Sec.~\ref{colliderconstr}, we derive the exclusion limits from the coloron searches in $t \bar t$ and $j j$ final states, $Z'$ searches in $\tau^+ \tau^-$ and vector leptoquark searches. Finally, we highlight the non-standard phenomenology of the vector-like lepton (and vector-like quarks) as the most novel aspect of the high-$p_T$ discussion.
The upshot of this section is that the $4321$ model predicts a vastly richer set of high-$p_T$ signatures than the simplified dynamical model of a vector leptoquark introduced in~\cite{Buttazzo:2017ixm}.

\subsection{Resonances spectrum}
\label{sec:resspec}

The $4321$ model predicts a plethora of new resonances around the TeV scale that are potential targets for direct searches with the ATLAS and CMS experiments. In this section we discuss the spectrum of new resonances and their couplings, focusing on the parameter space of the model preferred by the flavour anomalies and consistent with other low-energy data.

The starting point is the low-energy fit to the charged current anomalies in $R_{D^{(*)}}$. In the limit $s_{q_2} \gg V_{c b}$, the following approximate formula can be derived,
\beq
\Delta R_{D^{(*)}} \approx 0.2 \left (\frac{2~\textrm{TeV}}{M_{U}} \right)^{2} \left(\frac{g_4}{3.5} \right)^{2} \sin (2 \theta_{LQ}) \left(\frac{s_{\ell_3}}{0.8}\right)^2 \left(\frac{s_{q_3}}{0.8}\right) \left(\frac{s_{q_2}}{0.3}\right)~.\label{eq:fitscaling}
\eeq
To explain $R_{D^{(*)}}$, one needs (i) a rather low $\mathcal{G}_{4321} \to \mathcal{G}_{321}$ breaking scale, $M_U / g_4 \sim$ $\mathcal{O}$(TeV), (ii) large leptoquark flavour violation controlled by $\theta_{LQ}$ and (iii) sizable fermion mass mixings. Requiring, in addition, the couplings of the model, $g_4$, $\lambda_q$ and $\lambda_\ell$, to be perturbative, sets an upper limit on the masses of new vectors and fermions.

The spectrum of the new scalar resonances depends on the details of the scalar potential (see \app{scalspect}), 
which introduces extra free parameters that are less directly related to the flavour anomalies. In the following, we focus on the fermionic and vector resonances, postponing the discussion of the radial scalar excitations to \sect{sec:highptscalars}.

\subsubsection*{New vectors}

\begin{figure}[t]
\centering
\includegraphics[width=0.485\textwidth]{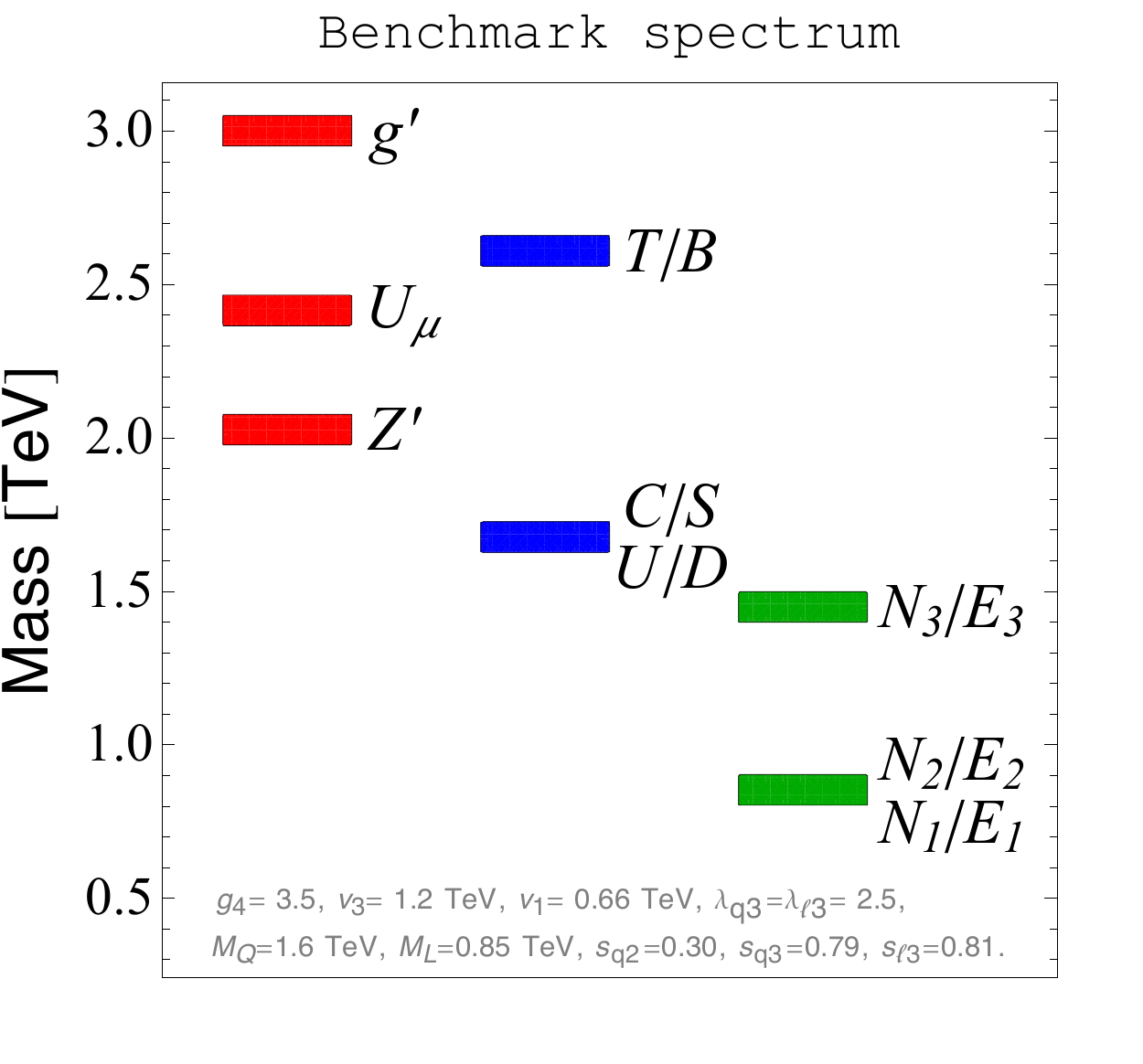} \; \;
\includegraphics[width=0.47\textwidth]{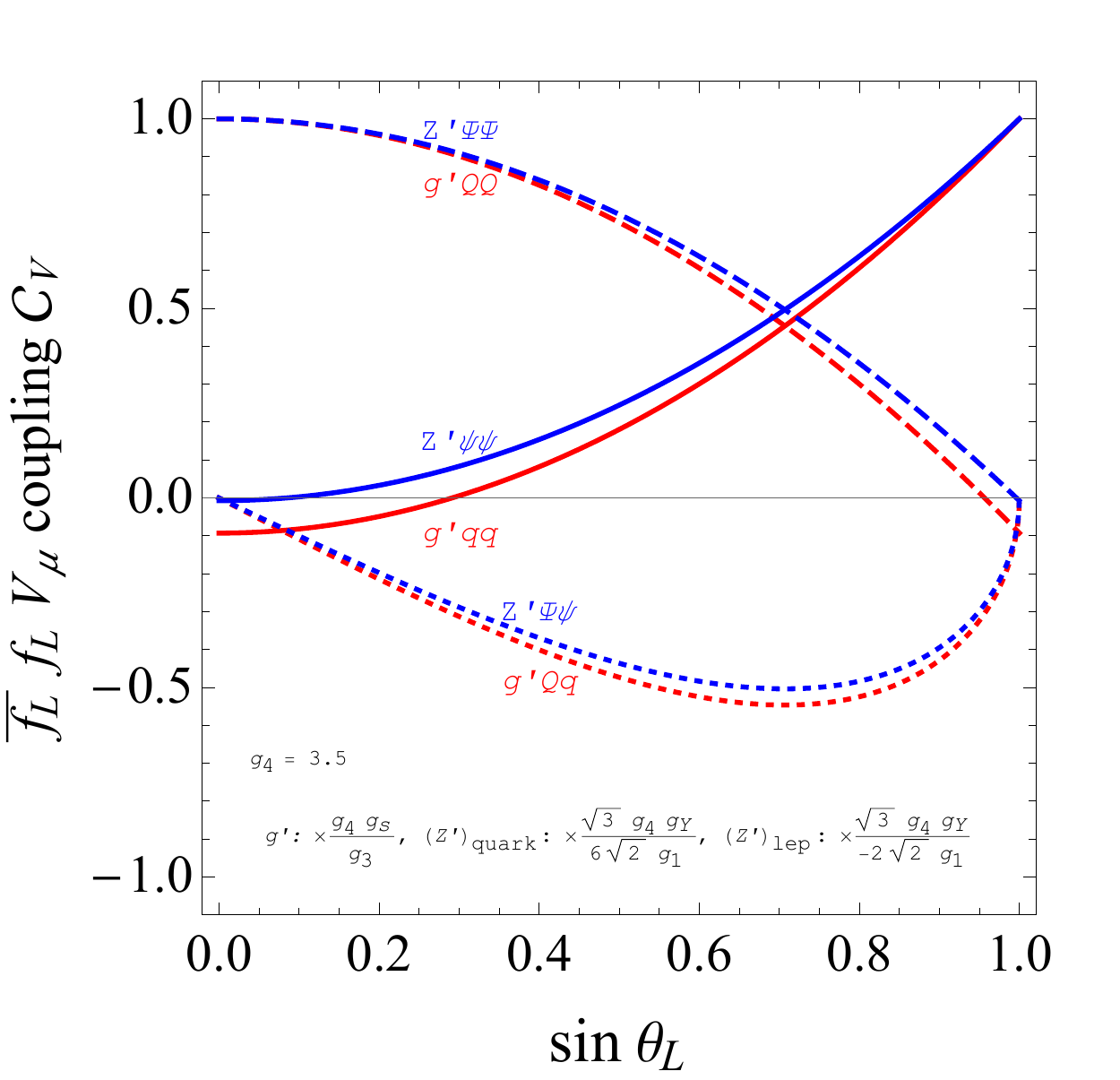}
\caption{\sf ({\bf Left panel}) A typical spectrum of new vectors and fermions. The benchmark point is: $g_4 = 3.5$, $v_3 = 1.2, \,v_1 = 0.66$~TeV and $v_{15}=0.3$~TeV, which fixes the masses of $g'_\mu$, $U_\mu$ and $Z'_\mu$, while $M_Q = 1.6$~TeV, $M_L=0.85$~TeV, $s_{q_3}=0.79$, $s_{\ell_3}=0.81$ and $s_{q_2}=0.3$, which sets the fermionic masses. ({\bf Right panel}) Normalized $V f_L \bar f_L$ couplings of the $g'$ (red) and $Z'$ (blue) to left-handed fermions as a function of the $\sin \theta_L$. Solid, dotted and dashed lines are for the {\it light-light}, {\it light-heavy} and {\it heavy-heavy} combinations, respectively. The coupling normalizations are, $\frac{g_4 g_s}{g_3}$ for $g'$ to quarks, and $\frac{\sqrt{3} g_4 g_Y}{6 \sqrt{2} g_1}$ ($\frac{\sqrt{3} g_4 g_Y}{-2 \sqrt{2} g_1}$) for the $Z'$ to quarks (leptons).\label{fig:spectrum}}
\end{figure}

Applying Eq.~\eqref{eq:fitscaling} to the perturbative parameter space of the $4321$ model, the implied mass scale of the new vectors $g'_\mu$, $U_\mu$ and $Z'_\mu$ is in the interesting range for direct searches at the LHC. Setting $\theta_{LQ}=\pi/4$ and maximising the left-handed fermion mixings for the third family, the spectrum can be further moved up by increasing $s_{q_2}$ and $g_4$ -- eventually limited by phenomenology (see e.g.~Eq.~\eqref{eq:DDbar_scaling}) and perturbativity, respectively. In the motivated limit, $v_{15} \ll v_1 \ll v_3$ (for the minimisation of the scalar potential see \app{scalpot}), and $g_1, g_3 \ll g_4$, the spectrum of the new vectors approximately follows the pattern $m_{g'} : m_{U_1} : m_{Z'} \approx \sqrt{2} : 1 : \frac{1}{\sqrt{2}}$. A typical benchmark point is illustrated in Fig.~\ref{fig:spectrum} (left panel).

The structure of the $V f \bar f$ interactions is discussed in length in \app{intmb}. Here we highlight the key aspects for the high-$p_T$ searches. The fermion mass mixing in the right-handed sector is neglected for the purposes of this discussion (the largest mixing being $s_{u^3_R} \lesssim 0.1$). All the $V f_R \bar f_R$ interactions are practically flavour diagonal, except for the leptoquark couplings to fermionic partners described by the $W$ matrix. The couplings to right-handed SM fermions are suppressed. 

In contrast, the fermion mass mixing in the left-handed sector plays a major role. These interactions are worked out in Eqs.~\eqref{eq:LLmbS}~to~\eqref{eq:LLmbE}. To illustrate the main implications, in Fig.~\ref{fig:spectrum} (right panel) we show the normalized $V f_L \bar f_L$ couplings for $Z'$ and $g'$ as a function of  $\sin \theta_L$, valid for any of the left-handed mixing angles.  Solid, dotted and dashed lines represent couplings to {\it light-light}, {\it light-heavy} and {\it heavy-heavy} combinations, where labels {\it light} and {\it heavy} denote a SM fermion and its partner, respectively. Red color is for $g'$ couplings ($\mathcal{C}_{g'}$) normalized as $\mathcal{L} \supset \mathcal{C}_{g'} \, \frac{g_4 g_s}{g_3} \, \bar \psi_q T^a \gamma^\mu P_L \psi_q \,  g'^a_\mu$, while blue is for $Z'$ couplings ($\mathcal{C}_{Z'}$) normalized as $\mathcal{L} \supset \mathcal{C}_{Z'} \,\frac{\sqrt{3} g_4 g_Y}{6 \sqrt{2} g_1} \, \left(\bar \psi_q \gamma^\mu P_L \psi_q - 3\, \bar \psi_\ell \gamma^\mu P_L \psi_\ell\right)\, Z'_\mu$~. It is worth noting that sizable couplings to SM fermions are generated only for large mixing angles. In practice, the third family mixings, $s_{q_3}$ and $s_{\ell_3}$, typically control the decay channels of new resonances, while $s_{q_2}$~($=s_{q_1}$) is relevant for their production mechanisms in $p p$ collisions.

%

\subsubsection*{New fermions}

The main features of the fermion spectrum are controlled by the fermion mass mixing constraints discussed in Sec.~\ref{sec:FermConst}. Relevant facts for the high-$p_T$ discussion are the following: $i)$ the components of an $SU(2)_L$ doublet are practically degenerate, $ii)$ partners of the first two families are close in mass, $iii)$ a partner of the third SM family is always heavier than the partners of the first two, and $iv)$ lepton partners are typically lighter than quark partners as required by consistency with loop-induced $\Delta F =2$ observables, see Sec.~\ref{sec:PhenoHad}.

Consistency with tree-level $\Delta C = 2$ transitions requires $s_{q_1} = s_{q_2}\equiv s_{q_{12}}$ as discussed in Sec.~\ref{sec:Dmix}. One the one hand, sizeable $s_{q_2}$ boosts the NP contribution to $R(D^{(*)})$. On the other hand, $s_{q_1}$ cannot be too large since it leads to an increased production cross section of new vector bosons in $pp$ collisions due to couplings to valence quarks. For example, for $s_{q_{12}} \lesssim 0.4$, the formula for the light quark partner's mass, $M_{U^i (D^i)} \approx {\hat M_Q}$, holds at $10 \%$ level, see Eq.~\eqref{eq:maseteskih}, while for the third family quark partner, $M_{C(U)} / M_T \approx \cos \theta_{q_3}.$ Note that perturbativity of $\lambda_q^i$, together with the requirement of fitting the $R(D^{(*)})$ anomaly, implies an upper limit on the $\hat M_Q$ to be not far above $\sim 1$~TeV, see Eq.~\eqref{eq:epsx}. Similar arguments hold for the lepton partners since $s_{\ell_3}$ is almost maximal, while $s_{\ell_1}, s_{\ell_2}$ are rather small. 

The typical spectrum of new fermions is illustrated in Fig.~\ref{fig:spectrum} (left panel) and will serve as a benchmark in the following discussion. 

\subsection{Decay channels}
\label{sec:DecCha}

The rich spectrum of new resonances, together with the peculiar structure of $V f \bar f$ interactions, leads to an interesting decay phenomenology in the $4321$ model. For example, cascade decays involving particles in Fig.~\ref{fig:spectrum} (left panel) are possible, predicting spectacular signatures in the detector. Let us survey the main decay modes of each new state separately. 

\subsubsection{Vector decays}

The dominant decay modes of the vector bosons are $1 \to 2$ processes induced by the $V f \bar f$ couplings listed in \app{intmb}. We start with the Lagrangian,
\begin{equation}
\mathcal{L} \supset \, \bar \chi^i \gamma^\mu\,(g_L P_L + g_R P_R)\, \mathcal{C}^{ijk} \psi^j  \,V^k_\mu + \textrm{h.c.}~,\label{eq:decLag}
\end{equation}
where $\chi$ and $\psi$ are Dirac fermions with color indices $i$ and $j$ and masses $m_\chi$ and $m_\psi$, respectively, while $\mathcal{C}^{ijk}$ is the color tensor and $k$ is the color index of vector $V_\mu$ with mass $M_V$. Chiral projectors in the spinor space are $P_{R/L} = (1\pm\gamma^5)/2$. The general formula for $V \to \chi \bar \psi$ partial decay width, following from this Lagrangian, is ($M_V > m_\chi + m_\psi$)
{\small
\begin{equation}
\Gamma(V \to \chi \bar \psi) = N_{\mathcal{C}} \frac{M_V}{48 \pi}\sqrt{\lambda_1\left(\frac{m_\chi}{M_V},\frac{m_\psi}{M_V}\right)} \left[ \left(|g_L|^2+|g_R|^2 \right) \lambda_2\left(\frac{m_\chi}{M_V},\frac{m_\psi}{M_V}\right) + 12 \frac{m_\chi m_\psi}{M^2_V} \mathcal{R}(g_L g^*_R)\right]~,~\label{eq:decForm}
\end{equation}
}
where $N_{\mathcal{C}}$ is the color factor and
\begin{equation}
\begin{split}
\lambda_1(x_1,x_2) &= 1 - 2\, (x_1^2 + x_2^2 + x_1^2 x_2^2) + x_1^4 + x_2^4~, \\
\lambda_2(x_1,x_2) &= 3 \,(1-x_1^2 - x_2^2)-\lambda_1(x_1,x_2)~.
\end{split}
\end{equation}
For the $Z'$ boson, $\mathcal{C}^{ijk} \to \delta^{i j}$ and $k$ index is trivial for quarks (or all indices trivial for leptons). The color factor is $N_{\mathcal{C}} = 3 \, (1)$ for $Z'$ decays to quarks (leptons). For $g'$ instead $\mathcal{C}^{i j a} \to T^a_{i j}$, where $a=1,...,8$, and the color factor for decays to quarks is $N_{\mathcal{C}} = 1/2$. Finally, for the $U_\mu$ leptoquark, 
$\mathcal{C}^{i j k} \to \delta^{i k}$, $j$ index is trivial, and the colour factor for decay to quark and lepton is $N_{\mathcal{C}} = 1$. The same formula Eq.~\eqref{eq:decForm} also applies in the case of $\psi \equiv \chi$ and real vector field $V_\mu^\dagger \equiv V_\mu$, provided that $g_L, g_R$ are real and there is no '+ h.c.' term in Eq.~\eqref{eq:decLag}.

Using these formulas, we calculate the total width and the leading branching ratios for $g'$, $U$, and $Z'$ vector bosons. The results for the benchmark spectrum from Fig.~\ref{fig:spectrum} (left panel) are shown in Table~\ref{tab:BRv}. In this context, the most relevant parameters are the two largest mixing angles $s_{q_3}$ and $s_{\ell_3}$, as well as, the masses of the vector-like fermions. 

\begin{table}[ht]
\centering
  \begin{centering}
    \begin{tabular}{c l c c c}
          {\sf Particle} & {\sf Decay mode}  & {\sf $\mathcal{B}$ (BP)} & {\sf $\Gamma / M$ (BP)}\\
      \hline
      \hline 
      \multirow{5}{*}{$U_\mu$}  & $q_3 \ell_3$    $=t \nu, b\tau$ & $\sim 0.3$   &      \multirow{5}{*}{$12 \%$} \\
                                                & $q_3 L_2$      $=t N_2, b E_2$  & $\sim 0.3$   \\
                                                & $q_3 L_3$      $=t N_3, b E_3$  & $\sim 0.1$   \\
                                                & $q_1 L_1$      $=j N_1, j E_1$  & $\sim 0.1$     \\
                                                & $Q_2 \ell_3$   $=C \nu, S \tau$  & $\sim 0.1$ \\
 \hline
      \multirow{3}{*}{$g'_\mu$}  &  $q_3  q_3$    $=t t, b b$ & $\sim 0.6$ &  \multirow{3}{*}{$10 \%$} \\
      					      &  $q_1 Q_1$   $=j U, j D$ & $\sim 0.2$   \\
                                                &  $q_2 Q_2$   $=j C, j S$ & $\sim 0.2$   \\
 \hline
      \multirow{3}{*}{$Z'_\mu$}   &  $L_1 L_1$   $=N_1 N_1, E_1 E_1$ & $\sim 0.4$   &  \multirow{3}{*}{$44 \%$} \\
                                                &  $L_2 L_2$   $=N_ 2 N_2, E_2 E_2$ & $\sim 0.4$   \\
                                                &  $\ell_3 \ell_3$  $=\tau \tau, \nu \nu$ & $\sim 0.1$  \\                                            
                                        
    \end{tabular}
  \end{centering}
  \caption{\sf Relevant decay channels of new vectors $U_\mu$, $g'_\mu$ and $Z'_\mu$. Branching ratios ($\mathcal{B}$) are calculated for the benchmark point (BP) corresponding to the spectrum shown in Fig.~\ref{fig:spectrum} (left panel). The last column shows the total decay width to mass ratio ($\Gamma / M$) for the BP.}
  \label{tab:BRv}
\end{table}

{\bf $g'$:} The coloron will decay most of the time to a pair of third family SM quarks, $t \bar t$ or $b \bar b$. It could, in principle, decay also to vector-like quark partners if these are kinematically accessible. For the benchmark point, sizeable decays are into {\it light - heavy} combination. The exclusion limits on the coloron from $p p \to j j$ and $p p \to t \bar t$ searches are explored in more detail in Sec.~\ref{colliderconstr:col}. 

{\bf $U_\mu$:} The vector leptoquark is expected to decay to $t \nu$ and $b \tau$ final states. Decay modes involving {\it light - heavy} combinations are also relevant if kinematically allowed. Examples include $U_\mu \to t N_2$ and $U_\mu \to b E_2$ decays.

{\bf $Z'$:} Decays of the $Z'$ boson are typically into a pair of third family SM leptons, $\tau \tau$ and $\nu_\tau \nu_\tau$, as well as, heavy vector-like lepton partners, which are required to be relatively light by $\Delta F = 2$ constraints as already discussed in Sec.~\ref{sec:PhenoHad}. It is worth noting that, for the benchmark point, $Z'$ has a rather large total decay width $\Gamma / M \sim 40\%$ (unlike $g', U$ with $\Gamma / M \sim 10\%$) signalling that the model is at the edge of perturbativity.  The extra decay modes to heavy lepton partners are welcome to avoid the bounds from $Z' \to \tau \tau$ as discussed in Sec.~\ref{colliderconstr:ta}. However, $Z'$ assisted production becomes the dominant production mechanism for heavy lepton partners, as discussed in Sec.~\ref{colliderconstr:Leptons}.

\begin{figure}[t]
\centering
  \begin{tabular}{c c c}
      \includegraphics[width=0.275\textwidth]{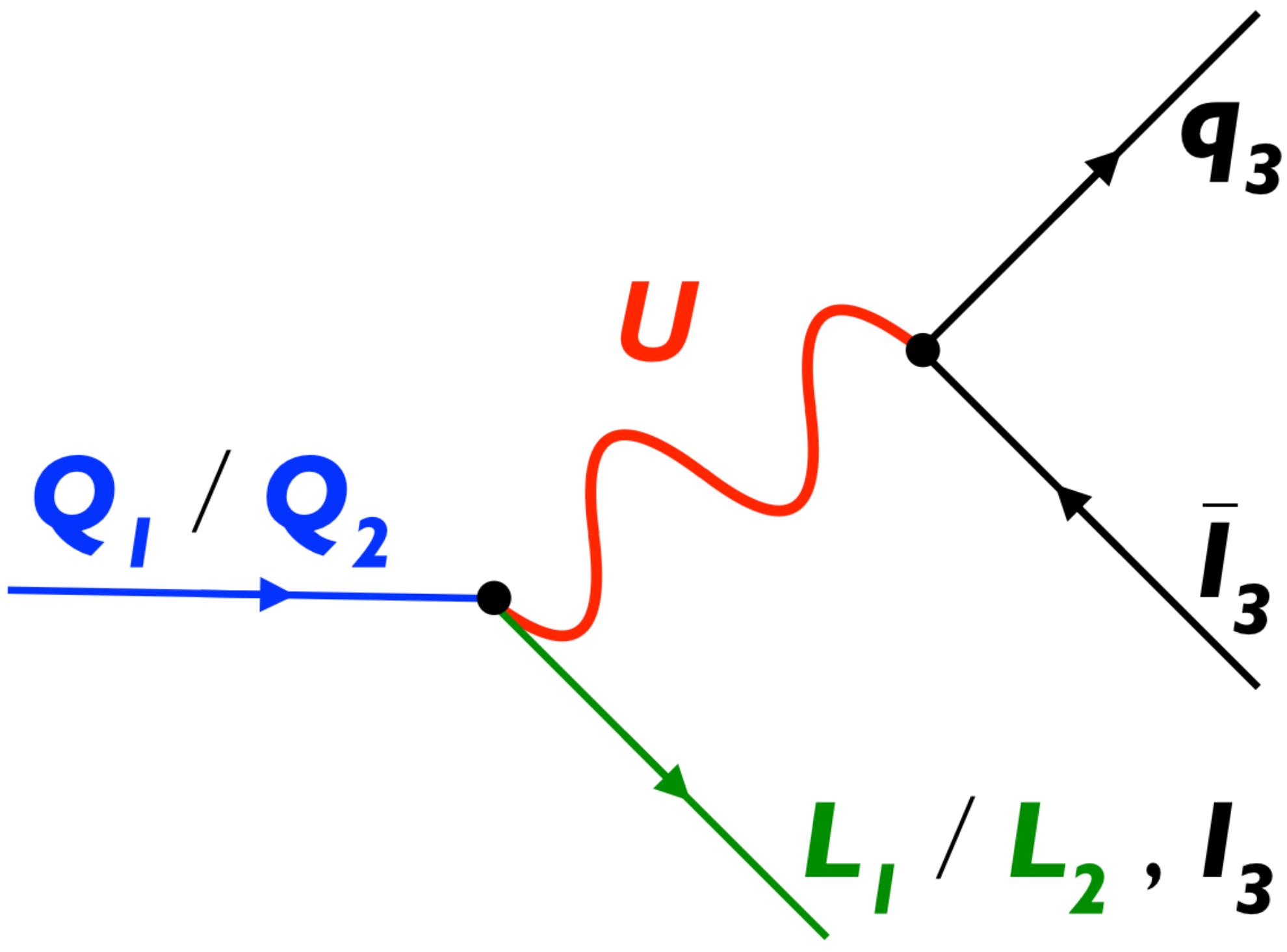} & \;
       \includegraphics[width=0.33\textwidth]{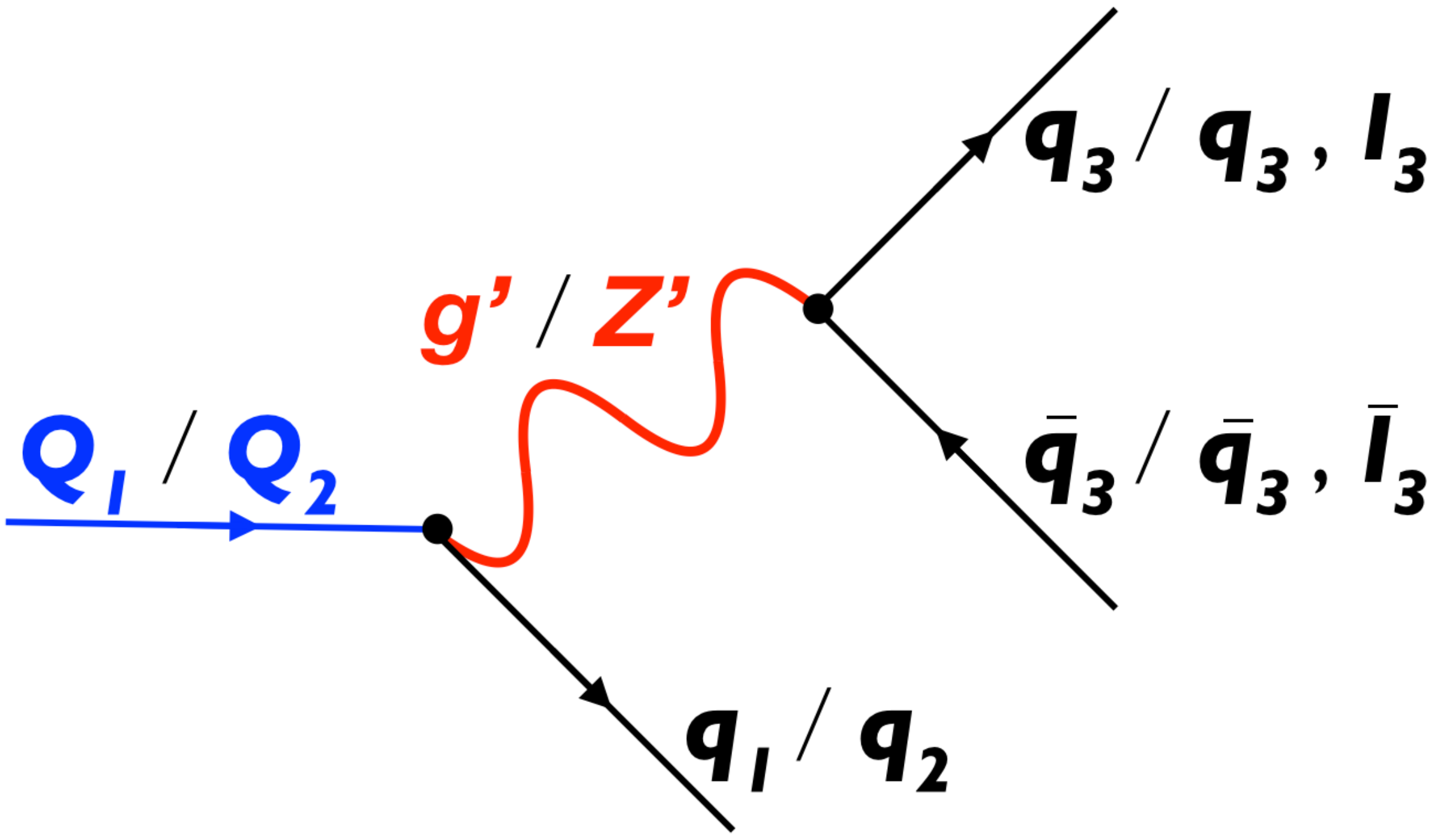} & \;
       \includegraphics[width=0.3\textwidth]{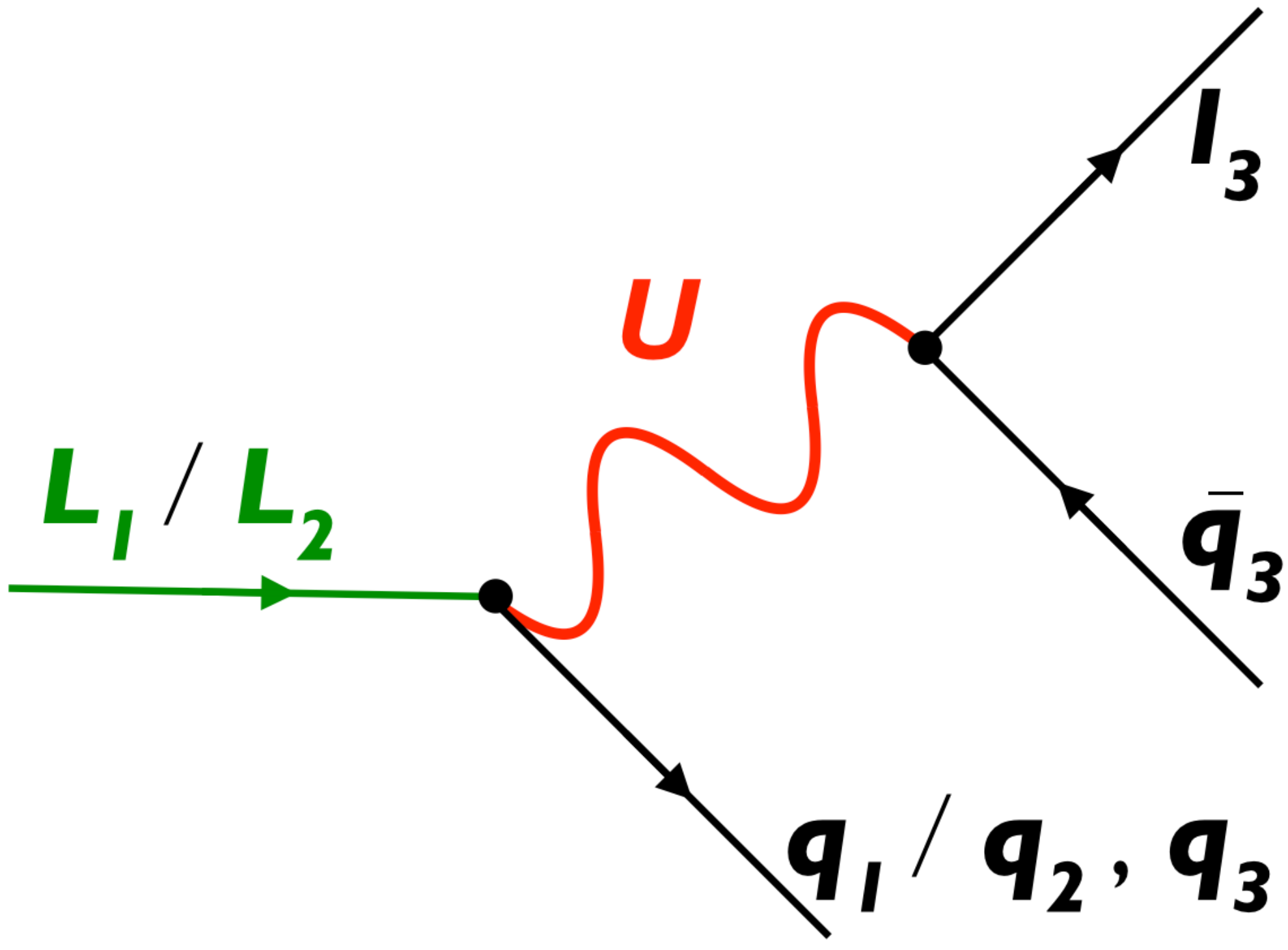}
    \\
        (a) & (b) & (c)
    \end{tabular}
 \includegraphics[width=0.5\textwidth]{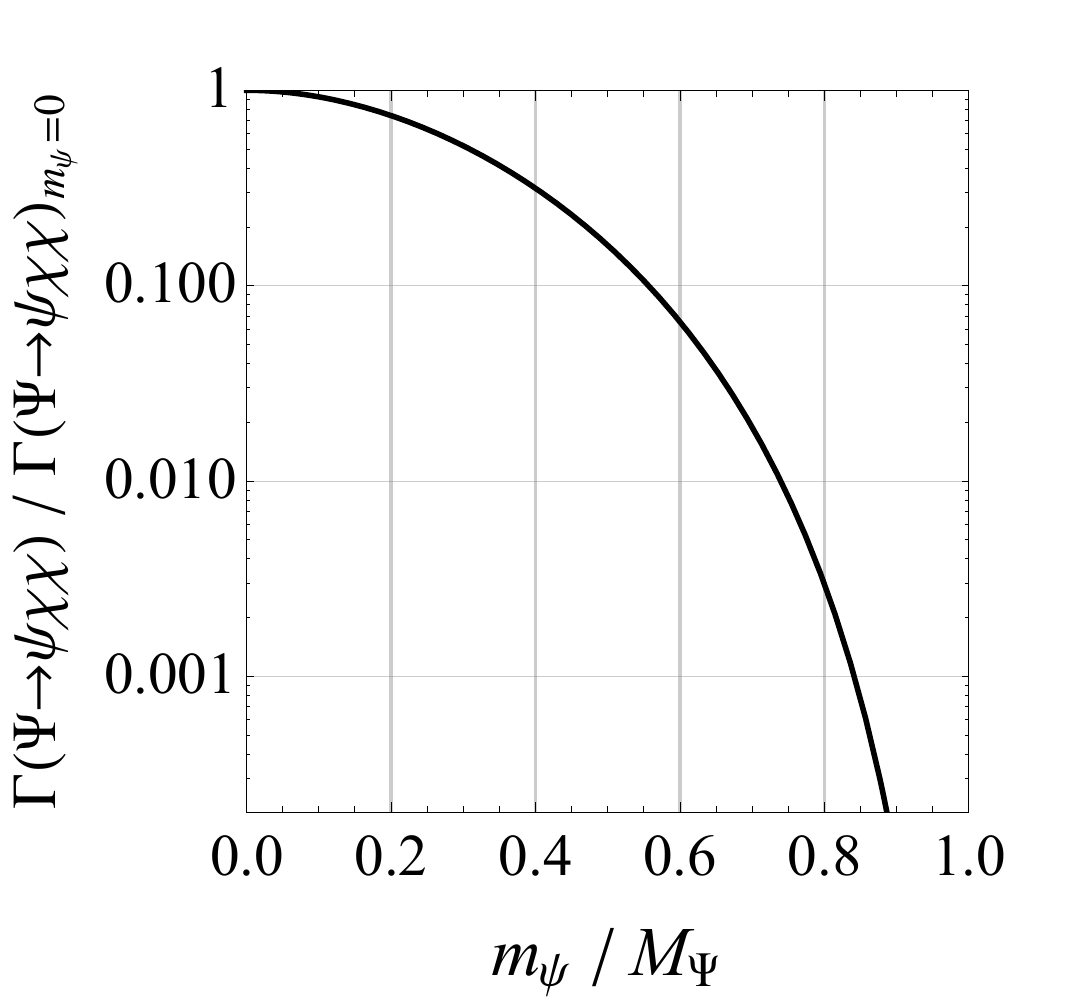}
\caption{\sf ({\bf Top panel}) Representative Feynman diagrams for dominant vector-like fermion decays. ({\bf Bottom panel}) Phase space suppression in a fermion decay to three fermions of which one is massive and two massless.  \label{fig:FeynmanProdVLFs}}
\end{figure}

\begin{table}[!]
\centering
  \begin{centering}
    \begin{tabular}{c c}
          {\sf Particle} & {\sf Decay mode} \\
   \hline
   \hline
           $N_1$, $E_1$       & $j (t \nu)$, $j (b \tau)$   \\
    \hline
  \multirow{2}{*}{$N_2$}   & $t (t \nu), t (b \tau)$          \\
                                         & $j (t \nu), j (b \tau)$           \\
    \hline
   \multirow{2}{*}{$E_2$}  & $b (t \nu), b (b \tau)$        \\
                                         & $j (t \nu), j (b \tau)$           \\

     \hline
     \hline
      \multirow{2}{*}{$U$}     & $N_1 (t \nu), N_1 (b \tau)$       \\
                                           & $j (t t), j (b b),  j (\tau \tau), j (\nu\nu)$     \\
      \hline
       \multirow{2}{*}{$D$}     & $E_1 (t \nu), E_1 (b \tau)$       \\
                                           & $j (t t), j (b b),  j (\tau \tau), j (\nu\nu)$     \\
         \hline                                  
        \multirow{3}{*}{$C$}   & $N_2 (t \nu), N_2 (b \tau)$       \\
                                           & $\nu (t \nu), \nu (b \tau)$       \\
                                           & $j (t t), j (b b),  j (\tau \tau), j (\nu\nu)$    \\
        \hline                                   
       \multirow{3}{*}{$S$}     & $E_2 (t \nu), E_2 (b \tau)$       \\
                                           & $\tau (t \nu), \tau (b \tau)$       \\
                                           & $j (t t), j (b b),  j (\tau \tau), j (\nu\nu)$   \\                                        
    \end{tabular}
  \end{centering}
  \caption{\sf Leading decay modes of vector-like fermion partners of the first and second family. The corresponding Feynman diagrams are shown in Fig.~\ref{fig:FeynmanProdVLFs} (top panel). See Sec.~\ref{sec:fermionDec} for more details.}
  \label{tab:BRf}
\end{table}

\subsubsection{Fermion decays}
\label{sec:fermionDec}

SM-like Yukawa interactions in Eq.~\eqref{LYUK1} induce a vector-like fermion decay to its SM partner and a Higgs, $W$ or $Z$ since the heavy fermion mass eigenstate has a projection over the $q'_L$ or $\ell'_L$ states. Working in the SM unbroken phase, the partial decay width for $L^a_1 \to H^a e_R$ is
\begin{equation}
\Gamma(L^a_1 \to H^a e_R) = \frac{M_L}{16 \pi} \left( \frac{m_e }{v} t_{\ell_1} \right)^2~,
\end{equation}
where $a=1,2$ denotes the component of an $SU(2)_L$ doublet and $t_{\ell_1} \equiv \tan \theta_{\ell_1}$. Analogous formulae hold for the other fermions. Being suppressed by the SM fermion mass squared, this decay channel is negligible for the fermion partners of the first and second family. Even for the charm quark partner, we find $\mathcal{B}(C \to \tilde H^0 c) < 10^{-7}$ in the interesting parameter range.\footnote{This is in contrast to the decays of $(T,B)$ due to the large top quark mass. The predictions for the branching ratios are $\mathcal{B}(T \to h t) \approx \mathcal{B}(T \to Z t) \approx 0.5$ and $\mathcal{B}(B \to W t)\approx1$. Recent dedicated experimental searches exclude $M_B < 1.35$~TeV~\cite{Aaboud:2018uek} and $M_T < 1.3$~TeV~\cite{Aaboud:2018xuw}. These are below the indicated limits from electroweak precision observables discussed in Sec.~\ref{sec:FermConst}. That is, the collider searches for the third family partners are less relevant for the spectrum on Fig.~\ref{fig:spectrum} (left panel).}  

In addition, a vector-like fermion decay to a SM fermion and a radial scalar excitation is, in principle, possible via Eq.~\eqref{LYUK2}. The precise details depend on the scalar potential, however, we expect scalar modes to be heavy enough such that on-shell $1 \to 2$ decay is kinematically forbidden.  

The dominant decay modes of the first and second family vector-like fermion partners are $1 \to 3$ processes induced via an off-shell $g'$, $U_\mu$ or $Z'_\mu$ mediator exchanged at tree-level. Typically, a heavy fermion will decay to three SM fermions of which (at least) two are third generation, or it will decay to another vector-like partner and two SM fermions (see representative Feynman diagrams in Fig.~\ref{fig:FeynmanProdVLFs} (top panel)).
To a good approximation, we can integrate out heavy vectors and work with the following effective Lagrangian,
\begin{equation}
\mathcal{L} \supset  \frac{1}{\Lambda_{\textrm{eff}}^2} \mathcal{C}^{i j k l} (\bar \Psi^i_L \gamma^\mu \psi^j_L) (\bar \chi^k_L \gamma_\mu \chi_L^l ) + \textrm{h.c.}~,
\end{equation}
where $\mathcal{C}^{i j k l}$ is the colour tensor, $\Psi$ is the decaying fermion (in this case $Q$ or $L$) with color index $i$ and mass $M_\Psi$, while $\psi$ is a massive/massless final state fermion with color index $j$ and mass $m_\psi$. Also, $\chi$ and $\bar \chi$ are two massless SM fermions with colour indices $k$ and $l$, respectively. We consider fermions to be triplets or singlets of colour. The partial decay width following from this Lagrangian is
\begin{equation}
\Gamma(\Psi \to \psi \chi \bar \chi) = N_{\mathcal{C}}\, \frac{1}{\Lambda_{\textrm{eff}}^4}\, \frac{M_\Psi^5}{1536 \pi^3} \, F\left(\frac{m^2_\psi}{m^2_\Psi}\right)
\end{equation} 
where $N_{\mathcal{C}}$ is the color factor depending on the $\mathcal{C}^{i j k l}$. For example, for $Q_1^a \to L_1^a  \ell_3^b q_3^b$, the $\mathcal{C}^{i j k l} \to \delta^{i l}$ with indices $j, k$ trivial and $N_{\mathcal{C}}=1$. Another example is $L_1^a \to q_1^a q_3^b \ell_3^b$, where $\mathcal{C}^{i j k l} \to \delta^{j k}$ with indices $i, l$ trivial and $N_{\mathcal{C}} = 3$. (Here, $SU(2)_L$ indices $a,b$ are fixed and not summed over.)  The phase space suppression for $\Psi \to \psi \chi \chi$ decay with massive $\psi$ and massless $\chi, \bar \chi$ is~\cite{vanRitbergen:1999fi}
\begin{equation}
F(x) = 1 - 8 x - 12 x^2 \ln(x) + 8 x^3 - x^4~.
\end{equation}
This function is plotted in Fig.~\ref{fig:FeynmanProdVLFs} (bottom panel), showing rather large suppression factors for sizable $m_\psi / M_\Psi$. Using these relations, we calculated the partial decay widths and identified the leading vector-like fermion decay modes in Table~\ref{tab:BRf}. The precise branching ratios depend strongly on the benchmark point. For example, for the selected BP, diagrams (a) and (b) from Fig.~\ref{fig:FeynmanProdVLFs} (top panel) lead to rates of similar sizes, which is, however, highly sensitive on the $M_L / M_Q$ ratio, see Fig.~\ref{fig:FeynmanProdVLFs} (bottom panel). It is also interesting to note that these resonances are rather narrow, $\Gamma / M \sim \mathcal{O}(10^{-4})$.

Loop-induced $1 \to 2$ decays can, in principle, compete with tree-level $1 \to 3$ decays. An example in the $4321$ model is $C \to t \gamma$ with the dipole operator generated by the heavy neutral lepton and vector letoquark in the loop. These decays are typically sub-leading in the relevant parameter space due to an extra suppression from the electroweak gauge coupling.

To sum up, the $4321$ model predicts drastically different signatures of light vector-like fermion partners from those currently being searched for by experiments (see e.g. Ref.~\cite{Sirunyan:2017lzl}).

\subsection{Collider constraints}
\label{colliderconstr}

In this section we investigate the most stringent current LHC limits on the $4321$ model, and propose novel (exotic) collider signatures for future searches. As a recap, the core implication of the $R(D^{(*)})$ anomaly is $i)$ a relatively light vector leptoquark and $ii)$ relatively light vector-like leptons -- to simultaneously pass the bounds from $\Delta F = 2$ transitions. 

\begin{figure}[t]
\centering
  \begin{tabular}{c c}
    \includegraphics[width=0.32\textwidth]{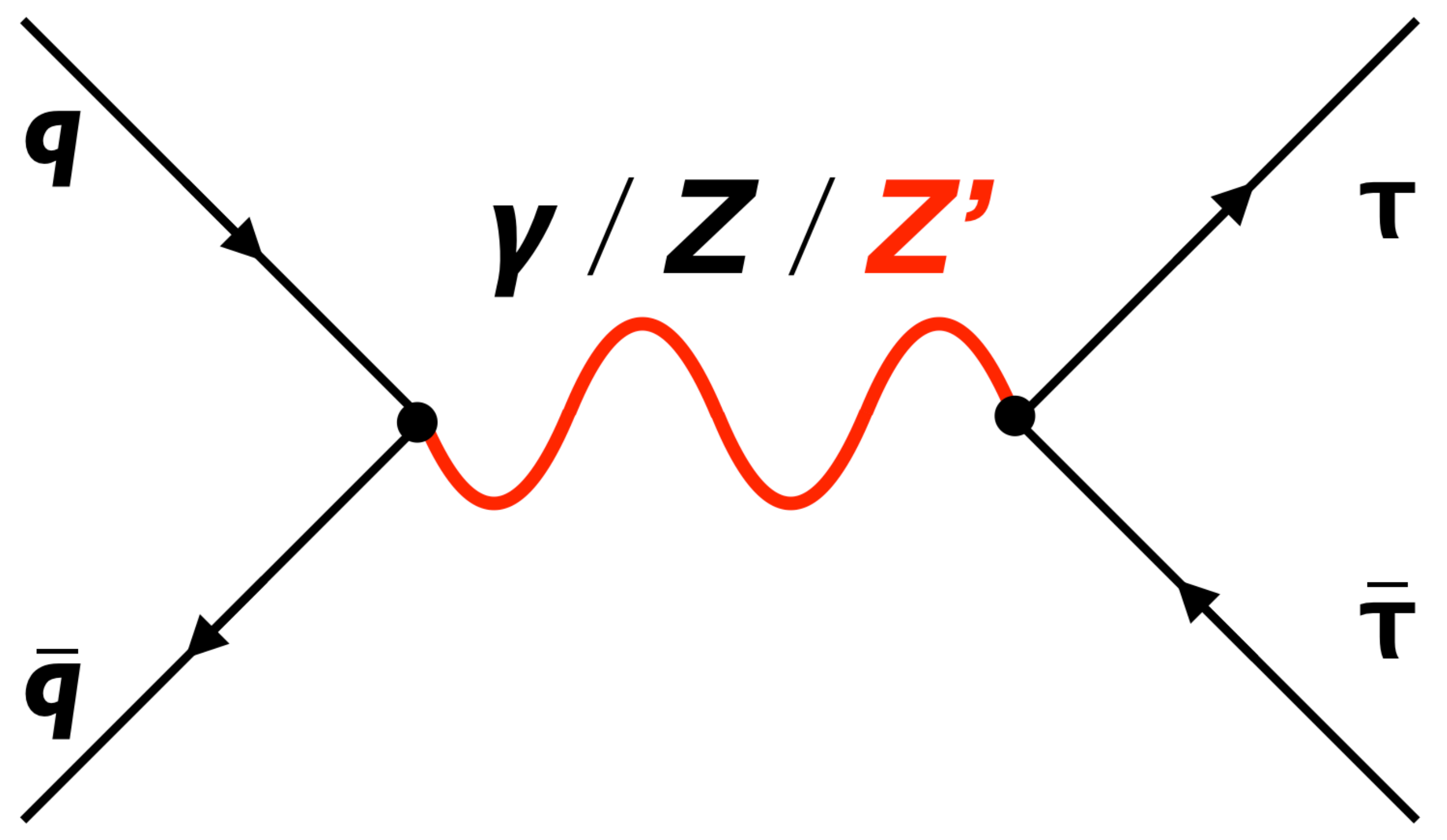} & \;\;\;\;\;
    \includegraphics[width=0.32\textwidth]{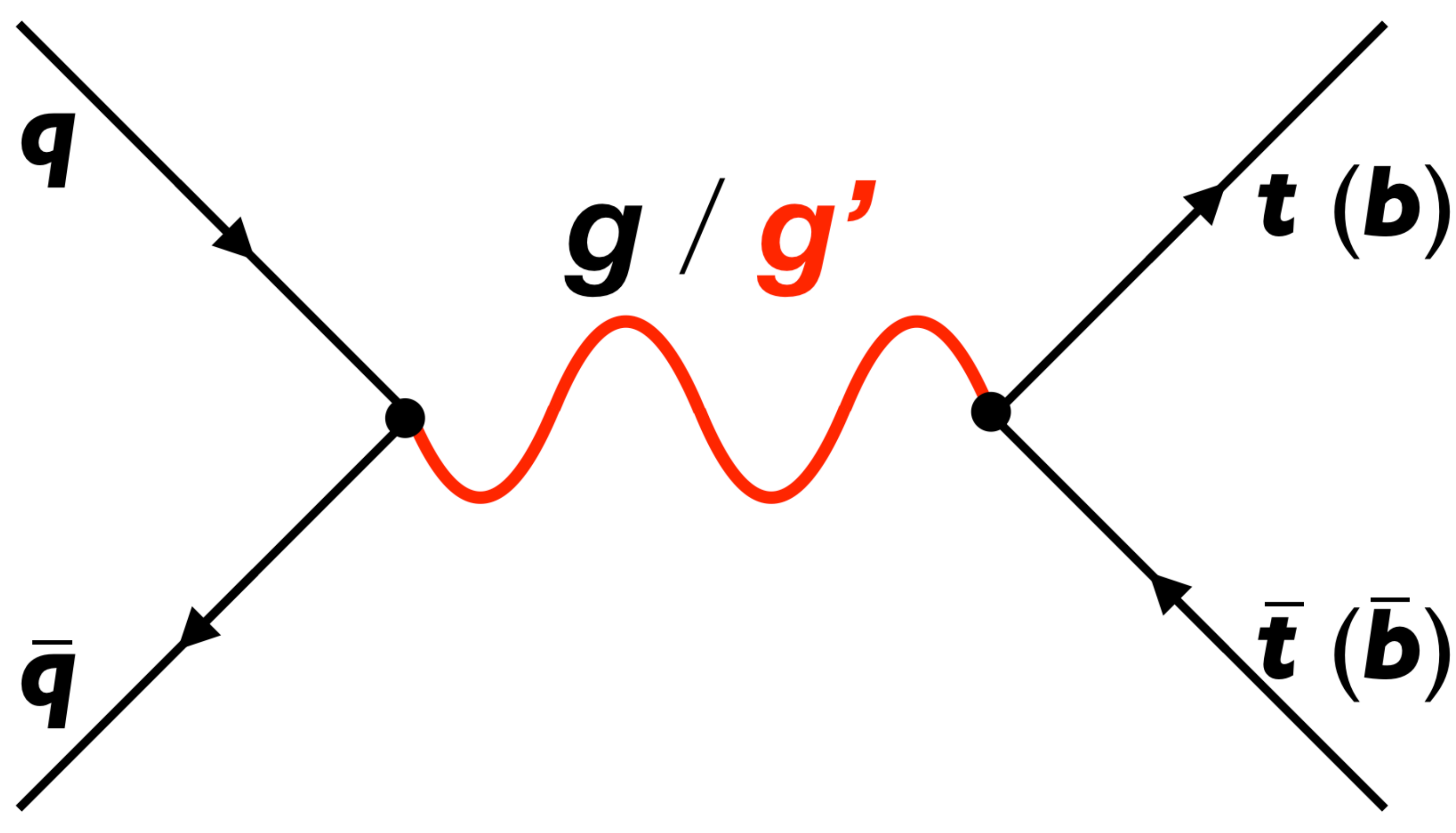} \\
        (a) & (b)
  \end{tabular}
\caption{\sf Feynman diagrams for the coloron and $Z'$ searches. \label{fig:FeynmanProdVectors}}
\end{figure}

By the model construction, the accompanying vector resonances are close in mass to the leptoquark, and are resonantly produced in $pp$ collisions. The strongest collider constraints are due to an $s$-channel coloron (or $Z'$) decaying to a pair of third family SM femions, see Fig.~\ref{fig:FeynmanProdVectors}. Such final state has $i)$ a large branching ratio and $ii)$ a simple topology. Although these topologies have been extensively exploited by experiments, a simple interpretation in terms of a narrow-width resonance fails to capture the effect, and a slight complication arises in properly including finite width and interference effects. By performing a dedicated recast of the existing dijet and $t \bar t$ searches, we show how to consistently extract bounds on the model's parameter space.

An essential ingredient of the $4321$ model is the existence of heavy SM fermion partners with masses below the vector boson spectrum -- with peculiar new decay channels leading to exotic final states with multiple jets and/or leptons -- a distinct smoking gun signature of the model. Here we provide a catalog of promising topologies and estimate their potential future impact.

\subsubsection{Coloron searches in $t \bar t$ and $b \bar b$ final states}
\label{colliderconstr:col}

\begin{figure}[t]
  \centering
  \includegraphics[width=0.45\textwidth]{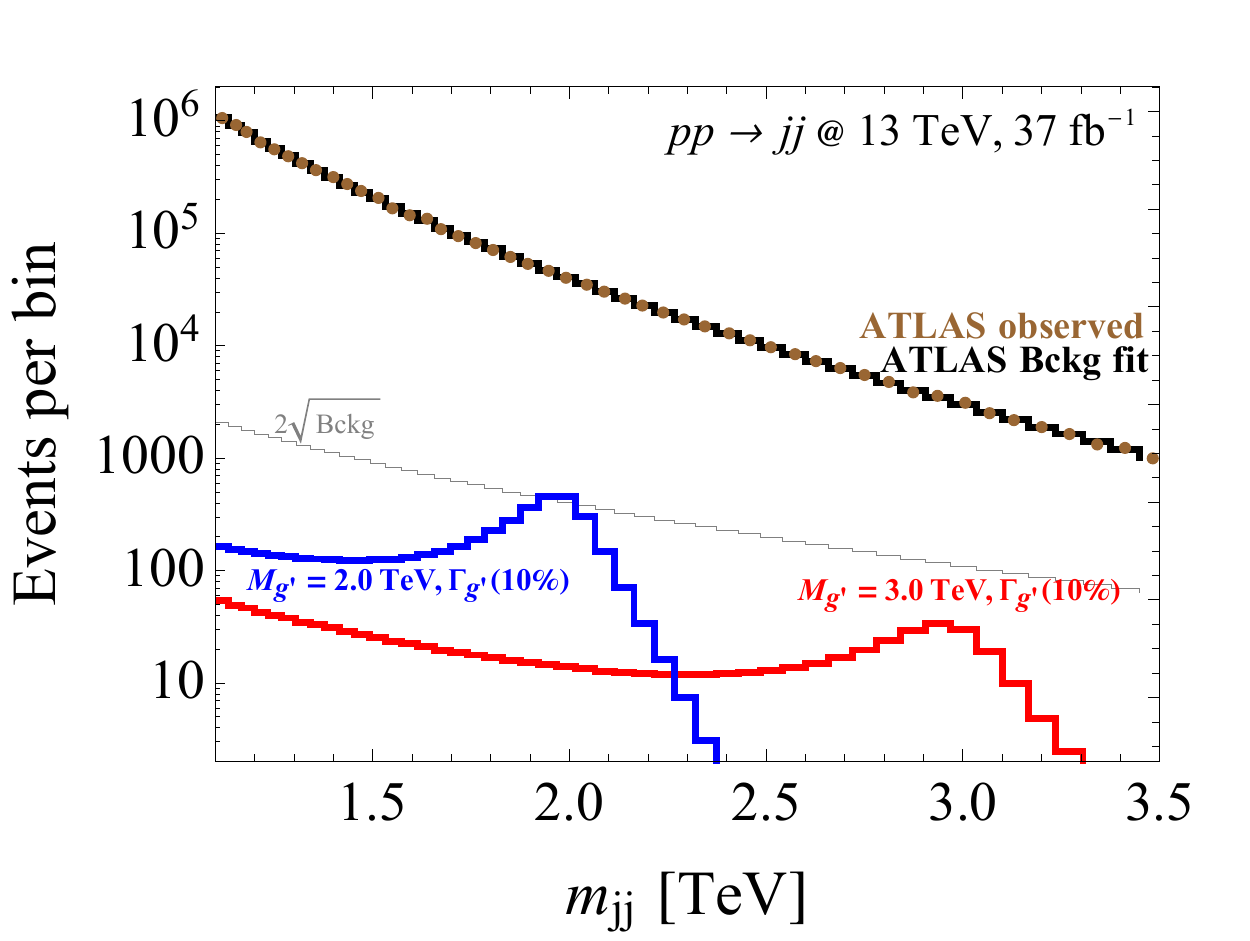} \; \; \; \;
  \includegraphics[width=0.45\textwidth]{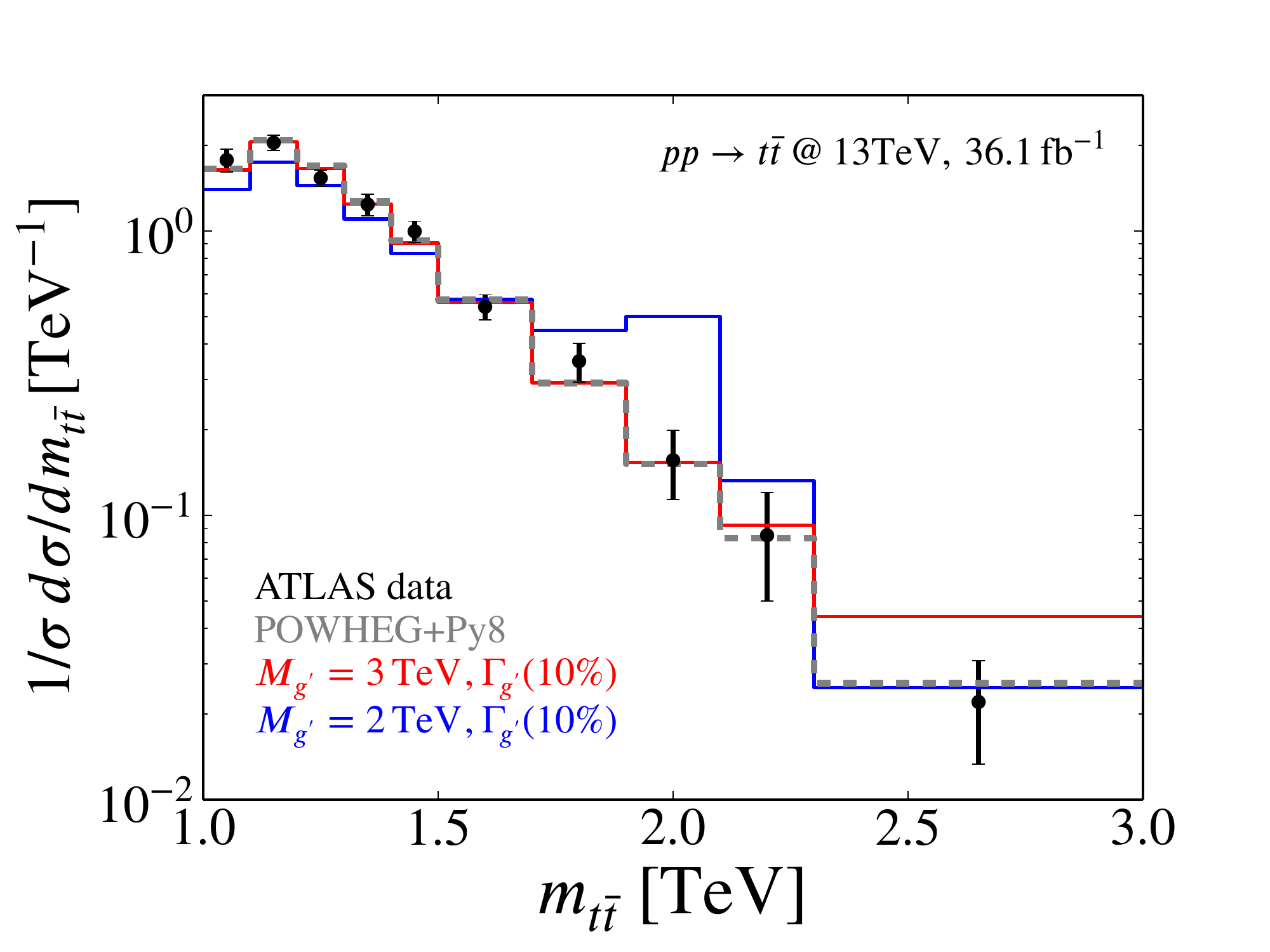}\\
    \includegraphics[width=0.45\textwidth]{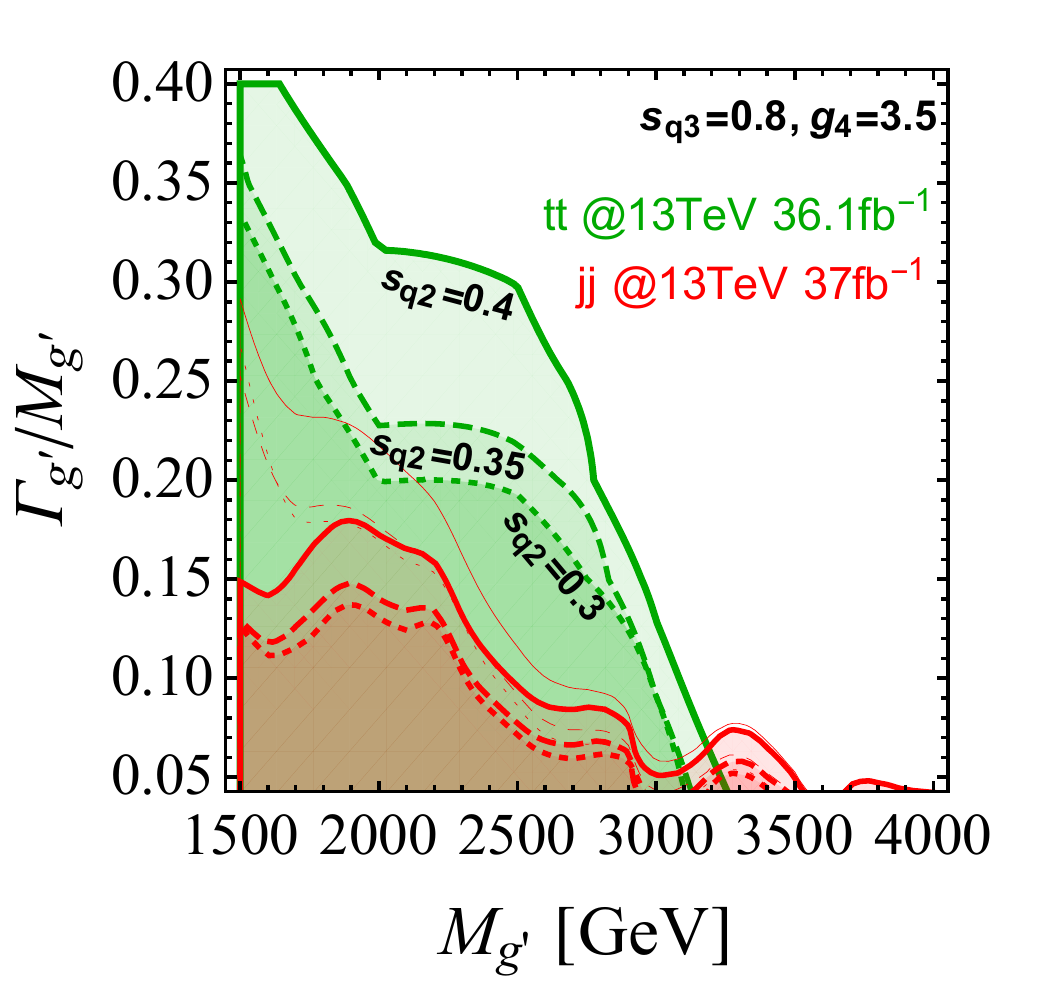}
    \includegraphics[width=0.45\textwidth]{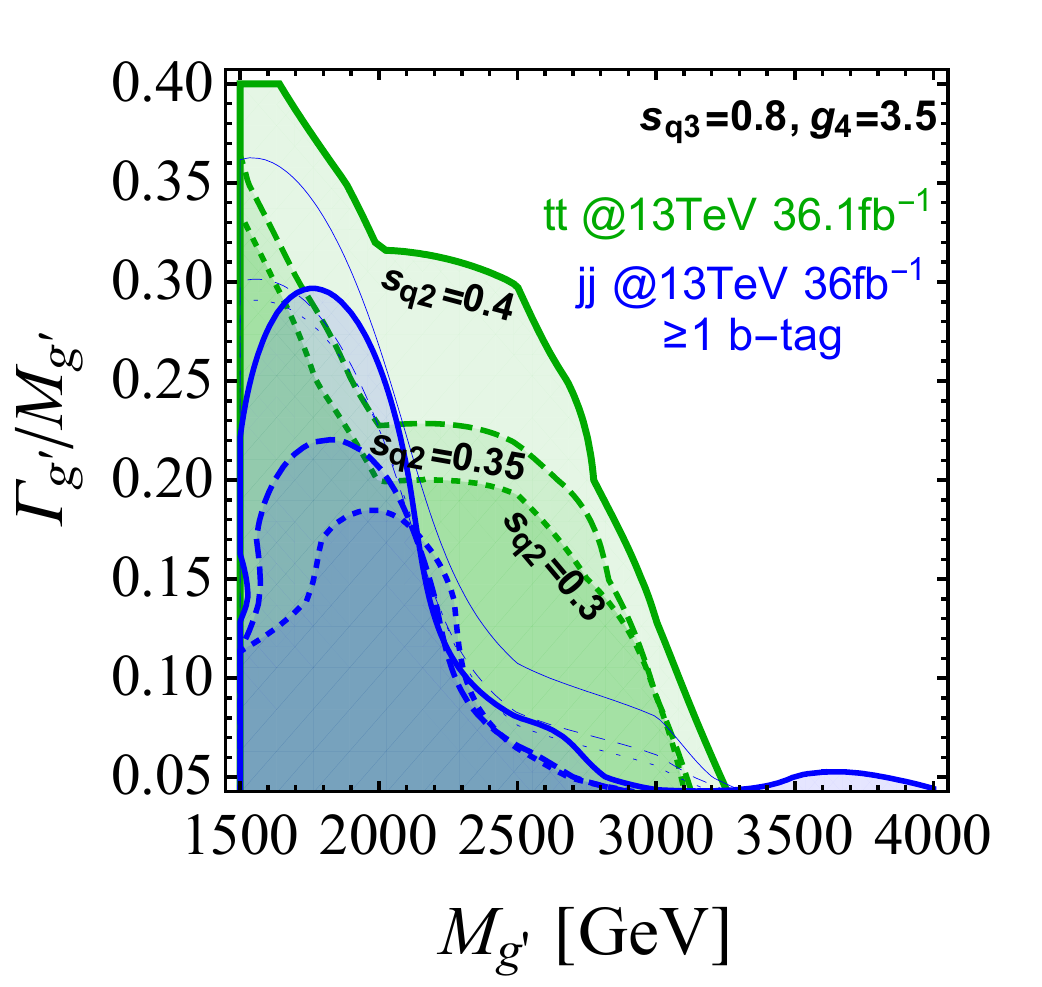}
    \caption{\sf ({\bf Top panel}) Coloron contribution to the $p p \to j j$ (left panel) and $p p \to \bar t t$ (right panel) invariant mass spectrum for two representative benchmark points. ({\bf Bottom panel}) Coloron exclusion limits in the mass-total width plane for $jj$ and $t\bar t$ for several representative $s_{q_2}$ benchmarks. \label{fig:coloron}}
\end{figure}

The dominant production mechanism of the colour octet $g^{\prime}$ in $pp$ collisions is resonant production from a quark-antiquark pair, $q \bar  q \to g^{\prime}$. There is no tree-level coupling between a single $g'$ and a $g g$ pair, see \app{gbvrtx}. Due to the flavour structure of the model, the couplings to light quarks are suppressed, however the PDF enhancement of valence quarks relative to third generation quarks in the proton ensures that this channel is nevertheless dominant. The interesting regimes of the model are when the width is rather large (but still calculable) or the resonance is narrow but rather heavy.

Existing analyses which are most sensitive to the coloron are an ATLAS $t \bar{t}$ invariant mass measurement~\cite{Aaboud:2018eqg}, an ATLAS dijet resonance search~\cite{Aaboud:2017yvp}, and an ATLAS dijet resonance search with one or two jets identified as $b$ jets~\cite{Aaboud:2018tqo}.  The relevant Feynman diagrams are shown in Fig.~\ref{fig:FeynmanProdVectors}; the largest contribution to the dijet process is through production of a left-handed $b\bar b$ pair. 

We calculate the model predictions for the $t\bar t$ process using {\tt Madgraph5\_aMC@NLO}~\cite{Alwall:2014hca}, implementing the coloron and its interactions in {\tt FeynRules}~\cite{Alloul:2013bka}, and using the default {\tt NNPDF2.3} leading order PDF set~\cite{Ball:2012cx}. The representative benchmark examples are shown in Fig.~\ref{fig:coloron} (top right panel). We use the measured unfolded, parton-level invariant mass distribution, which allows direct comparison to parton level predictions, and involves a cut of $p_T>500$ GeV for the leading top quark, and $p_T>350$ GeV for the second leading top quark. Exclusion regions are then calculated from the measured $t\bar t$ invariant mass spectrum~\cite{Aaboud:2018eqg} requiring $\Delta \chi^2 > 6.2$.
The excluded regions found in this way, for 3 different values of $s_{q_2}$, are shown in green in Fig.~\ref{fig:coloron} (bottom panel). As shown in Fig.~\ref{fig:spectrum} (right panel), the coloron coupling to left-handed valence quarks depends on $s_{q_2}$ and leads to the reduced coloron production for $s_{q_2} \sim 0.3$.

Additionally, exclusion regions are calculated from an ATLAS dijet resonance search~\cite{Aaboud:2017yvp}. The search involves dijet events with $m_{jj}>1.1$ TeV, for which the transverse momentum of the leading (subleading) jet is greater than 440 (60) GeV, and the rapidity difference between the jets is less than 0.6.
Cross sections differential in the invariant mass are calculated for the process $p p \to b \bar b$ using MSTW PDF sets~\cite{Martin:2009iq}, including the effects of interference between the SM and coloron-mediated diagrams. Note that $g' \to b \bar{b}$ is by far the dominant coloron dijet decay.) We estimated the signal acceptance in the relevant invariant mass region to be about $35\%$.

Following an ATLAS method, we determine whether bumps could be seen in the total invariant mass spectrum by fitting the background with a curve $f$ defined as
\begin{equation}
\label{eq:fitfunction}
f(z)=p_1 (1-z)^{p_2} z^{p_3} z^{p_4 \log(z)},
\end{equation}
where $z=m_{jj}/\sqrt{s}$.\footnote{In Ref.~\cite{Aaboud:2017yvp}, the analysis in fact made use of a novel fit method with a sliding window, such that in each section of the spectrum defined by the window, a new three-parameter fit was made. However, they compared both methods and found compatible results between this sliding window method and the traditional global four-parameter fit described here, so we use the four-parameter fit model for simplicity.} In each case, this parameterised curve is binned and added to the binned new physics contribution, and the $\chi^2$ calculated by comparison with the ATLAS measured data, assuming poissonian errors on the data. Each of the curve parameters $p_i$ is allowed to vary independently to minimise the $\chi^2$ value, and this minimum $\chi^2$ is used to determine whether the coloron parameter point is ruled out if $\Delta \chi^2 > 6.2$. The resulting exclusion regions are shown in red in Fig.~\ref{fig:coloron} (bottom left panel), for three different $s_{q_2}$ values.

A recent ATLAS search~\cite{Aaboud:2018tqo} looks for bumps, in a very similar way, in the invariant mass spectrum of dijet events for which one or both of the leading jets pass $b$-tagging requirements. Since our coloron-mediated dijet signal is made up almost entirely of $b\bar b$ pair production events, this is clearly an important search. In the ``high-mass region'' for which the invariant mass of the dijet pair is $m_{jj}>1.2$ TeV, the analysis requires that the transverse momentum of the leading (subleading) jet is greater than 430 (80) GeV. Both leading jets are additionally required to have pseudorapidity $|\eta|< 2.0$, and the rapidity difference between them is required to be less than 0.8. 
We use the fitting method described above (Eq.~\eqref{eq:fitfunction}) to extract exclusion regions from the measured invariant mass spectrum requiring $\geq 1$ $b$-tag.\footnote{The $\geq 1$ $b$-tag selection was chosen rather than the 2 $b$ tag selection because the signal efficiency becomes very small for the 2 $b$-tag selection.} The $b$-tag efficiency for the signal events is taken from Fig.~ 2 (a) of~\cite{Aaboud:2018tqo}. The exclusion regions found in this way are shown in blue in Fig.~\ref{fig:coloron} (bottom right panel), for three different $s_{q_2}$ values.

Some discussion of the different shapes and reaches of the $t\bar{t}$ and dijet exclusions shown in Fig.~\ref{fig:coloron} (bottom panel) is in order. The green $t \bar t$ regions exclude even large widths, because the $t\bar t$ predictions exist for the SM. This means that even for large widths, when the signal is spread over many bins, the discrepancy from the SM can still be apparent. The sensitivity falls off sharply around coloron masses of 3 TeV, because the spectrum is only measured up to $m_{t\bar t}=3$ TeV, and in the last bin the error on the data is already rather large. By contrast, for the dijet bump hunts, the SM background must be simply fitted to the data. So if the coloron has a very large width, such that its effects are spread over many bins, then the signal can be hidden within the background fit, and the bump hunt is no longer sensitive.  This is why the red and blue dijet regions do not reach to such large widths as the green $t\bar t$ regions. The thin red lines represent dijet limits when fixing the background to the SM-only fitted value, rather than profiling. Indeed, these show similar behaviour to $t\bar t$ exclusions.

For low coloron masses, the blue region found from the bump hunt with $b$-tags reaches larger widths than the red region found from the bump hunt without $b$-tags. This is because the $b$-tag requirement increases the signal over background ratio for dijet invariant masses below around 2.5 TeV. This advantage disappears for larger coloron masses because the signal $b$-tagging efficiency decreases for higher invariant masses. Finally, we would like to point out that a different $b$-tagging (and misidentification) operating point choice might be more optimal for our signal. In particular, one might try a tighter $b$-tagging requirement with rather severe background rejection rate.

\subsubsection{$Z'$ search in $\tau^+ \tau^-$ final state}
\label{colliderconstr:ta}

Production of high-$p_T$ $\tau^+ \tau^-$ pairs in $p p$ collisions (e.g.~\cite{Aaboud:2017sjh}) has been identified as a generic signature of models addressing $R(D^{(*)})$ anomalies~\cite{Faroughy:2016osc}. The $4321$ model is not an exception, as the effect comes from an on-shell $Z'$ boson, i.e. $p p \to Z' \to \tau^+ \tau^-$. 

Let us, for a moment, assume that the $Z'$ exclusively decays to SM fermions (unlike in the chosen benchmark). For large $s_{q_3} \approx s_{\ell_3}$ (and small $s_{q_2}$) the $Z'$ decay width is saturated by decays to $b \bar b$, $t \bar t$, $\nu \bar \nu$ and $\tau^+ \tau^-$, with a branching ratio $\mathcal{B}(Z' \to \tau^+ \tau^-) \sim 3/8$. The total $Z'$ decay width, for $g_4 \sim 3$, is at the level of $\Gamma / M \sim 10 \%$. For a small $s_{q_2}$, the dominant production mechanism is from $b_L \bar b_L \to Z'$, followed by $u_R \bar u_R$ fusion (a factor of $\sim 4$ smaller). Increasing $s_{q_2}$ leads to sizeable increase in the production cross section from the valence quarks, $\bar u_L u_L $ and $\bar d_L d_L$. The observed upper limit on the narrow resonance $\sigma \times \mathcal{B} (Z' \to \tau^+ \tau^-)$ is about $\lesssim10$~fb for the $Z'$ masses in the 1 - 3~TeV range (see Figure 7 (c) in the latest ATLAS search done at 13~TeV with 36~fb$^{-1}$~\cite{Aaboud:2017sjh}). If we assume (conservatively) that the $Z'$ is produced exclusively from bottom-bottom fusion, without even considering extra $u \bar u$ and $d \bar d$ channels, these constraints imply that $m_Z' \gtrsim 1.8$~TeV. This illustrates the tension with the present data if the $Z'$ exclusively decays to SM fermions.

On the other hand, the reference benchmark point easily avoids the $\tau^+ \tau^-$ bound due to extra open decay channels to vector-like lepton partners (see Table~\ref{tab:BRv}). Interestingly enough, these states are also required to be light for the consistency with $\Delta F = 2$  observables. The extra decay channels ensure $i)$ a diluted branching ratio to $\tau^+ \tau^-$ and $ii)$ a large total decay width ($\Gamma / M \sim 40 \%$) which reduces the effectiveness of the search (see Fig.~[4] in Ref.~\cite{Faroughy:2016osc}). Finally, we note that the limits from $Z' \to \mu^+ \mu^-$ decay are irrelevant due to the small $s_{\ell_2} \sim 0.1$.

\subsubsection{Leptoquark signatures}

Vector leptoquarks are copiously produced in pairs via QCD interactions. For more details on their phenomenology at hadron colliders, we refer the reader to Sec.~2.2 of Ref.~\cite{Dorsner:2018ynv}. We compute the leptoquark pair production cross-section at LO in QCD in $p p$ collisions at 13~TeV, 
using the {\tt FeynRules} model implementation of Ref.~\cite{Dorsner:2018ynv} with {\tt MadGraph5\_aMC@NLO} (see also Ref.~\cite{Blumlein:1996qp}). The results are shown in Fig.~\ref{fig:LQcrosssections} with a solid black line. It is worth noting the fast drop of the cross section with the leptoquark mass.

Vector leptoquark decays to $t \nu$ or $b \tau$ final states with large branching ratios. (Other relevant decays are listed in Table~\ref{tab:BRv}.) A dedicated analysis targeting the simplified dynamical model of Ref.~\cite{Buttazzo:2017ixm} has recently been performed by CMS~\cite{Sirunyan:2018kzh}, excluding pair-produced leptoquarks with masses $m_{U_1} < 1.53$~TeV, under the assumption of $\mathcal{B}(U_1 \to t \nu) = \mathcal{B}(U_1 \to b \tau) = 0.5$. This limit is also shown in Fig.~\ref{fig:LQcrosssections} (grey region). For the benchmark point in Table~\ref{tab:BRv}, this bound is slightly relaxed due to somewhat smaller branching ratio. The first lesson is that direct bounds on leptoquarks cannot compete with those indirectly inferred from e.g. coloron exclusions.

\begin{figure}[t]
\centering
 \begin{tabular}{c c}
    \includegraphics[width=0.3\textwidth]{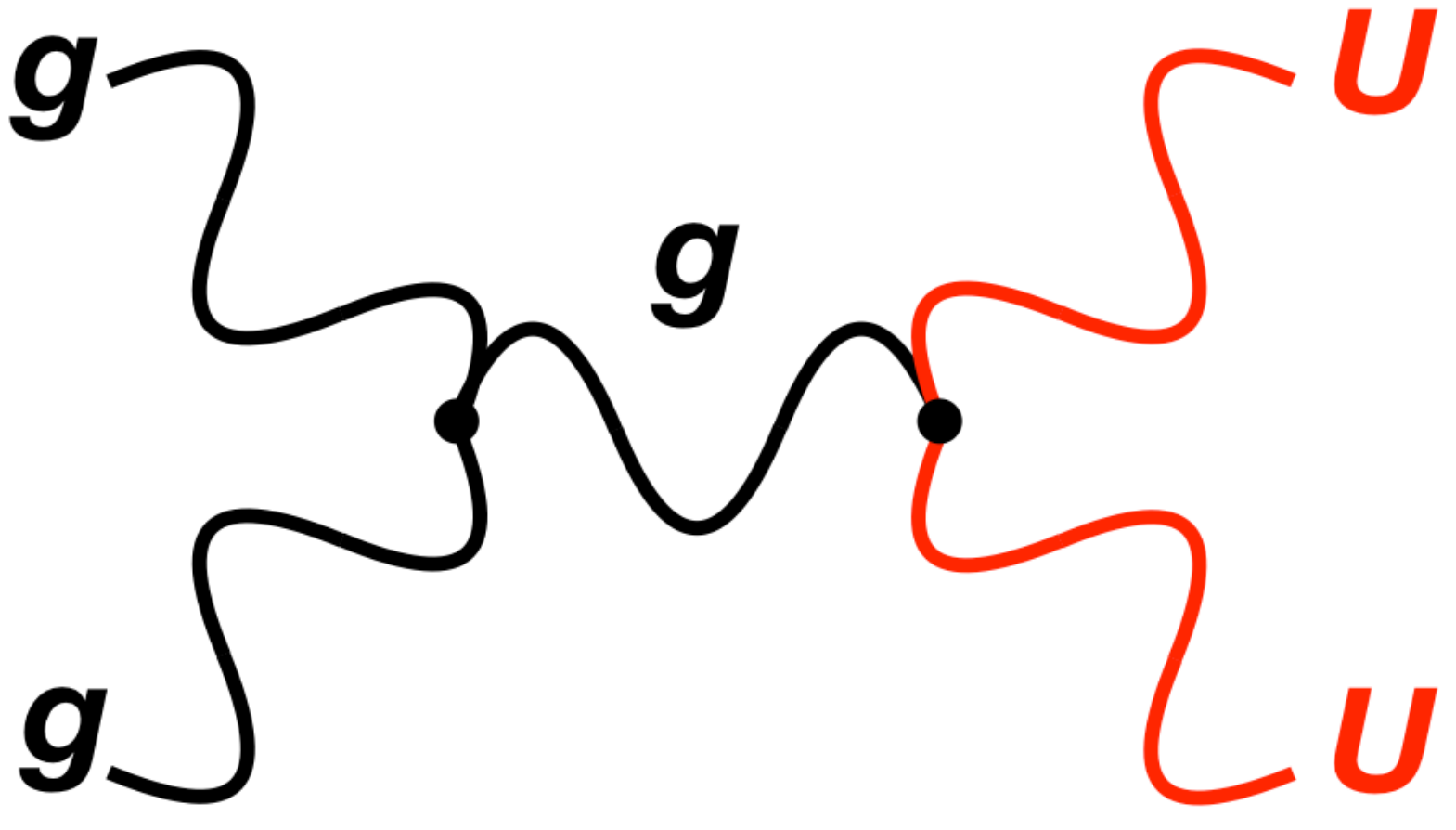} & \;\;\;\;\;
    \includegraphics[width=0.33\textwidth]{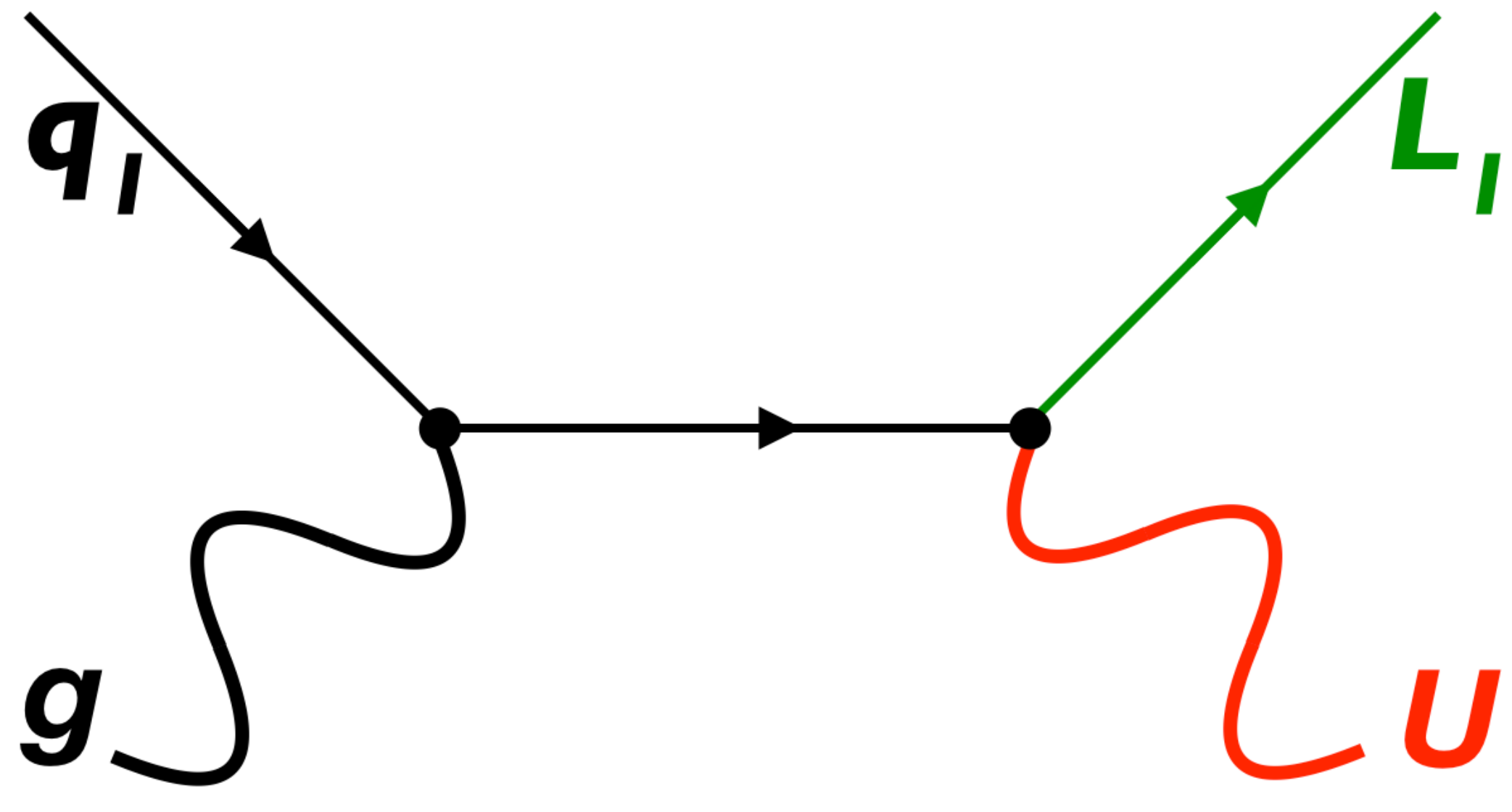} \\
        (a) & (b)
  \end{tabular}
\caption{\sf Representative Feynman diagrams for (a) vector leptoquark pair production and (b) vector leptoquark production in association with an $L_1$ lepton partner. \label{fig:FeynmanVecLQ}}
\end{figure}

\begin{figure}[t]
\centering
    \includegraphics[width=0.5\textwidth]{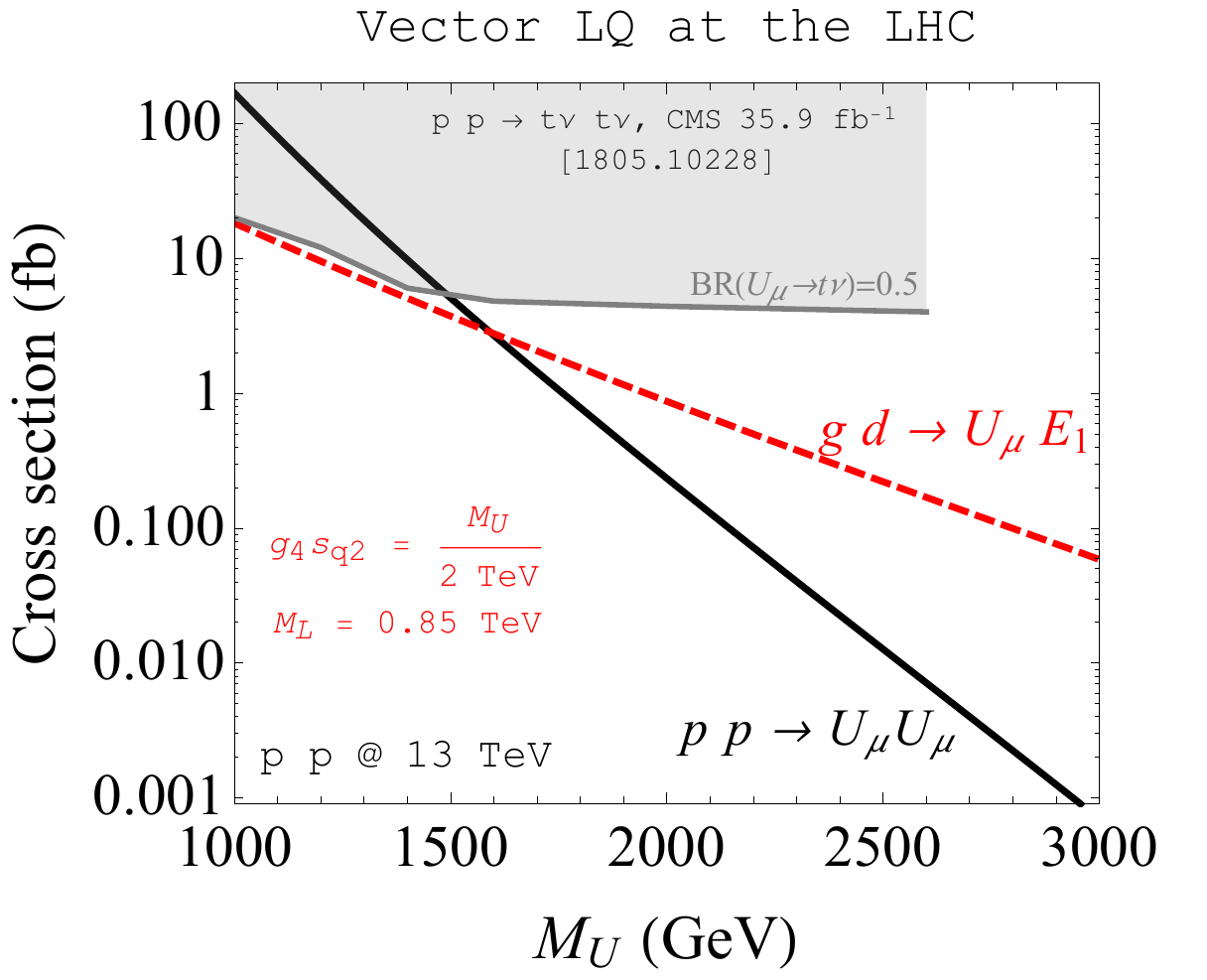} 
\caption{\sf Vector leptoquark ($U_\mu$) production cross section (in fb) at the 13 TeV LHC. The solid black line shows the leptoquark pair production cross section via QCD interactions.  The dashed red line represents $p p \to E_1 + U_\mu$ production cross section, fixing $g_4 s_{q_2} = M_U / 2$~TeV to fit the $R(D^{(*)})$ anomaly, and with the $E_1$ lepton partner mass set to $M_L = 0.85$~TeV. The grey region is excluded by the CMS leptoquark search in the $(t\nu)(t\nu)$ final state, assuming $\mathcal{B}(U_\mu \to t \nu)=0.5$~\cite{Sirunyan:2018kzh}.\label{fig:LQcrosssections}}
\end{figure}

In fact, as the experimental searches are moving forward, the dominant mechanism for on-shell leptoquark production will instead become $g \,q_1 \to L_1 \,U_\mu$, see Fig.~\ref{fig:FeynmanVecLQ}. As shown in Fig.~\ref{fig:LQcrosssections} (red dashed line), the cross section for the single leptoquark production in association with a vector-like lepton $E_1$ dominates over the leptoquark pair production for large $M_U$. In this calculation, we fix $M_L = 0.85$ and $g_4 s_{q_2} = M_U / (2~$TeV), as indicated by $R(D^{(*)})$ anomaly. The present excluded mass reach from the CMS search~\cite{Sirunyan:2018kzh} is already nearing the point where this channel becomes dominant, and suggests reconsideration of the working strategy to search for our leptoquark.

In addition to extending the scope to a novel production channel, we also suggest searching in new decay modes as listed in Table~\ref{tab:BRv}. For example, there is a significant branching fraction to $U_\mu \to q_3 L_2 \to q_3 (q_3 \ell_3 q_3)$ -- clearly calling for a dedicated experimental analysis.

%

\subsubsection{Vector-like lepton production}
\label{colliderconstr:Leptons}

\begin{figure}[t]
\centering
\includegraphics[width=0.47\textwidth]{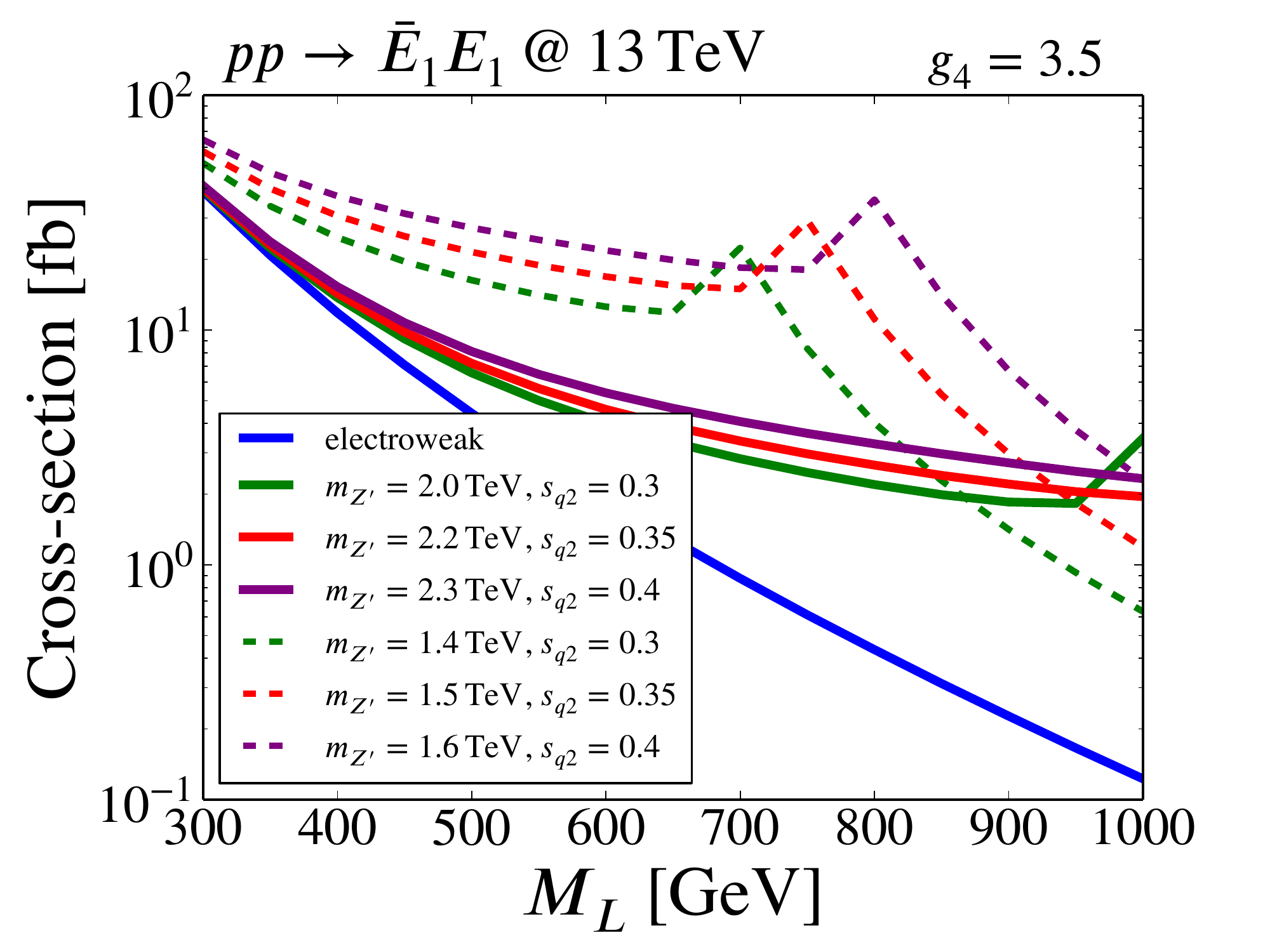}
\includegraphics[width=0.47\textwidth]{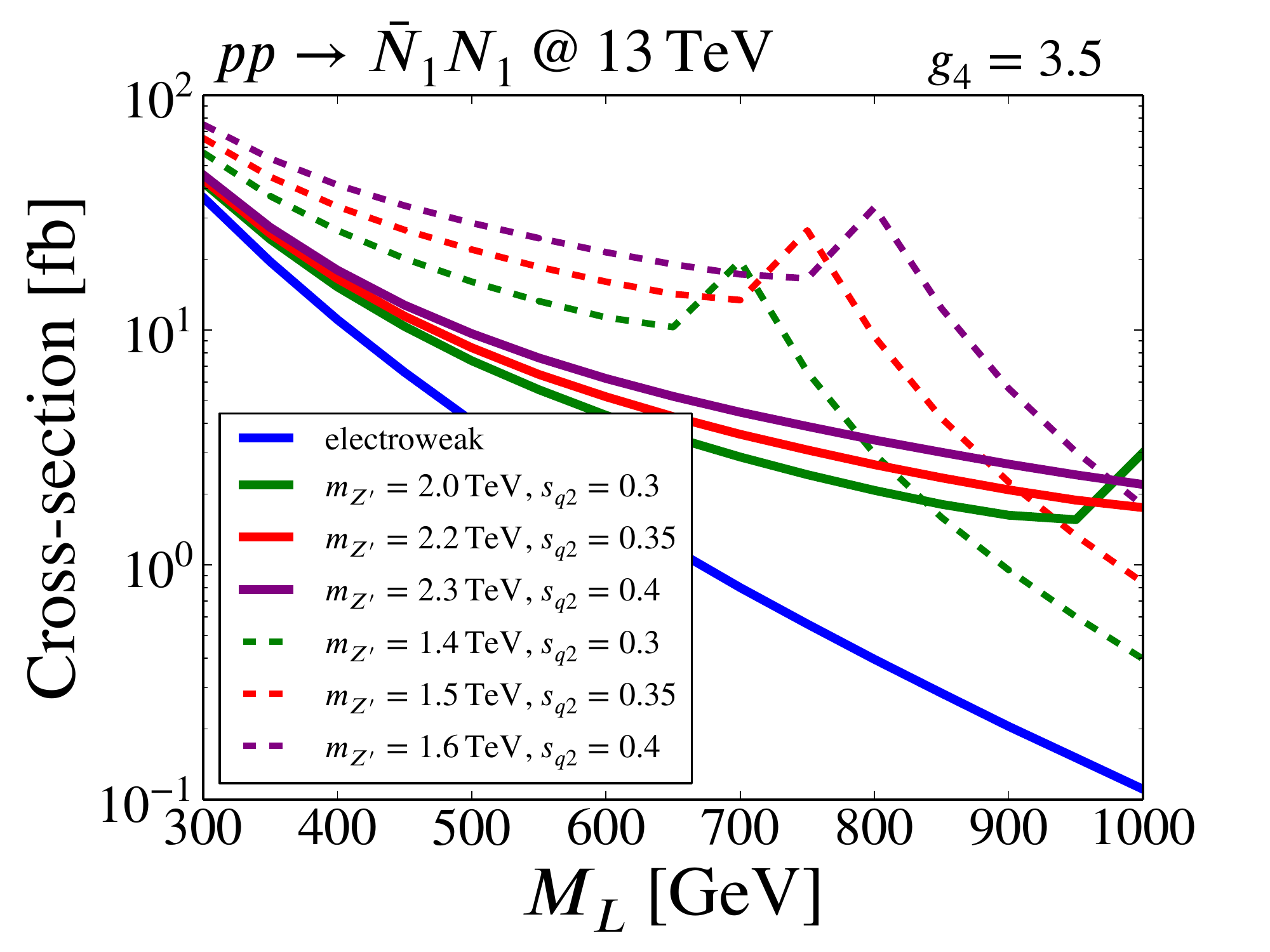}\\
\includegraphics[width=0.47\textwidth]{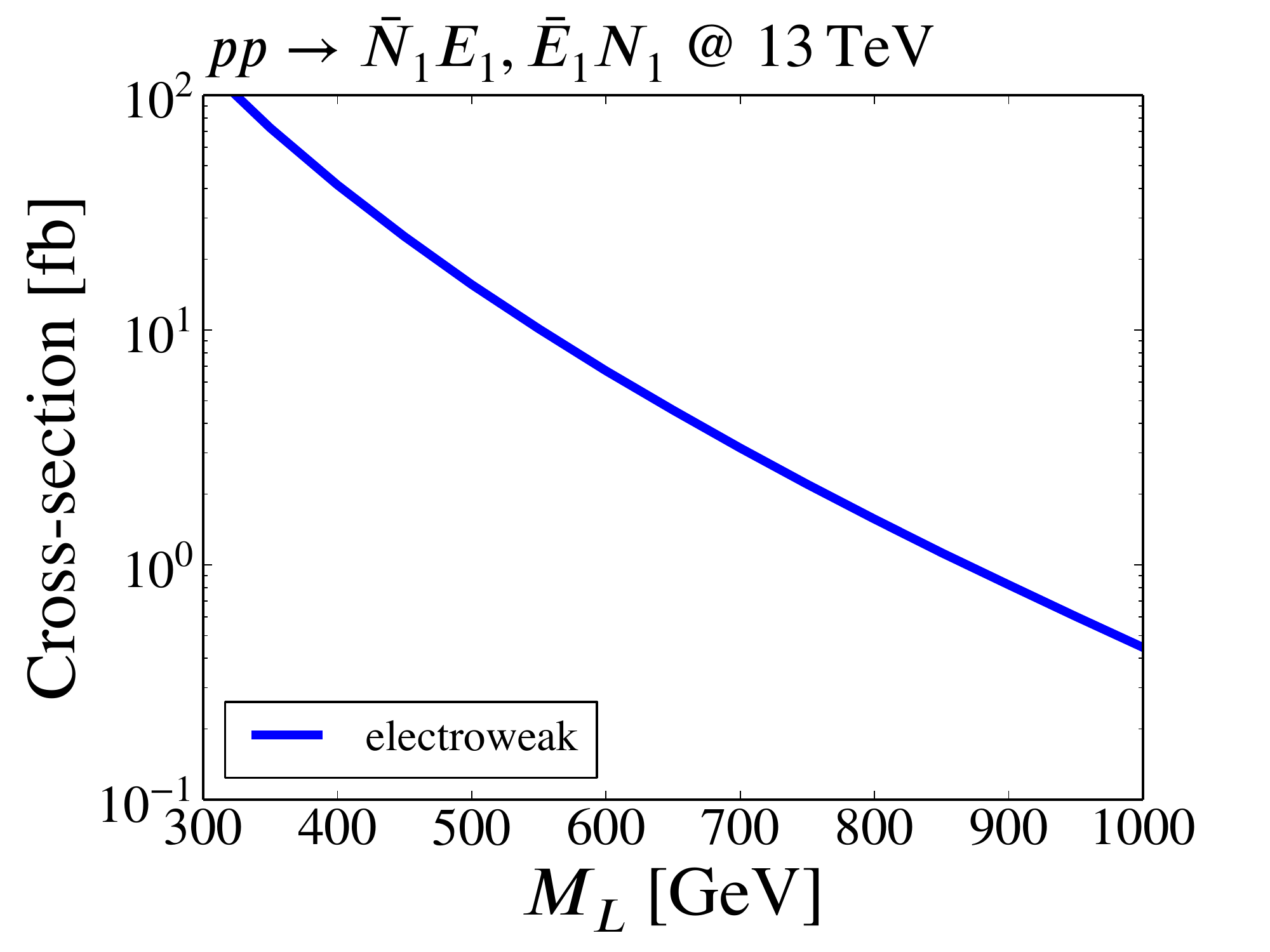}
\caption{\sf Vector-like lepton pair production cross section in $pp$ collisions at 13 TeV as a function of the lepton mass for several benchmark points. \label{fig:vllprod}}
\end{figure}

Vector-like leptons are pair produced in proton-proton collisions via electroweak interactions. In addition, a sizeable contribution to the total cross section comes from $s$-channel $Z'$ exchange. A quantitative estimate of this effect is illustrated in Fig.~\ref{fig:vllprod}, where we plot the total cross section for $p p \to E E$, $p p \to E N$ and $p p \to N N$ (where $E_1$ is the first generation charged lepton partner and $N_1$ is the first generation neutrino partner) as a function of the vector-like lepton mass $M_L$ for several motivated benchmark points. The cross sections were calculated using {\tt MadGraph5\_aMC@NLO} (with the $Z^{\prime}$ and vector-like leptons implemented using {\tt FeynRules}), including both electroweak production and the $Z^{\prime}$-assisted process. At each vector-like lepton mass, the $Z^{\prime}$ width was recalculated, taking into account all the kinematically accessible final states. 

Let us analyse Fig.~\ref{fig:vllprod} in more detail. While the charged current process is fixed by the electroweak interactions, it is important to notice that the neutral current processes $i)$ receive increased contribution for large  $s_{q_2}$, and that $ii)$ the cross section exhibits a plateau for $M_L < M_{Z'}/2$, that is when $Z' \to L \bar L$ is kinematically open. Neutral current processes are basically dominated by the $Z^{\prime}$-assisted production in the interesting range of parameters.  As discussed in Sec.~\ref{sec:multimulti}, these processes lead to distinct collider signatures which already set an upper limit on the total production 
of $\mathcal{O}(10)$ fb. Cross sections below $\lesssim 10$~fb are obtained for relatively heavy $\hat M_L \gtrsim 0.8$~TeV and relatively small mixing $s_{q_2} \lesssim 0.4$.\footnote{We also note that $p p \to L_1 \bar L_1$ can be induced via vector leptoquark exchanged in t-channel, e.g. $d \bar d \to E_1 \bar E_1$ and $u \bar u \to N_1 \bar N_1$. The cross section due to this diagram scales with $s_{q_2}^4$ and is relevant only for large $s_{q_2}$. We have checked that for $s_{q_2} \lesssim 0.4$, the $Z'$-assisted production dominates.} Having the $Z'$ mass below $2 \hat M_L$ also helps to reduce the yield, however, this scenario is disfavoured by $Z' \to \tau^+ \tau^-$ searches, see Sec.~\ref{colliderconstr:ta}.

\subsubsection{Vector-like quark production}
\label{sec:VLQproduction}

The QCD induced cross section for $p p \to Q \bar Q$ is completely determined by $m_Q$, and it is dominated by the gluon fusion subprocess, $g g \to  Q \bar Q$, and the sub-leading quark fusion, $q \bar q \to Q \bar Q$. Specific to this model is an extra contribution to the quark fusion subprocess, $q \bar q \to g' / Z' \to Q \bar Q$, which depends on the $g' / Z'$ masses and interactions. Due to the flavour structure of the model, $g'$ and $Z'$ could decay to a pair of heavy partners of same flavour, or to a heavy-light combination.

\begin{figure}[t]
\centering
\includegraphics[width=0.47\textwidth]{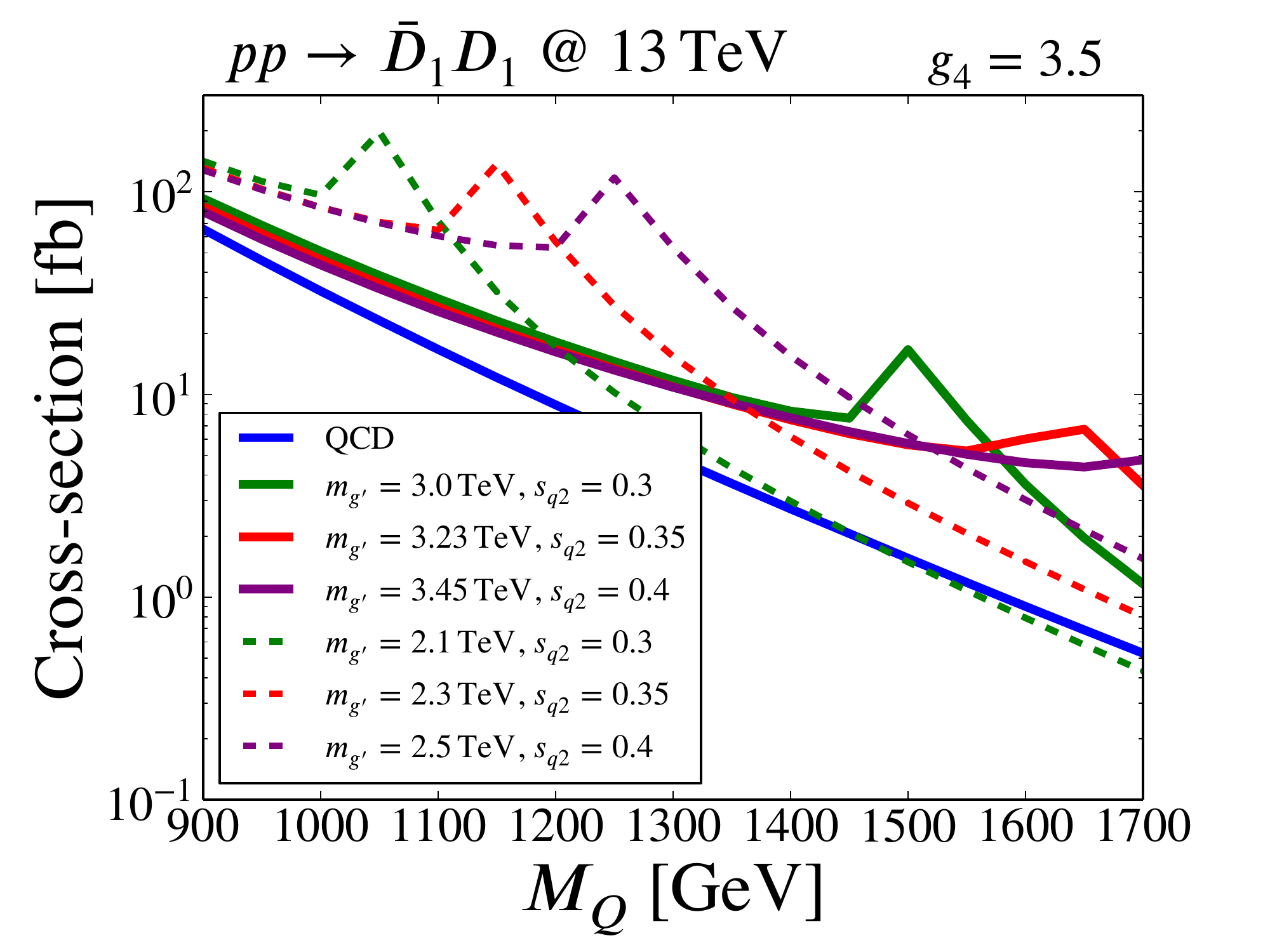}
\includegraphics[width=0.47\textwidth]{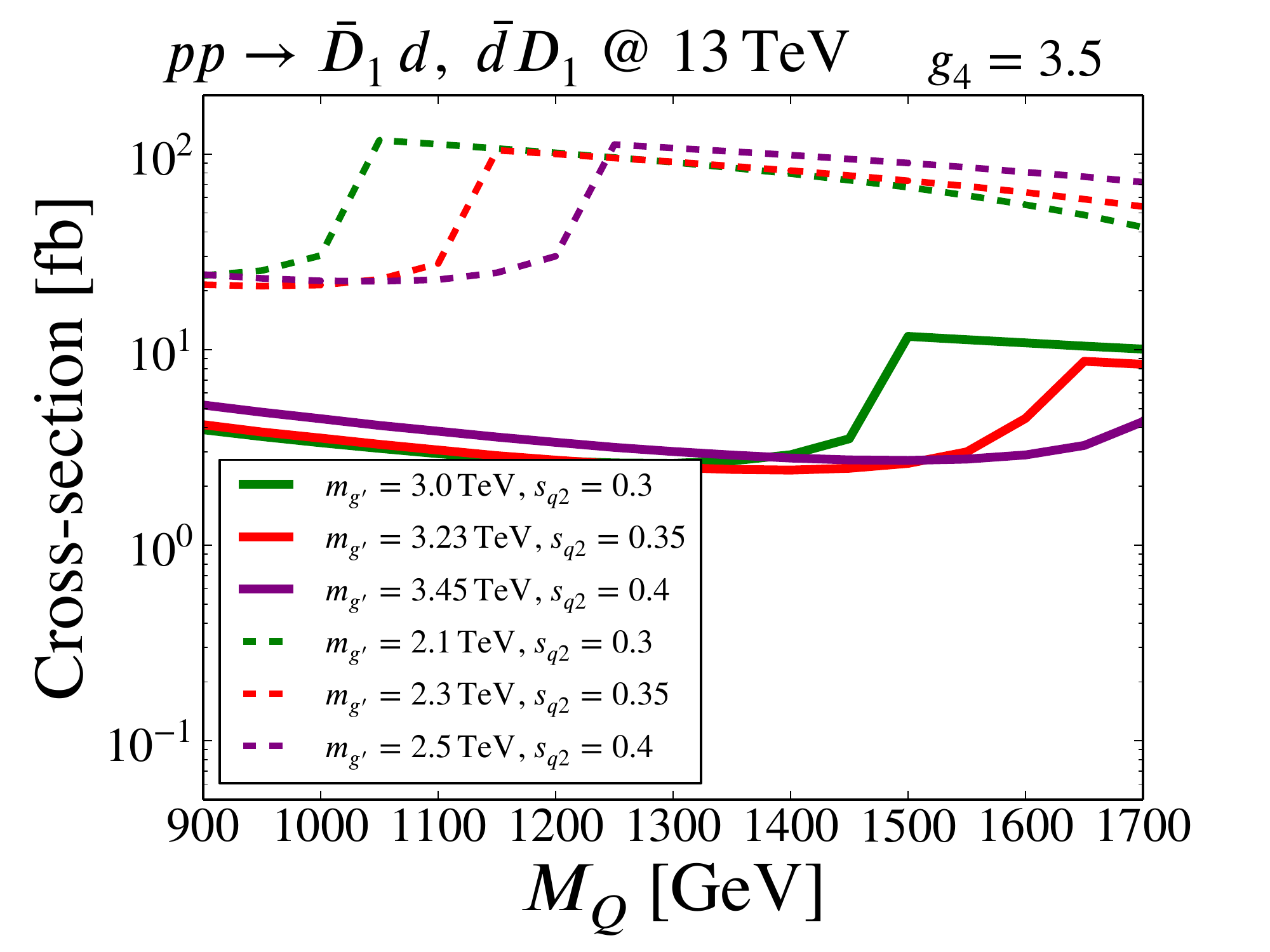}
\caption{\sf ({\bf Left panel})Vector-like quark pair production cross section in $pp$ collisions at 13 TeV as a function of the quark mass for several benchmark points. ({\bf Right panel}) Same for the single vector-like quark production in association with a light quark.} \label{fig:vqprod}
\end{figure}

To investigate the importance of the $g'$ assisted production, we plot the total $p p \to Q \bar Q$ and $p p \to Q \bar q, q \bar Q$ cross sections as a function of $m_Q$ in Fig.~\ref{fig:vqprod} for several benchmark $g^{\prime}$ masses and $s_{q_2}$ mixing. We do not include the $Z^{\prime}$-mediated process here as it is highly subdominant to the coloron-mediated process. Again, the cross sections were calculated using {\tt MadGraph5\_aMC@NLO} (the $g^{\prime}$ and vector-like quarks were implemented using {\tt FeynRules}), with the coloron width varying as a function of the vector-like quark mass. 

Let us discuss the main implications of Fig.~\ref{fig:vqprod}. The $Q \bar Q$ production (left panel) is dominated by the $g'$ diagram, and shows a plateau for $2 \hat M_Q < M_{g'}$, i.e. when $g' \to  Q \bar Q$ decay is kinematically opened, while it drops fast for larger $\hat M_Q$. On the contrary, single production of a vector-like quark in association with a light quark (right panel) increases when the $g' \to Q \bar Q$ decays is forbidden, due to the jump in $\mathcal{B}(g' \to Q q)$.
The benchmark point from Fig.~\ref{fig:spectrum} (left panel) has suppressed cross section for pair production as this process has a more constraining signature, see Sec~\ref{sec:multimulti}.

\subsubsection{Multi-leptons plus multi-jets}
\label{sec:multimulti}

As shown in Figs.~\ref{fig:vllprod} and~\ref{fig:vqprod}, vector-like fermions are produced with sizeable cross sections in the interesting parameter range ($1-10$~fb), and decay dominantly to three SM fermions of the third family, $t$, $b$, $\tau$ and $\nu_\tau$. Thus, the signature in the detector contains multiple jets and leptons and is rich with $b$-tags, hadronic $\tau$-tags, etc. While the extraction of precise limits requires a dedicated experimental analysis, we estimate the potential sensitivity in the current and near-future datasets, by comparing with the existing R-parity conserving (RPC) and R-parity violating (RPV) supersymmetry (SUSY) searches.

Using 36 fb$^{-1}$ of $13$ TeV $pp$ collision data, the ATLAS collaboration has searched for signatures involving multiple b-jets, high missing transverse momentum and either (at least) three isolated leptons, or two isolated same-sign leptons~\cite{Aaboud:2017dmy}. Following this general selection, the upper limits are set on the signal regions based on the number of $b$-jets, jets, leptons and $E_T^{\textrm{miss}}$, which are then interpreted in terms of simplified SUSY benchmarks. As an example, pair production of gluinos, each decaying to a top pair and a neutralino, can be qualitatively compared to our $p p \to N_2 N_2 \to (t t \nu) (t t\nu)$. Interpreting naively the exclusion limits, that is, neglecting any differences in acceptances between our model and the SUSY benchmarks, we conclude that the signal rate for this process is $\lesssim 5$~fb. This search is already starting to probe the interesting parameter space, see Fig.~\ref{fig:vllprod} (top right panel).
Another relevant RPC example involves pair production of stops, each decaying to $t$, $W^\pm$ and neutralino, and sets an upper limit on the cross section $\lesssim 10$~fb. Finally, the limit from RPV searches on gluino pair production, where each decays to $t b j$, implies an upper limit of $\lesssim 15$~fb.

\begin{figure}[t]
\centering
 \includegraphics[width=0.5\textwidth]{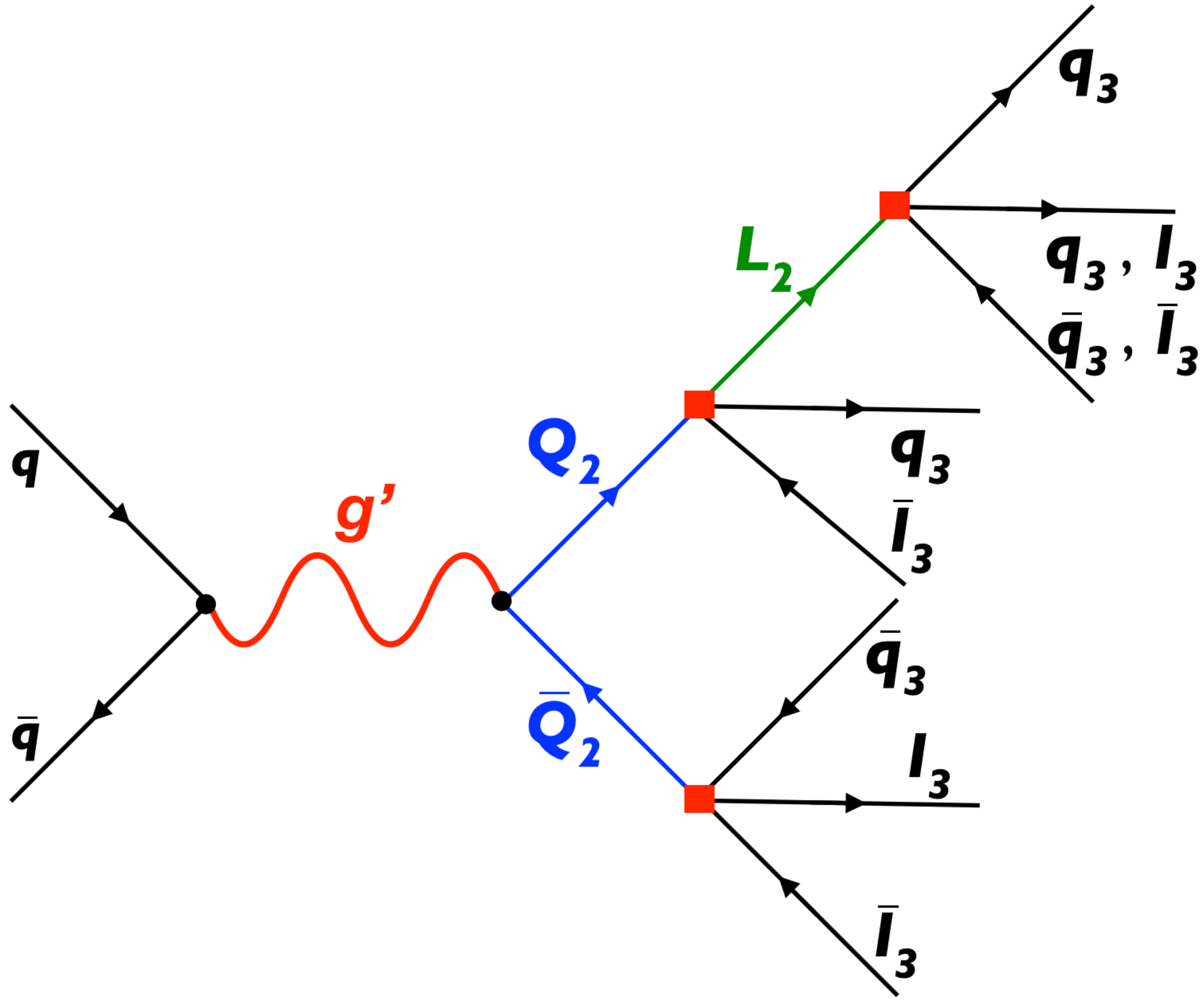}
\caption{\sf An example of the cascade decay process at the LHC leading to heavy-flavoured multi-lepton + multi-jet final state signature. \label{fig:FeynmanProdCascade}}
\end{figure}

In addition to these final states, the $4321$ model predicts even more exotic multi-lepton plus multi-jet signatures due to cascade decays among particles shown in Fig.~\ref{fig:spectrum} (left panel). An example of such process is illustrated in Fig.~\ref{fig:FeynmanProdCascade}. In this example, a pair of vector-like quarks is created by an $s$-channel coloron, and one of them decays to vector-like lepton which eventually decays to three SM fermions. The final state contains $3 q_3 + 5 \ell_3$, or $5 q_3 + 3 \ell_3$, where $q_3 = t,b$ and $\ell_3 = \nu_\tau, \tau$.

To sum up, the $4321$ model predicts a plethora of novel signatures and calls for a dedicated experimental effort.

\subsubsection{Scalars}
\label{sec:highptscalars}
The radial modes of the Higgs fields which break the 4321 symmetry to the SM group may also be produced at the LHC. They consist of a singlet, a colour triplet and a colour octet, as described in Sec.~\ref{scalspect}. A scalar singlet at the TeV scale is not easy to find at the LHC, so here we focus on the coloured states.

The colour octet can be pair produced via QCD and via an $s$-channel coloron, and will, if kinematically allowed, decay with close to 100\% branching ratio to a vector-like quark and a SM quark. It may also decay to a $\bar t t$ pair, but this coupling is suppressed by the right handed top-vector like quark mixing. The colour octet's mass is very dependent on parameters of the scalar sector but can easily be as heavy as $2-3$ TeV. A very similar model has been investigated in Ref.~\cite{Bai:2018jsr}, and the pair production cross-section of the colour octet scalar was found to be $O(10^{-2})$ fb for a 2 TeV octet scalar. Depending on the width of the coloron and the mass of the scalar, the coloron-assisted production may be larger than this, but will not be more than $10^{-1}$ fb at 2 TeV. This is more than an order of magnitude smaller than the vector-like quark pair production cross section investigated in Sec.~\ref{sec:VLQproduction} above, and will produce a similar signature with extra jets. 

The colour triplet will also be pair produced via QCD and coloron-assisted production, and will decay to a vector-like quark and a SM lepton, or a vector-like lepton and a SM quark. The signature will appear as vector-like quark pair production with extra leptons (or missing energy),vector-like lepton pair production with extra jets, or a combination of single vector-like lepton and vector-like quark single production with extra leptons and jets. Again, the mass depends on the scalar parameters, but can be as large as a few TeV. The QCD pair production cross-section for this state can be found in Ref.~\cite{Diaz:2017lit}, and is again $O(10^{-2})$ fb at 2 TeV, while the coloron assisted production is parameter-dependent but no larger than  $O(10^{-1})$ fb. This is therefore subdominant to other vector-like lepton and vector-like quark production modes in the model.



\section{Conclusions}
\label{sec:concl}

The evidence for flavour anomalies in semi-leptonic $B$ decays is growing with time.  
Although they might eventually disappear, they represent at the moment one of the few 
hints of NP, and deserve exploration. 
From a beyond the SM perspective they are certainly unexpected: the common 
expectation was that NP in flavour had to appear first in rare processes where the SM 
contribution is suppressed, while the anomalous data in $b \to c \tau \nu$ charged currents 
suggest a sizeable deviation in a channel where the SM contributes at tree level. 
Even allowing for the most conservative estimate of the scale of NP based on perturbative unitarity 
arguments \cite{DiLuzio:2017chi}, something has to happen below 9 TeV in order to unitarize 
the four-fermion operator responsible for the NP contribution to $b \to c \tau \nu$. 
How is it possible to reconcile this with a plethora of long-standing indirect probes (flavour, electroweak, etc.) 
and the impressively growing amount of LHC data at high-$p_T$? 

The main challenge, from a model building point of view, is to find 
a model which can offer a coherent explanation for the anomalies in both charged 
and neutral currents, while remaining compatible with all other constraints. 
A non-trivial point when addressing this question is to show that this is actually possible 
in a UV complete construction which allows for a connection between the 
anomalies and other observables. This is what we have achieved in this paper, in the context of the 
4321 model introduced in Ref.~\cite{DiLuzio:2017vat}.

The main virtue of the model is that it is renormalizable and fully calculable. 
It is however fair to say that the phenomenological constraints push the model in the direction of 
largish couplings, although still in the perturbative domain. Depending on the size 
of the anomaly in charged currents, the model can be extrapolated in the UV 
over several decades of energy. 

The $4321$ model belongs to a class of $B$-anomalies solutions which involve new dynamics in purely left-handed currents (see e.g.~\cite{Buttazzo:2017ixm}). The effect in $R(D^{(*)})$ is obtained either from $i)$ a pure third family interaction via the CKM, or 
$ii)$ a large direct 3-2 flavour violation. The second option points to a larger effective NP scale (by a factor of order $1/ \sqrt{V_{cb}} \approx 5$) alleviating the problems with indirect constraints in precisely measured $Z$-pole observables and leptonic $\tau$ decays \cite{Feruglio:2016gvd,Feruglio:2017rjo}, as well as direct searches at the LHC (see e.g.~\cite{Faroughy:2016osc,Greljo:2017vvb}); 
but introduces a potential problem with FCNC due to the $SU(2)_L$ nature of left-handed interactions. In the context of our model, one has not only to address tree-level $\Delta F = 2$  effects due to $g'$ and $Z'$, 
but also a large 3-2 leptoquark transition which feeds in at one loop in $\Delta F = 2$ observables.

A crucial phenomenological ingredient of this construction in order to make the size of the anomalies compatible with indirect and direct constraints is a generalisation of the Cabibbo-GIM mechanism of the SM, which allows a large 3-2 leptoquark transition while suppressing FCNC at tree level and at one loop. 
We have computed these 
and shown that they are under control as long as 
lepton partners, which play a crucial role in suppressing the contribution to the box diagrams 
via a GIM-like mechanism, 
are light enough. The latter are predicted to be the lowest-lying 
states of the NP spectrum (cf.~Fig.~\ref{fig:spectrum}) and hence 
represent a clear target for LHC. 
This is a nice example of the complementarity between indirect and direct searches, which is 
possible only thanks to the fact that the model is UV complete and hence the observables are calculable. 

The $4321$ model predicts a very rich phenomenology at high-$p_T$. 
While the low-energy phenomenology can be matched, to a large extent, to the simplified dynamical single-mediator model of Ref.~\cite{Buttazzo:2017ixm}, the high-$p_T$ physics 
is very different from that of the simplified model. Since
the fit to the $R(D^{(*)})$ anomaly requires the vector leptoquark to be around the TeV scale, 
the same is true for the whole spectrum of new vectors and fermions. 
The main high-$p_T$ signatures are, in fact, not related to the vector leptoquark, but to $i)$ the coloron decaying to $t \bar t$ and $b \bar b$, $ii)$ $Z' \to \tau^+ \tau^-$ and $iii)$ production of heavy vector-like leptons and quarks. 
As a proof of principle, we have identified a benchmark point which  fits the low-energy data well and is safe from the present LHC exclusions.

One of the key predictions of the $4321$ model is the peculiar decay phenomenology of the new fermions, and a possibility of cascade decays leading to exotic signatures with multiple leptons and jets in the detector. 
Such novel signatures, exemplified by Fig.~\ref{fig:FeynmanProdCascade}, require a dedicated experimental effort which, if performed, could significantly improve the limits in the near future.

\section*{Acknowledgments}
We thank Riccardo Barbieri, Stefano Bertolini, 
Matthias K\"onig, Jorge Martin Camalich, Federico Mescia, Richard Ruiz, Martin Schmaltz, Luca Silvestrini, Yi-Ming Zhong and, in particular, Gino Isidori 
for very helpful discussions. 
L.D.L.~and A.G.~would like to thank the CERN Theory group for hospitality and financial support during the 
final stages of this work. 
The work of J.F.~was supported in part by the Swiss National Science Foundation (SNF) under contract 200021-159720. 
The work of S.R.~was supported in part by the BMBF grant DESY, 05H15UMCA1
Verbundprojekt 05H2015 - Physik bei h\"ochsten Energien mit dem ATLAS-Experiment am LHC: Physik mit dem ATLAS-Experiment
und Pr\"azisionsberechnungen f\"ur die Collider-, Higgs- und Flavorphysik am LHC.

\appendix

\section{Anatomy of the 4321 model}
\label{details4321}

In this Appendix we provide a detailed account of several theoretical aspects of the 4321 model, 
including: the minimisation of the scalar potential (\app{scalpot}), 
the scalar spectrum (\app{scalspect}),  
the Yukawa interactions of the radial modes (\app{app:radmode}),
the gauge boson spectrum within the minimal scalar sector (\app{gbspec}) 
and beyond (\app{app:gaugescpectgt}), the vector-fermion interactions in the 
mass basis (\app{intmb}), the relevant tri-linear gauge vertices (\app{gbvrtx}), 
the renormalisation group equations (\app{RGE}), 
and the list of $SU(4)$ generators and structure constants (\app{SU4gt}).

\subsection{Scalar potential}
\label{scalpot}

The scalar sector comprises the representations:
$\Omega_3 \sim \left( \mathbf{\bar 4}, \mathbf{3}, \mathbf{1}, 1/6 \right)$, 
$\Omega_1 \sim \left( \mathbf{\bar 4}, \mathbf{1}, \mathbf{1}, -1/2 \right)$, 
$\Omega_{15} \sim \left( \mathbf{15}, \mathbf{1}, \mathbf{1}, 0 \right)$
and $H \sim (\mathbf{1},\mathbf{1},\mathbf{2},1/2)$. 
Note that $\Omega_{15}$ is taken to be a real field. 
Given the hierarchy $\vev{\Omega_{3}} > \vev{\Omega_{1}} \gg \vev{\Omega_{15}} \gg \vev{H}$ 
suggested by phenomenology, we simplify the problem by first considering the 
$\Omega_{3,1}$ system in isolation and comment later on about 
the inclusion of the other fields. 
We represent $\Omega_3$ and $\Omega_1$ respectively as a $4 \times 3$ matrix and a $4$-vector 
transforming as $\Omega_3 \to U^*_4 \Omega_3 U_{3'}^T$ and $\Omega_1 \to U^*_4 \Omega_1$ under 
$SU(4) \times SU(3)'$. 
The most general scalar potential involving $\Omega_{3}$ and $\Omega_{1}$ can be written as 
\begin{align}
\label{eqscalpot}
V_{\Omega_{3},\Omega_{1}}
= & \, \mu_3^2 \, \Tr (\Omega_3^\dag \Omega_3) 
+ \lambda_1 ( \Tr (\Omega_3^\dag \Omega_3) - \tfrac{3}{2} v_3^2 )^2 
+ \lambda_2 \Tr ( \Omega_3^\dag \Omega_3 - \tfrac{1}{2} v_3^2 \mathbb{1}_3 )^2 \nonumber \\
+ & \mu_1^2 \abs{\Omega_1}^2 +
\lambda_3 ( \abs{\Omega_1}^2 - \tfrac{1}{2} v_1^2 )^2 
+ \lambda_4 ( \Tr (\Omega_3^\dag \Omega_3) - \tfrac{3}{2} v_3^2 ) ( \abs{\Omega_1}^2 - \tfrac{1}{2} v_1^2 ) 
 \nonumber \\
+ & \lambda_5 \Omega_1^\dag \Omega_3 \Omega_3^\dag \Omega_1 + \lambda_6 \left( \left[ \Omega_3 \Omega_3 \Omega_3 \Omega_1 \right]_1 + \text{h.c.} \right) \, ,
\end{align}
where $\mathbb{1}_3$ denotes the $3\times 3$ identity matrix and we have used a relative rephasing 
between the fields $\Omega_{1}$ and $\Omega_{3}$ in order to remove the phase of $\lambda_6$.   
Note that the non-trivial invariants $\Omega_1^\dag \Omega_3 \Omega_3^\dag \Omega_1$
and 
\beq 
\left[ \Omega_3 \Omega_3 \Omega_3 \Omega_1 \right]_1 \equiv 
\epsilon_{\alpha\beta\gamma\delta} \epsilon^{abc} (\Omega_3)^\alpha_a (\Omega_3)^\beta_b (\Omega_3)^\gamma_c (\Omega_1)^\delta \, ,
\eeq
are required in order to avoid extra global symmetries in the scalar potential leading to 
unwanted massless Goldstone bosons (GBs).  
The scalar potential in \eq{eqscalpot} is written in such a way that in the limit $\mu_3 = \mu_1 = 0$ and $\lambda_6 = 0$, 
the configuration 
\beq  
\label{vevconf}
\vev{\Omega_3} = 
\tfrac{1}{\sqrt{2}}
\left(
\begin{array}{ccc}
v_3 & 0 & 0 \\
0 & v_3 & 0 \\ 
0 & 0 & v_3 \\
0 & 0 & 0
\end{array}
\right) \, , \qquad
\vev{\Omega_1} = 
\tfrac{1}{\sqrt{2}}
\left(
\begin{array}{c}
0 \\ 
0 \\ 
0 \\
v_1
\end{array}
\right) \, , 
\eeq
is (by construction) a stationary point. 
For $\lambda_6 \neq 0$, the stationary equations are satisfied by 
\beq 
\mu_3^2 = - 3 \lambda_6 v_1 v_3 \, , \qquad \mu_1^2 = - 3 \lambda_6 \frac{v^3_3}{v_1} \, . 
\eeq
By imposing that the second derivatives of the 
potential (evaluated at the stationary point) are positive definite, 
we can make sure that the configuration 
in \eq{vevconf} is a local minimum\footnote{Determining the global minimum is a non-trivial mathematical problem. 
Nevertheless, in the limit $v_1 \to 0$ the configuration in \eq{vevconf} is the 
global minimum of the potential for $\lambda_2 > 0$ and $\lambda_1 > - \frac{1}{3} \lambda_{2}$ \cite{Georgi:2016xhm}.} 
and compute in turn the scalar spectrum. 

The decomposition of the scalar multiplets under the unbroken $\mathcal{G}_{\rm 321}$ symmetry reads 
\begin{align} 
\Omega_3 &\to S_3 \sim (\mathbf{1},\mathbf{1},0) \oplus T_3 \sim (\mathbf{3},\mathbf{1},2/3) \oplus O_3 \sim (\mathbf{8},\mathbf{1},0) \, ,\\ 
\Omega_1 &\to S_1 \sim (\mathbf{1},\mathbf{1},0) \oplus T^*_1 \sim ( \mathbf{\bar 3},\mathbf{1},-2/3) \, .
\end{align} 
More explicitly, the SM fragments are embedded into $\Omega_{3,1}$ as 
\beq
\Omega_3 = 
\begin{pmatrix} 
\tfrac{1}{\sqrt{2}} v_3 \mathbb{1}_3 + \chi 
\\
T_3 
\end{pmatrix} \, , \qquad
\Omega_1 = 
\begin{pmatrix} 
T^*_1 \\
\tfrac{1}{\sqrt{2}} v_1 + S_1 
\end{pmatrix} \, , 
\eeq
where $\chi = \frac{1}{\sqrt{6}} S_3 I_3 + O^a t^a$ and $t^a$ ($a=1,\ldots ,8)$ are the Gell-Mann matrices. 

\subsection{Scalar spectrum}
\label{scalspect}

The spectrum of the scalar excitations is readily obtained by 
evaluating the second derivatives of the scalar potential on the stationary point.   
Sorting the multiplets according to the SM quantum numbers, 
we obtain the following: 

\vspace{2mm}
\noindent $\bullet$ Octet sector 
\begin{align}
\mathcal{M}^2_{\Re O_3} &= 2 \left( \lambda_2 v_3^2 - 3 \lambda_6 v_1 v_3 \right) \, , \\
\mathcal{M}^2_{\Im O_3} &= 0 \, .  
\end{align}
The null eigenvalue corresponds to the GB eaten by the coloron, while the positivity of the non-zero 
eigenvalue yields the condition $\lambda_2 v_3 > 3 \lambda_6 v_1$. 

\vspace{2mm}
\noindent $\bullet$ Triplet sector 
\beq 
\mathcal{M}^2_{T} = 
\begin{pmatrix} 
\frac{1}{2} \lambda_5 v_1^2 - 3 \lambda_6 v_1 v_3 & \frac{1}{2} \lambda_5 v_1 v_3 - 3 \lambda_6 v_3^2 \\
\frac{1}{2} \lambda_5 v_1 v_3 - 3 \lambda_6 v_3^2 & \frac{1}{2} \lambda_5 v_3^2 - 3 \lambda_6 \frac{v_3^3}{v_1}
\end{pmatrix} \, ,   
\eeq
defined on the basis $(T_3,T_1)$. Upon diagonalization 
\begin{align}
\mathcal{M}^2_{T_{R}} &= \left( \frac{1}{2} \lambda_5 - 3 \lambda_6 \frac{v_3}{v_1} \right) \left(v_1^2 + v_3^2 \right)  \, , \\
\mathcal{M}^2_{T_{\rm GB}} &= 0 \, ,
\end{align}
where 
\beq 
\label{TRTGB}
\begin{pmatrix} 
T_{R} \\ 
T_{\rm GB}  
\end{pmatrix}
= 
\frac{1}{\sqrt{v_3^2 + v_1^2}}
\begin{pmatrix} 
v_1 & v_3 \\
- v_3 & v_1
\end{pmatrix}
\begin{pmatrix} 
T_3 \\ 
T_1 
\end{pmatrix} \, .
\eeq
The GB mode is associated with the longitudinal component of the leptoquark. In the limit 
$v_3 \gg v_1$ one has 
$T_{\rm GB} \simeq - T_3 + \frac{v_1}{v_3} T_1$ and $T_{R} \simeq T_1 + \frac{v_1}{v_3} T_3$. The positivity of the 
non-zero eigenvalue yields $\lambda_5 v_1 > 6 \lambda_6 v_3$. 

\vspace{2mm}
\noindent $\bullet$ Singlet sector 
\begin{footnotesize}
\begin{multline} 
\mathcal{M}^2_{S} = \\
\left(
\begin{array}{cccc}
 \frac{1}{2} \left( 3 \lambda_1 + \lambda_2 \right) v_3^2+3 \lambda_6 v_1 v_3 &
 \frac{1}{2} \left( 3 \lambda_1 + \lambda_2 \right) v_3^2-3 \lambda_6 v_1 v_3 & 
 \sqrt{\frac{3}{2}} \left( 3 \lambda_6 v_3^2 + \frac{1}{2} \lambda_4 v_1 v_3 \right) & 
 \frac{1}{2} \sqrt{\frac{3}{2}} \lambda_4 v_1 v_3 \\
 \frac{1}{2} \left( 3 \lambda_1 + \lambda_2 \right) v_3^2-3 \lambda_6 v_1 v_3 & 
 \frac{1}{2} \left( 3 \lambda_1 + \lambda_2 \right) v_3^2+3 \lambda_6 v_1 v_3 & 
 \frac{1}{2} \sqrt{\frac{3}{2}} \lambda_4 v_1 v_3 & 
 \sqrt{\frac{3}{2}} \left( 3 \lambda_6 v_3^2 + \frac{1}{2} \lambda_4 v_1 v_3 \right) \\
 \sqrt{\frac{3}{2}} \left( 3 \lambda_6 v_3^2 + \frac{1}{2} \lambda_4 v_1 v_3 \right) &
 \frac{1}{2} \sqrt{\frac{3}{2}} \lambda_4 v_1 v_3 & 
 \lambda_3 v_1^2 & 
 \lambda_3 v_1^2-3\lambda_6\frac{v_3^3}{v_1} \\
 \frac{1}{2} \sqrt{\frac{3}{2}} \lambda_4 v_1 v_3 & 
 \sqrt{\frac{3}{2}} \left( 3 \lambda_6 v_3^2 + \frac{1}{2} \lambda_4 v_1 v_3 \right) & 
 \lambda_3 v_1^2-3\lambda_6\frac{v_3^3}{v_1} & 
 \lambda_3 v_1^2 \\
\end{array}
\right) \, ,
\end{multline}
\end{footnotesize}
defined in the basis $(S_3,S^*_3,S_1,S^*_1)$. 
It turns out that $\text{Rank} \ \mathcal{M}^2_{S} =3$ and the zero mode corresponds to the 
eigenvector 
\beq 
S_{\rm GB} = \frac{1}{\sqrt{v_1^2 + \frac{2}{3} v_3^2}} \left( \frac{v_3}{\sqrt{3}}  S_3 - \frac{v_3}{\sqrt{3}} S^*_3 
- \frac{v_1}{\sqrt{2}} S_1 +  \frac{v_1}{\sqrt{2}} S_1^* \right) \, ,
\eeq
which is associated to the longitudinal degree of freedom of the $Z'$. In the limit $v_3 \gg v_1$ one has 
$S_{\rm GB} \simeq i \Im S_3 - \sqrt{\frac{3}{2}} \frac{v_1}{v_3} i \Im S_1$. 
One of the three non-zero eigenvalues has a simple form 
\beq 
\mathcal{M}^2_{S_0} = 3 \lambda_6 v_3 \left( \frac{3}{2} v_1 + \frac{v_3^2}{v_1} \right) \, ,
\eeq
and it is associated to the eigenstate 
\beq 
S_0 = \frac{1}{\sqrt{v_3^2 + \frac{3}{2} v_1^2}} \left( - \frac{v_1}{\sqrt{2}}  S_3 + \frac{v_1}{\sqrt{2}} S^*_3 
- \frac{v_3}{\sqrt{3}} S_1 +  \frac{v_3}{\sqrt{3}} S_1^* \right) \, ,  
\eeq
while the two remaining non-zero eigenvalues have a complicated analytical expression 
and we do not report them explicitly (we just mention that they are different from zero in the $\lambda_6 \to 0$ limit). 
Note that in the $\lambda_6 \to 0$ limit one recovers an extra (physical) GB, which can be understood 
in terms of an additional global $U(1)$ emerging in \eq{eqscalpot}. 
Hence the presence of the $\lambda_6$ term in the scalar potential is crucial 
for a proper description of the scalar spectrum 
and it also 
affects in a non-trivial way the determination of the accidental global symmetries of the Lagrangian 
(cf.~the discussion below \Table{fieldcontent}).  

The inclusion in the scalar potential of the other two representations $\Omega_{15}$ and $H$, 
which are assumed to take 
the VEVs $\vev{\Omega_{15}} = T_{15} v_{15}$ 
and $\vev{H} = \tfrac{1}{2} (0, v)^T$, with $v = 246$ GeV, can be safely considered as a perturbation.  
The reason is because their VEVs are subleading for phenomenological reasons 
and they do not alter the pattern of global symmetries 
of the scalar potential. 
Finally, the decomposition of $\Omega_{15}$ under $\mathcal{G}_{321}$ yields: 
$\Omega_{15} \to (\mathbf{1},\mathbf{1},0) \oplus (\mathbf{3},\mathbf{1},2/3) \oplus (\mathbf{\bar 3},\mathbf{1},-2/3) 
\oplus (\mathbf{8}, \mathbf{1}, 0)$, whose mixing with the states contained in $\Omega_{3,1}$ is parametrically suppressed 
by the ratio $v^2_{15} / v^2_{3,1}$ and hence they play a subleading role for phenomenology.  

\subsection{Radial modes}
\label{app:radmode}

The radial modes discussed in the previous section have non-trivial consequences for low-energy 
phenomenology. In particular, the most relevant one is the state 
$T_R \sim (\mathbf{3},\mathbf{1},2/3)$ which induces one-loop FCNC
via its Yukawa interactions, 
beyond the ``model-independent'' contribution of the vector leptoquark containing the 
corresponding GB mode as a longitudinal degree of freedom. 
Working in the phenomenological limit $v_{15} \ll v_{1,3}$, which allows us to decouple 
the contribution of $\Phi$ to the radial modes originating from $\Omega_{1,3}$, 
the relevant interaction terms after $\mathcal{G}_{4321}$ breaking are 
readily extracted from \eq{LYUK2}
\begin{align}
\label{Lmixradmode}
\mathcal{L}_{\rm mix}&\supset
\left(-c_{\beta_T}\,T_R+s_{\beta_T}\,T_{\rm GB}\right)\,\lambda_q\,\bar q_L^{\,\prime}\,L_R^\prime 
+ \left(s_{\beta_T}\,T_R^*+c_{\beta_T}\,T_{\rm GB}^*\right)\,\lambda_\ell \,\bar\ell_L^{\,\prime}\,Q_R^\prime
\nonumber \\
&+\left(M_Q\,\bar Q_L^\prime+\frac{\lambda_q\,v_3}{\sqrt{2}}\,\bar q_L^\prime\right)Q_R^\prime 
+ \left(M_L\,\bar L_L^\prime+\frac{\lambda_\ell\,v_1}{\sqrt{2}}\,\bar\ell_L^\prime\right)L_R^\prime+\text{h.c.} \,,
\end{align}
where we have defined $\tan\beta_T=v_3/v_1$ and used \eq{TRTGB} after 
the redefinition $T_3 \to - T_3$. 
The mass parameters $M_{Q,L}$ are instead
\beq
\label{MQLdef}
M_Q=M+\frac{\lambda_{15}\, v_{15}}{2\sqrt{6}}\,,  \qquad M_L=M-\frac{3\,\lambda_{15}\,v_{15}}{2\sqrt{6}}\,.
\eeq
Following the discussion about the flavour structure of the 4321 model (cf.~\sect{sec:cabibbo}) we assume 
$M_{L,Q} \equiv \hat M_{L,Q}$ to be proportional to the identity, and the following textures for the $\lambda_{q,\ell}$ matrices:
\begin{align}
\lambda_q\equiv\hat\lambda_q=\mathrm{diag}(\lambda_{q_{12}},\lambda_{q_{12}},\lambda_{q_3})\,,\qquad\lambda_\ell\equiv\hat \lambda_\ell\cdot W=\mathrm{diag}(0,\lambda_{\ell_2},\lambda_{\ell_3})\cdot W\,,
\end{align}
with $W$ a unitary matrix. The mass eigenstates of the fermion fields (denoted without a prime) 
and prior to EW 
symmetry breaking are 
\begin{align}
\begin{aligned}
Q_L^{\prime\;i}&=s_{q_i}\,q_L^i+c_{q_i}\,Q_L^i\,,&&\qquad&Q_R&=Q_R^\prime\,,\\
W_{ij}\,L_L^{\prime\;j}&=s_{\ell_i}\,\ell_L^i+c_{\ell_i}\,L_L^i\,,&&\qquad&L_R&=W\,L_R^\prime\,,\\
q_L^{\prime\;i}&=c_{q_i}\,q_L^i-s_{q_i}\,Q_L^i\,,&&\qquad&\ell_L^{\prime\;i}&=c_{\ell_i}\,\ell_L^i-s_{\ell_i}\,L_L^i\,,
\end{aligned}
\end{align}
where
\begin{align}
\begin{aligned}
s_{q_i}&=\frac{\hat\lambda_q^i\,v_3}{\sqrt{2}\,M_{Q_i}}\,, &\qquad&& c_{q_i}&=\frac{\hat M_Q}{M_{Q_i}}\,,\\
s_{\ell_i}&=\frac{\hat\lambda_\ell^i\,v_1}{\sqrt{2}\,M_{L_i}}\,, &\qquad&& c_{\ell_i}&=\frac{\hat M_L}{M_{L_i}}\,,\\
M_{Q_i}&={\sqrt{\frac{|\hat\lambda_q^i|^2\,v_3^2}{\sqrt{2}}+\hat M_Q^2}}\,, &\qquad&& M_{L_i}&={\sqrt{\frac{|\hat\lambda_\ell^i|^2\,v_1^2}{\sqrt{2}}+\hat M_L^2}}\,.
\end{aligned}
\end{align}
Hence, in terms of mass eigenstates, \eq{Lmixradmode} reads 
\begin{align}
\label{Lmixradmode2}
\mathcal{L}_{\rm mix} &\supset\left(s_{\beta_T}\,T_R^*+c_{\beta_T}\,T_{\rm GB}^*\right)\,\hat\lambda_\ell^i\,\left(c_{\ell_i}\,\bar\ell_L^{\,i}-s_{\ell_i}\,\bar L_L^i\right)\,W_{ij}\,Q_R^j \nonumber \\
&\quad+\left(-c_{\beta_T}\,T_R+s_{\beta_T}\,T_{\rm GB}\right)\,\hat\lambda_q^i\,\left(c_{q_i}\,\bar q_L^i-s_{q_i}\,\bar Q_L^i\right)\,W_{ji}^*\,L_R^j \nonumber \\
&\quad+M_L^i\,\bar L_L^i\,L_R^i+M_Q^i\,\bar Q_L^i\,Q_R^i+\text{h.c.} \,.
\end{align}
The latter equation can be re-parametrised in a more compact notation as 
(ignoring heavy-heavy interactions which are not relevant for $\Delta F=2$) 
\begin{align}
\label{Lmixradmode3}
\mathcal{L}_{\rm mix} &\supset\left(T_{\rm GB}^*+t_{\beta_T}\,T_R^*\right)\,\Big(\frac{4\,G_F}{\sqrt{2}}\,C_U\,\hat M_L^2\Big)^{1/2}\,s_{\ell_i}\,\bar\ell_L^{\,i}\,W_{ij}\,Q_R^j \nonumber \\
&+\left(T_{\rm GB}-t_{\beta_T}^{-1}\,T_R\right)\,\Big(\frac{4\,G_F}{\sqrt{2}}\,C_U\,\hat M_Q^2\Big)^{1/2}\,s_{q_i}\,\bar q_L^{\,i}\,W_{ji}^*\,L_R^j \nonumber\\
&+M_L^i\,\bar L_L^i\,L_R^i+M_Q^i\,\bar Q_L^i\,Q_R^i+\text{h.c.} \,.
\end{align}

\subsection{Gauge boson spectrum}
\label{gbspec}

Let us introduce the following notation:  
given the extended gauge group $\mathcal{G}_{\rm 4321}$
we denote respectively the gauge fields by $H^\alpha_\mu, G'^a_\mu, W^i_\mu, B'_\mu$, the gauge couplings by
$g_4, g_3, g_2, g_1$ and the generators by $T^\alpha, T^a, T^i, Y'$
(with indices $\alpha = 1, \dots, 15$, $a = 1, \dots, 8$, $i=1,2,3$). 
In order to determine the gauge boson spectrum 
we start from the definition of the covariant derivatives acting on the scalar fields 
$\Omega_{3,1,15}$   
(in the following, 
$A=9,\ldots,14$ spans over the $SU(4) / (SU(3)_4 \times U(1)_4)$ coset and we neglect EW symmetry breaking effects):
\begin{align}
D_\mu \Omega_3 
&= \partial_\mu \Omega_3 + i g_4 H_\mu^a T^{a\ast} \Omega_3 + i g_4 H_\mu^A T^{A\ast} \Omega_3 
+ i g_4 H^{15}_\mu T^{15\ast} \Omega_3 - i g_3 G'^a_\mu T^a \Omega_3 - \tfrac{1}{6} i g_1 B'_\mu \Omega_3 \, ,  \nonumber \\
D_\mu \Omega_1 
&= \partial_\mu \Omega_1 + i g_4 H_\mu^a T^{a\ast} \Omega_1 + i g_4 H_\mu^A T^{A\ast} \Omega_1 
+ i g_4 H^{15}_\mu T^{15\ast} \Omega_1 + \tfrac{1}{2} i g_1 B'_\mu \Omega_1 \, , \nonumber \\
D_\mu \Omega_{15} &= \partial_\mu \Omega_{15} - i g_4 \left[ T^a, \Omega_{15} \right] H^a_\mu  - i g_4 \left[ T^A, \Omega_{15} \right] H^A_\mu  - i g_4 \left[ T^{15}, \Omega_{15} \right] H^{15}_\mu
\, .
\end{align}
The gauge boson masses are extracted from the (canonically normalized) 
kinetic term of the scalar fields: 
\begin{align}
\label{GBmass}
& \Tr \left( D_\mu \vev{\Omega_3} \right)^\dag D^\mu \vev{\Omega_3}  
= 
\tfrac{1}{2} 
\left( 
\begin{array}{cc}
H^a_\mu & G'^a_\mu
\end{array}
\right) 
\left( 
\begin{array}{cc}
g_4^2 & - g_4 g_3 \\
- g_4 g_3 & g_3^2
\end{array}
\right) \frac{v_3^2}{2}
\left( 
\begin{array}{c}
H^{b\mu} \\ 
G'^{b\mu}
\end{array}
\right)
+ \tfrac{1}{2} \left( g_4^2 v_3^2 \right) H_\mu^A H^{\mu A}
\nonumber \\
&  \qquad\qquad\qquad\qquad\quad\ \   + \tfrac{1}{2} 
\left( 
\begin{array}{cc}
H^{15}_\mu & B'_\mu
\end{array}
\right) 
\left( 
\begin{array}{cc}
\frac{1}{4} g_4^2 & - \frac{1}{2 \sqrt{6}} g_4 g_1 \\
- \frac{1}{2 \sqrt{6}} g_4 g_1 & \frac{1}{6} g_1^2
\end{array}
\right) \frac{v_3^2}{2}
\left( 
\begin{array}{c}
H^{15 \mu} \\ 
B'^\mu
\end{array}
\right) \, , \\
\label{GBmass}
&\left( D_\mu \vev{\Omega_1} \right)^\dag D^\mu \vev{\Omega_1} 
=  \tfrac{1}{2} \left( g_4^2 v_1^2 \right) H_\mu^A H^{A \mu}  \nonumber \\
& \qquad\qquad\qquad\qquad\  + \tfrac{1}{2} 
\left( 
\begin{array}{cc}
H^{15}_\mu & B'_\mu
\end{array}
\right) 
\left( 
\begin{array}{cc}
\frac{3}{4} g_4^2 & - \frac{3}{2 \sqrt{6}} g_4 g_1 \\
- \frac{3}{2 \sqrt{6}} g_4 g_1 & \frac{1}{2} g_1^2
\end{array}
\right) \frac{v_1^2}{2}
\left( 
\begin{array}{c}
H^{15 \mu} \\ 
B'^\mu
\end{array}
\right) \, , \\
&\tfrac{1}{2} \Tr (D_\mu \vev{\Omega_{15}})^\dag (D^\mu \vev{\Omega_{15}})  
= \tfrac{1}{2} \left(\tfrac{1}{3} g_4^2 v_{15}^2\right) H^A_\mu H^{A\mu}  \, .
\end{align}
Putting together all the contributions we get the massive gauge boson spectrum 
\begin{align}
M^2_{U} &= \frac{1}{4} g_4^2 (v_1^2 + v_3^2 + \frac{4}{3} v_{15}^2) \, , \\
M^2_{g'} &= \frac{1}{2}  (g_4^2 + g_3^2) v_3^2 \, , \\
M^2_{Z'} &= \frac{1}{4} \left( \frac{3}{2} g_4^2 + g_1^2 \right) \left(v_1^2 + \frac{1}{3} v_3^2 \right) \, , 
\end{align}
corresponding to the mass eigenstates 
\begin{align}
\label{defU}
&U_\mu^{1,2,3} = \frac{1}{\sqrt{2}} \left( H^{9,11,13}_\mu - i H^{10,12,14}_\mu \right) \, , \\
\label{gptransf}
& g'^a_\mu = \frac{g_4 H^a_\mu - g_3 G'^a_\mu}{\sqrt{g_4^2 + g_3^2}} \, , \\
\label{Zptransf}
& Z'_\mu = \frac{g_4 H^{15}_\mu - \sqrt{\frac{2}{3}} g_1 B'_\mu}{\sqrt{g_4^2 + \frac{2}{3} g_1^2}} \, .
\end{align}
The combinations orthogonal to \eqs{gptransf}{Zptransf}
correspond instead to the massless $SU(3)_c \times U(1)_Y$ degrees of freedom of $\mathcal{G}_{\rm 321}$ 
prior to electroweak symmetry breaking 
\begin{align}
\label{gtransf} 
&g^a_\mu = \frac{g_3 H^a_\mu + g_4 G'^a_\mu}{\sqrt{g_4^2 + g_3^2}} \, , \\
\label{Btransf} 
&B_\mu = \frac{\sqrt{\frac{2}{3}} g_1 H^{15}_\mu + g_4 B'_\mu}{\sqrt{g_4^2 + \frac{2}{3} g_1^2}} \, .
\end{align}
The matching among the couplings $g_4$, $g_3$, $g_1$ and $g_Y$, $g_s$ 
is readily obtained by acting on a field which transforms trivially under $SU(4)$. 
Let us consider, for instance, $u'_R = (\mathbf{1},\mathbf{3},\mathbf{1},2/3)$. 
The covariant derivative can be decomposed as follows
\begin{align}
D_\mu u'_R 
&= \partial_\mu u'_R - i g_3 G'^{a}_\mu T^{a} u'_R - \frac{2}{3} i g_1 B'_\mu u'_R \nonumber \\
& \supset \partial_\mu u'_R - i \frac{g_4 g_3}{\sqrt{g_4^2 + g_3^2}} g^{a}_\mu T^{a} u'_R 
- \frac{2}{3} i \frac{g_4 g_1}{\sqrt{g_4^2 + \frac{2}{3} g_1^2}} B_\mu u'_R
 \, ,
\end{align}
where in the last step we projected on the $SU(3)_c \times U(1)_Y$ 
fields: $G'^a_\mu \to \frac{g_4}{\sqrt{g_4^2 + g_3^2}} g^a_\mu$ and $B'_\mu \to 
\frac{g_4}{\sqrt{g_4^2 + \frac{2}{3} g_1^2}} B_\mu$. 
Hence, the matching with the SM gauge couplings reads 
\begin{align}
\label{matchinggsgs}
g_s &= \frac{g_4 g_3}{\sqrt{g_4^2 + g_3^2}} 
\, , \\ 
\label{matchinggsgY}
g_Y &= \frac{g_4 g_1}{\sqrt{g_4^2 + \frac{2}{3} g_1^2}} 
\, .  
\end{align}
Evolving the SM gauge couplings up to $\mu=2$ TeV we obtain 
$g_s = 1.02$ and $g_Y = 0.363$. 
Since $g_s \leq g_{4,3}$ and $g_Y \leq \sqrt{\tfrac{3}{2}} g_{4}, g_{1}$,  
the hierarchy $g_s \gg g_Y$ also implies $g_{4,3} \gg g_Y \simeq g_1$. 
In the 
limit $v_3 \gg v_1 \gg v_{15}$ 
and $g_4 \gg g_3 \gg g_1$,
one has for instance $M_{g'} \simeq \sqrt{2} M_U$ and $M_{Z'} \simeq \tfrac{1}{\sqrt{2}} M_U$. 

\subsection{Gauge boson spectrum beyond minimal scalar sector}
\label{app:gaugescpectgt}

Given the tight relations between the vector boson masses within the minimal scalar sector, 
one might wonder whether it is possible to parametrically decouple the $Z'$ and $g'$ from the 
leptoquark mass scale by considering different scalar representations responsible for the 
$\mathcal{G}_{4321}$ breaking. To this end, we have considered 
the contribution of all the possible one- and two-index tensor representations of 
$SU(4) \times SU(3)'$ to the gauge boson mass spectrum such that 
$SU(4) \times SU(3)' \to G \supseteq SU(3)_c$. Defining $r_{g'}=M_{g'} / M_U$ and $r_{Z'}=M_{Z'} / M_U$, 
and working in the phenomenologically motivated limit $ g_4 \gg g_3, g_1$, we find the 
results displayed in \Table{tablegt}.     
The best option for simultaneously maximising both the $g'$ and 
$Z'$ masses is a $(\mathbf{10,6})$, which yields $r_{g'}=\sqrt{2}$ and $r_{Z'}=1$. 
None of them really allows for a sizeable decoupling from the mass scale of the leptoquark. 

\begin{table}[htp]
\begin{center}
\begin{tabular}{|c|c|c|}
\hline
$SU(4) \times SU(3)'$ & $r_{g'}$ & $r_{Z'}$ \\
\hline
$(\mathbf{4,1})$ & 0 & $\sqrt{3/2}$ \\
$(\mathbf{4,\overline{3}})$ & $\sqrt{2}$ & $1/\sqrt{2}$ \\
$(\mathbf{6,\overline{3}})$ & 1 & 1 \\
$(\mathbf{10,1})$ & 0 & $\sqrt{3}$ \\
$(\mathbf{10,\overline{3}})$ & 1 & 1 \\
$(\mathbf{10,\overline{6}})$ & $\sqrt{2}$ & 1 \\
$(\mathbf{15,1})$ & 0 & 0 \\
$(\mathbf{15,\overline{3}})$ & $1/\sqrt{2}$ & $\sqrt{2}$ \\
$(\mathbf{15,8})$ & $3/2$ & 0 \\
\hline
\end{tabular}
\end{center}
\caption{\sf 
Vector mass ratios for different scalar representations. 
}
\label{tablegt}
\end{table}

\subsection{Fermion diagonalization to the mass basis}
\label{app:rot_massbasis}

As discussed in~\app{scalpot}, one can choose appropriate scalar potential parameters such that a global
minimum is obtained for the VEV configurations of $\vev{\Omega_{3,1}}$ defined in \eq{vevconf},  
$\vev{\Omega_{15}}=v_{15}\,T^{15}$  
and $\langle H^\dagger H\rangle=v^2/2$, with the hierarchies $v_3>v_1>v_{15}>v = 246$~GeV. 
After SSB, mixing among the chiral and the vector-like fermions is induced. Using the flavour basis
defined by Eq.~\eqref{eq:Yukawas} and employing the Yukawa textures assumed in~\eqref{lamqfs}, 
the $6\times6$ fermion mass matrices read
\begin{align}\label{eq:mass_matrices}
\begin{aligned}
\mathcal{M}_u&=
\begin{pmatrix}
V^\dagger\, \hat Y_u\,\frac{v}{\sqrt{2}} &\hat\lambda_q\,\frac{v_3}{\sqrt{2}}\\[2pt]
0 & \hat M_Q
\end{pmatrix}\,,&&&
\mathcal{M}_d&=
\begin{pmatrix}
\hat Y_d\,\frac{v}{\sqrt{2}} &\hat\lambda_q\,\frac{v_3}{\sqrt{2}}\\[2pt]
0 & \hat M_Q
\end{pmatrix}\,,\\[5pt]
\mathcal{M}_N&=
\begin{pmatrix}
0 &\hat\lambda_\ell\,\frac{v_1}{\sqrt{2}}\\[2pt]
0 & \hat M_L
\end{pmatrix}\,,&&&
\mathcal{M}_e&=
\begin{pmatrix}
\hat Y_e\,\frac{v}{\sqrt{2}} &\hat\lambda_\ell\,W^\dagger\,\frac{v_1}{\sqrt{2}}\\[2pt]
0 & \hat M_L
\end{pmatrix}\,,
\end{aligned}
\end{align}
with $\hat Y_{u,d,e}$ and $\hat \lambda_{q,\ell}$ diagonal, $V$ and $W$ unitary matrices, and
\begin{align}
\label{hatMQLdef}
\hat M_Q=\hat M+\frac{\lambda_{15}\, v_{15}}{2\sqrt{6}}\,,  &&& \hat M_L=\hat M-\frac{3\,\lambda_{15}\,v_{15}}{2\sqrt{6}}\,,
\end{align}
being proportional to the identity matrix. The mass matrices in Eqs.~\eqref{eq:mass_matrices} can be readily diagonalised by means of the unitary transformations: $\psi_x^\prime=U_x\,\psi_x$, where $\psi_x$ ($x=q,u,d,\ell,e,N$) denotes 6-dimensional fields containing both chiral and vector-like fermions and the unprimed fields denote the mass eigenstates.\footnote{Note that since we do not include a $\nu_R$ field, the vector $\psi_{N}$ is actually 3-dimensional (namely its components only contain $N_R \subset \Psi_R$).  For notational simplicity we use 6-dimensional vectors.}  The chosen flavour structure is such that in the limit $W\to\mathbb{1}$ the mixing is family-specific, i.e.~each vector-like family mixes with only one chiral family (up to CKM rotations). At leading order, the resulting mixing matrices read
\begin{align}\label{eq:RotMass}
\begin{aligned}
U_q&\approx\mathcal{R}_{14}(\theta_{q_1})\,\mathcal{R}_{25}(\theta_{q_2})\,\mathcal{R}_{36}(\theta_{q_3})\,, &\qquad&&U_\ell&\approx\mathcal{R}_{14}(\theta_{\ell_1})\,\mathcal{R}_{25}(\theta_{\ell_2})\,\mathcal{R}_{36}(\theta_{\ell_3})\,,\\
U_u&\approx\mathcal{R}_{14}(\theta_{u_R})\,\mathcal{R}_{25}(\theta_{c_R})\,\mathcal{R}_{36}(\theta_{t_R})\,, &\qquad&&U_e&\approx
\begin{pmatrix}
\mathbb{1} & 0\\[5pt]
0 & W
\end{pmatrix}
\mathcal{R}_{14}(\theta_{e_R})\,\mathcal{R}_{25}(\theta_{\mu_R})\,\mathcal{R}_{36}(\theta_{\tau_R})\,,\\
U_d&\approx\mathcal{R}_{14}(\theta_{d_R})\,\mathcal{R}_{25}(\theta_{s_R})\,\mathcal{R}_{36}(\theta_{b_R})\,, &\qquad&&U_N&\approx
\begin{pmatrix}
0 & 0\\[5pt]
0 & W
\end{pmatrix}
\,,
\end{aligned}
\end{align}
where we adopted a flavour basis for the SM $SU(2)_L$ fermion multiplets in which
\begin{align}
\label{SU2Lfb}
q^i=
\begin{pmatrix}
V^*_{ji}\,u_L^j\\
d_L^i
\end{pmatrix}\,,
\,\qquad\qquad
\ell^\alpha=
\begin{pmatrix}
\nu_L^\alpha\\
e_L^\alpha
\end{pmatrix}\,,
\end{align}
with $V$ the CKM matrix. The mixing angles are defined in terms of the Lagrangian parameters as
\begin{align}\label{eq:mixingangles}
\begin{aligned}
\sin\theta_{q_i}&=\frac{\lambda_i^q\,v_3}{\sqrt{|\lambda_i^q|^2\, v_3^2+2\,\hat M_Q^2}}\,, &\qquad\qquad&& \cos\theta_{q_i}&=\frac{\sqrt{2}\,\hat M_Q}{\sqrt{|\lambda_i^q|^2 \,v_3^2+2\,\hat M_Q^2}}\,,\\[5pt]
\sin\theta_{\ell_i}&=\frac{\lambda_i^\ell\,v_1}{\sqrt{|\lambda_i^\ell|^2\, v_1^2+2\,\hat M_L^2}}\,, &&& \cos\theta_{\ell_i}&=\frac{\sqrt{2}\,\hat M_L}{\sqrt{|\lambda_i^\ell|^2 \,v_1^2+2\,\hat M_L^2}}\,,\\[5pt]
\sin\theta_{u_R^i}&=\frac{m_{u_i}}{M_{Q_i}}\,\,\tan\theta_{q_i}\,, &\qquad\qquad&& \sin\theta_{d_R^i}&=\frac{m_{d_i}}{M_{Q_i}}\,\tan\theta_{q_i}\,,\\[5pt]
\sin\theta_{e_R^i}&=\frac{m_{e_i}}{M_{L_i}}\,\tan\theta_{\ell_i}\,, &\qquad\qquad&& \cos\theta_{f_R^i}&=1 \quad (f=u,d,e)\,,
\end{aligned}
\end{align}
with $m_i$ and $M_i$ the physical fermion masses. These read (up to corrections of $\mathcal{O}\left(m_i^2/M_i^2\right)$)
\begin{align}\label{eq:ferm_masses}
\begin{aligned}
M_{L_i}&=\sqrt{\frac{|\lambda_i^\ell|^2\, v_1^2}{2}+\hat M_L^2}\,, &\qquad\qquad&&
M_{Q_i}&=\sqrt{\frac{|\lambda_i^q|^2\, v_3^2}{2}+\hat M_Q^2}\,,\\[2pt]
m_{f_i}&\approx|\hat Y_f^i|\,\cos\theta_{f_i}\frac{v}{\sqrt{2}} &(f=u,d,e)\,.
\end{aligned}
\end{align}

\subsection{Vector-fermion interactions in the mass basis}
\label{intmb}

The interaction terms of the massive gauge bosons with the fermions in the interaction basis, are readily obtained from the action of the covariant derivative on the fermion fields: 
\begin{align}
\mathcal{L}_L &= \frac{g_4}{\sqrt{2}} \bar{Q}'_L \gamma^\mu L'_L \, U_\mu + \textrm{h.c.}~\nonumber \\
& + g_s \left( \frac{g_4}{g_3} \, \bar{Q}'_L \gamma^\mu T^a Q'_L - \frac{g_3}{g_4} \,\bar{q}'_L \gamma^\mu T^a q'_L \right) g'^a_\mu ~\nonumber \\
&+g_Y \left( \sqrt{\frac{3}{2}}\,\frac{g_4}{g_1}\,Y(Q'_L) \, \bar{Q}'_L \gamma^\mu Q'_L - \sqrt{\frac{2}{3}}\,\frac{g_1}{g_4}\,Y(q'_L) \, \bar{q}'_L \gamma^\mu q'_L \right) Z'_\mu ~\nonumber\\
&+ g_Y \left( \sqrt{\frac{3}{2}}\,\frac{g_4}{g_1}\,Y(L'_L) \, \bar{L}'_L \gamma^\mu L'_L - \sqrt{\frac{2}{3}}\,\frac{g_1}{g_4}\,Y(\ell'_L) \, \bar{\ell}'_L \gamma^\mu \ell'_L \right) Z'_\mu~,\label{eq:coupL}\\[10pt]
\mathcal{L}_R &= \frac{g_4}{\sqrt{2}} \bar{Q}'_R \gamma^\mu L'_R \, U_\mu + \textrm{h.c.}~\nonumber \\
& + g_s \left( \frac{g_4}{g_3} \, \bar{Q}'_R \gamma^\mu T^a Q'_R - \frac{g_3}{g_4}  \,\bar{u}'_R \gamma^\mu T^a u'_R -\frac{g_3}{g_4} \,\bar{d}'_R \gamma^\mu T^a d'_R  \right) g'^a_\mu ~\nonumber \\
&+ g_Y \left( \sqrt{\frac{3}{2}}\,\frac{g_4}{g_1}\,Y(Q'_R) \,\bar{Q}'_R \gamma^\mu Q'_R - \sqrt{\frac{2}{3}}\,\frac{g_1}{g_4}\, Y(u'_R) \, \bar{u}'_R \gamma^\mu u'_R -\sqrt{\frac{2}{3}}\,\frac{g_1}{g_4}\, Y(d'_R) \, \bar{d}'_R \gamma^\mu d'_R \right) Z'_\mu ~\nonumber\\
&+g_Y \left( \sqrt{\frac{3}{2}}\,\frac{g_4}{g_1}\,Y(L'_R) \, \bar{L}'_R \gamma^\mu L'_R - \sqrt{\frac{2}{3}}\,\frac{g_1}{g_4}\,
Y(e'_R) \, \bar{e}'_R \gamma^\mu e'_R \right) Z'_\mu~,\label{eq:coupR}
\end{align}
where we left implicit the SM hypercharges: 
$Y(Q'_{L}) = Y(Q'_{R}) = Y(q'_{L}) = \tfrac{1}{6}$, $Y(u'_R) = \tfrac{2}{3}$, $Y(d'_R) = -\tfrac{1}{3}$, 
$Y(L'_L) = Y(L'_R) = Y(\ell'_L) = -\tfrac{1}{2}$ and $Y(e'_R) = -1$.   

To express the interactions above in the fermion mass basis, we collect the fields in 6-dimensional multiplets, $\psi_x$ ($x=q,u,d,\ell,e$), and apply the unitary transformations defined in~\eq{eq:RotMass}. Neglecting right-handed rotations, suppressed by the small masses of the SM fermions, we have
\begin{align}
\begin{aligned}
\mathcal{L}_U&=\frac{g_4}{\sqrt{2}}\,U_\mu\left[\beta\,\bar \psi_q\gamma^\mu \psi_\ell+W\,\bar Q_R\gamma^\mu L_R\right]+h.c.\,,\\
\mathcal{L}_{g^\prime}&= g_s\,\frac{g_4}{g_3}\,g^{\prime\,a}_\mu\left[\kappa_q\,\bar \psi_q\gamma^\mu\,T^a\, \psi_q+\kappa_u\,\bar \psi_u\gamma^\mu\,T^a\, \psi_u+\kappa_d\,\bar \psi_d\gamma^\mu\,T^a\, \psi_d\right]\,,\\
\mathcal{L}_{Z^\prime}&=\frac{g_Y}{2\sqrt{6}}\,\frac{g_4}{g_1}\,Z_\mu^\prime\left[\xi_q\,\bar \psi_q\gamma^\mu \psi_q+\xi_u\,\bar \psi_u\gamma^\mu \psi_u+\xi_d\,\bar \psi_d\gamma^\mu \psi_d-3\,\xi_\ell\,\bar \psi_\ell\gamma^\mu \psi_\ell-3\,\xi_e\,\bar \psi_e\gamma^\mu \psi_e\right]\,,\label{eq:LLmbS}
\end{aligned}
\end{align}
with ($A,B=4,5,6$ and $\alpha,\beta=1,\dots,6$)
\begin{align}
\begin{aligned}
\beta^{\alpha\beta}&= \big[U_q\big]^*_{A\alpha}\big[W\big]_{AB} \big[U_\ell\big]_{B\beta}\,,\\
\kappa_q^{\alpha\beta}&= \big[U_q\big]^*_{A\alpha} \big[U_q\big]_{A\beta}-\frac{g_3^2}{g_4^2}\,\delta_{\alpha\beta}\,,&\qquad&&\kappa_u\approx\kappa_d&\approx
\begin{pmatrix}
0 & 0\\
0 & \mathbb{1}_{3\times3}
\end{pmatrix}
-\frac{g_3^2}{g_4^2}\,\mathbb{1}_{6\times6}\,,\\
\xi_q^{\alpha\beta}&= \big[U_q\big]^*_{A\alpha} \big[U_q\big]_{A\beta}-\frac{2\,g_1^2}{3\,g_4^2}\,\delta_{\alpha\beta}\,,&&& \xi_u\approx\xi_d&\approx
\begin{pmatrix}
0 & 0\\
0 & \mathbb{1}_{3\times3}
\end{pmatrix}
-\frac{2\,g_1^2}{3\,g_4^2}\,\mathbb{1}_{6\times6}\,,\\
\xi_\ell^{\alpha\beta}&= \big[U_\ell\big]^*_{A\alpha} \big[U_\ell\big]_{A\beta}-\frac{2\,g_1^2}{3\,g_4^2}\,\delta_{\alpha\beta}\,,&&& \xi_e&\approx
\begin{pmatrix}
0 & 0\\
0 & \mathbb{1}_{3\times3}
\end{pmatrix}
-\frac{2\,g_1^2}{3\,g_4^2}\,\mathbb{1}_{6\times6}\,.
\end{aligned}
\end{align}
Note in particular that the $W$ matrix cancels by unitarity in the $Z^\prime$ and $g^\prime$ interactions. This is a key result of the assumed flavour structure. Assuming $W=\mathcal{R}_{56}(\theta_{LQ})$ and no CP violation in the mixing angles, the left-handed couplings can be rewritten as
\begin{align}\label{eq:Beta}
\beta&\approx
\left(
    \begin{array}{c;{5pt/5pt}r}
        \begin{matrix}
        s_{q_1}s_{\ell_1} & 0 & 0\\
        0 & c_{\theta_{LQ}}\,s_{q_2}\,s_{\ell_2} & s_{\theta_{LQ}}\,s_{q_2}\,s_{\ell_3}\\
        0 & -s_{\theta_{LQ}}\,s_{q_3}\,s_{\ell_2} & c_{\theta_{LQ}}\,s_{q_3}\,s_{\ell_3}\\
        \end{matrix}
        & 
        \begin{matrix}
        -s_{q_1}c_{\ell_1} & 0 & 0\\
        0 & -c_{\theta_{LQ}}\,s_{q_2}\,c_{\ell_2} & -s_{\theta_{LQ}}\,s_{q_2}\,c_{\ell_3}\\
        0 & s_{\theta_{LQ}}\,s_{q_3}\,c_{\ell_2} & -c_{\theta_{LQ}}\,s_{q_3}\,c_{\ell_3}\\
        \end{matrix}
    \\ \hdashline[5pt/5pt]
        \begin{matrix}
        -c_{q_1}s_{\ell_1} & 0 & 0\\
        0 & -c_{\theta_{LQ}}\,c_{q_2}\,s_{\ell_2} & -s_{\theta_{LQ}}\,c_{q_2}\,s_{\ell_3}\\
        0 & s_{\theta_{LQ}}\,c_{q_3}\,s_{\ell_2} & -c_{\theta_{LQ}}\,c_{q_3}\,s_{\ell_3}\\
        \end{matrix}
    &
       \begin{matrix}
        c_{q_1}c_{\ell_1} & 0 & 0\\
        0 & c_{\theta_{LQ}}\,c_{q_2}\,c_{\ell_2} & s_{\theta_{LQ}}\,c_{q_2}\,c_{\ell_3}\\
        0 & -s_{\theta_{LQ}}\,c_{q_3}\,c_{\ell_2} & c_{\theta_{LQ}}\,c_{q_3}\,c_{\ell_3}\\
        \end{matrix}    
    \end{array}
\right)\,,\\[5pt]
\kappa_q&\approx
\left(
    \begin{array}{c;{5pt/5pt}r}
        \begin{matrix}
        s_{q_1}^2\phantom{c_{q_1}} & 0\phantom{c_{q_1}} & \phantom{c}0\phantom{c}\\
        0\phantom{c_{q_1}} & s_{q_2}^2\phantom{c_{q_1}} & \phantom{c}0\phantom{c}\\
        0\phantom{c_{q_1}} & 0\phantom{c_{q_1}} & \phantom{c}s_{q_3}^2\phantom{c}\\
        \end{matrix}
        & 
        \begin{matrix}
        -c_{q_1}s_{q_1} & 0 & 0\\
        0 & -c_{q_2}\,s_{q_2} & 0\\
        0 & 0 & -c_{q_3}\,s_{q_3}\\
        \end{matrix}
    \\ \hdashline[5pt/5pt]
        \begin{matrix}
        -c_{q_1}s_{q_1} & 0 & 0\\
        0 & -c_{q_2}\,s_{q_2} & 0\\
        0 & 0 & -c_{q_3}\,s_{q_3}\\
        \end{matrix}
    &
       \begin{matrix}
        c_{q_1}^2\phantom{c_{q_1}} & 0\phantom{c_{q_1}}  & \phantom{c}0\phantom{c} \\
        0\phantom{c_{q_1}}  & c_{q_{2}}^2\phantom{c_{q_1}}  & \phantom{c}0\phantom{c} \\
        0\phantom{c_{q_1}}  & 0\phantom{c_{q_1}}  & \phantom{c}c_{\ell_3}^2\phantom{c} \\
        \end{matrix}    
    \end{array}
\right)-\frac{g_3^2}{g_4^2}\,\mathbb{1}_{6\times6}\,,\\[5pt]
\xi_q&\approx
\left(
    \begin{array}{c;{5pt/5pt}r}
        \begin{matrix}
        s_{q_1}^2\phantom{c_{q_1}} & 0\phantom{c_{q_1}} & \phantom{c}0\phantom{c}\\
        0\phantom{c_{q_1}} & s_{q_2}^2\phantom{c_{q_1}} & \phantom{c}0\phantom{c}\\
        0\phantom{c_{q_1}} & 0\phantom{c_{q_1}} & \phantom{c}s_{q_3}^2\phantom{c}\\
        \end{matrix}
        & 
        \begin{matrix}
        -c_{q_1}s_{q_1} & 0 & 0\\
        0 & -c_{q_2}\,s_{q_2} & 0\\
        0 & 0 & -c_{q_3}\,s_{q_3}\\
        \end{matrix}
    \\ \hdashline[5pt/5pt]
        \begin{matrix}
        -c_{q_1}s_{q_1} & 0 & 0\\
        0 & -c_{q_2}\,s_{q_2} & 0\\
        0 & 0 & -c_{q_3}\,s_{q_3}\\
        \end{matrix}
    &
       \begin{matrix}
        c_{q_1}^2\phantom{c_{q_1}} & 0\phantom{c_{q_1}}  & \phantom{c}0\phantom{c} \\
        0\phantom{c_{q_1}}  & c_{q_{2}}^2\phantom{c_{q_1}}  & \phantom{c}0\phantom{c} \\
        0\phantom{c_{q_1}}  & 0\phantom{c_{q_1}}  & \phantom{c}c_{\ell_3}^2\phantom{c} \\
        \end{matrix}    
    \end{array}
\right)-\frac{2\,g_1^2}{3\,g_4^2}\,\mathbb{1}_{6\times6}\,,\\[5pt]
\xi_\ell&\approx
\left(
    \begin{array}{c;{5pt/5pt}r}
        \begin{matrix}
        s_{\ell_1}^2\phantom{c_{\ell_1}} & 0\phantom{c_{\ell_1}} & \phantom{c}0\phantom{c}\\
        0\phantom{c_{_1}} & s_{\ell_2}^2\phantom{c_{\ell_1}} & \phantom{c}0\phantom{c}\\
        0\phantom{c_{\ell_1}} & 0\phantom{c_{\ell_1}} & \phantom{c}s_{\ell_3}^2\phantom{c}\\
        \end{matrix}
        & 
        \begin{matrix}
        -c_{\ell_1}s_{\ell_1} & 0 & 0\\
        0 & -c_{\ell_2}\,s_{\ell_2} & 0\\
        0 & 0 & -c_{\ell_3}\,s_{\ell_3}\\
        \end{matrix}
    \\ \hdashline[5pt/5pt]
        \begin{matrix}
        -c_{\ell_1}s_{\ell_1} & 0 & 0\\
        0 & -c_{\ell_2}\,s_{\ell_2} & 0\\
        0 & 0 & -c_{\ell_3}\,s_{\ell_3}\\
        \end{matrix}
    &
       \begin{matrix}
        c_{\ell_1}^2\phantom{c_{\ell_1}} & 0\phantom{c_{\ell_1}}  & \phantom{c}0\phantom{c} \\
        0\phantom{c_{\ell_1}}  & c_{\ell_{2}}^2\phantom{c_{\ell_1}}  & \phantom{c}0\phantom{c} \\
        0\phantom{c_{\ell_1}}  & 0\phantom{c_{\ell_1}}  & \phantom{c}c_{\ell_3}^2\phantom{c} \\
        \end{matrix}    
    \end{array}
\right)-\frac{2\,g_1^2}{3\,g_4^2}\,\mathbb{1}_{6\times6}\,.\label{eq:LLmbE}
\end{align}
We remind the reader that these expressions hold in the flavour basis for the $SU(2)_L$ doublets defined in \eq{SU2Lfb}.

\subsection{Tri-linear gauge boson vertices}
\label{gbvrtx}

The interactions among gauge bosons are obtained from the gauge kinetic term 
\beq 
\label{Lgauge}
\mathcal{L}_{\text{gauge}} = 
-\frac{1}{4} H^{\alpha}_{\mu\nu} H^{\alpha, \mu\nu} 
-\frac{1}{4} G'^{a}_{\mu\nu} G'^{a, \mu\nu}
-\frac{1}{4} W^{i}_{\mu\nu} W^{i, \mu\nu}
-\frac{1}{4} B'_{\mu\nu} B'^{\mu\nu} \, ,
\eeq
with the field strengths defined as 
\begin{align}
H^{\alpha}_{\mu\nu} &= \partial_{[\mu} H^\alpha_{\nu]} + g_4 f^{\alpha\beta\gamma} H^\beta_\mu H^\gamma_\nu \, , \\
G'^{\alpha}_{\mu\nu} &= \partial_{[\mu} G'^a_{\nu]} + g_3 f^{abc} G'^b_\mu G'^c_\nu \, , \\
W^{i}_{\mu\nu} &= \partial_{[\mu} W^i_{\nu]} + g_2 \epsilon^{ijk} W^j_\mu W^k_\nu \, , \\
B'_{\mu\nu} &= \partial_{[\mu} B'_{\nu]} \, .  
\end{align}
Let us first prove that the $Z'gg$ and $g'gg$ couplings are zero 
(this is important in view of LHC resonance searches). 
The former statement simply follows from the fact that $f^{ab15}=0$ (cf.~\Table{SU4f}); 
while to prove the latter we keep from \eq{Lgauge} only trilinear terms either in $H^a$ or $G'$ 
and use the projections on the mass eigenstates (cf.~\eq{gptransf} and (\ref{gtransf})):
\begin{align}
\label{Lgaugeexp}
\mathcal{L}_{\text{gauge}} \supset&
- \frac{1}{2} g_4 f^{abc} \partial_{[\mu} H^a_{\nu]} H^{b,\mu} H^{c,\nu} 
- \frac{1}{2} g_3 f^{abc} \partial_{[\mu} G'^a_{\nu]} G'^{b,\mu} G'^{c,\nu}  \nonumber \\
=& -\frac{1}{2} \frac{f^{abc}}{\sqrt{g_4^2 + g_3^2}} \left[ 
g_4 \partial_{[\mu} (g_4 g'^a_{\nu]} + g_3 g^a_{\nu]}) (g_4 g'^{b,\mu} + g_3 g^{b,\mu}) (g_4 g'^{c,\nu} + g_3 g^{c,\nu}) 
\right. \nonumber \\
& \left. + g_3 \partial_{[\mu} (-g_3 g'^a_{\nu]} + g_4 g^a_{\nu]}) (-g_3 g'^{b,\mu} + g_4 g^{b,\mu}) (-g_3 g'^{c,\nu} + g_4 g^{c,\nu}) 
\right] \nonumber \\ 
\supset& -\frac{1}{2} \frac{f^{abc}}{\sqrt{g_4^2 + g_3^2}} \left[ 
g_4^2 g_3^2 
\left( 
\partial_{[\mu} g'^a_{\nu]} g^{b,\mu} g^{c,\nu} + \partial_{[\mu} g^a_{\nu]} g'^{b,\mu} g^{c,\nu} + \partial_{[\mu} g^a_{\nu]} g^{b,\mu} g'^{c,\nu} \right) \right. \nonumber \\
& \left. 
- g_3^2 g_4^2 
\left( 
\partial_{[\mu} g'^a_{\nu]} g^{b,\mu} g^{c,\nu} + \partial_{[\mu} g^a_{\nu]} g'^{b,\mu} g^{c,\nu} + \partial_{[\mu} g^a_{\nu]} g^{b,\mu} g'^{c,\nu} \right)
\right] = 0 \, ,
\end{align}
where in the last step we only kept $g'gg$ terms. 

Other non-zero trilinear couplings of phenomenological relevance are for instance 
those of two leptoquarks to a $U(1)_Y$ 
gauge boson or to a $Z'$ (see e.g.~\cite{Biggio:2016wyy}). 
To determine them, we first compute the $U^\dag U H^{15}$ gauge vertex from  
\begin{align}
\label{CCdec1}
\mathcal{L}_{\text{gauge}}
& \supset 
-\frac{1}{2} g_4 f^{\alpha\beta\gamma} \partial_{[\mu} H^\alpha_{\nu]}  H^{\beta,\mu} H^{\gamma,\nu} \nonumber \\ 
& \supset 
-\frac{1}{2} g_4 f^{15AB} \left[ 
\partial_{[\mu} H^{15}_{\nu]}  H^{A,\mu} H^{B,\nu}
- \partial_{[\mu} H^{A}_{\nu]}  H^{15,\mu} H^{B,\nu}
-\partial_{[\mu} H^{B}_{\nu]}  H^{A,\mu} H^{15,\nu}
\right]
\, ,
\end{align}
where $A,B = 9, \ldots , 14$.  
There are only three non-zero $f^{15AB}$ (cf.~\Table{SU4f}), 
namely $f^{15,9,10} = f^{15,11,12} = f^{15,13,14} = \sqrt{\frac{2}{3}}$.
Let us focus e.g.~on the $f^{15,9,10}$ contribution, yielding
\begin{align}
\label{CCdec2}
\mathcal{L}_{\text{gauge}} &\supset
-\frac{1}{2} g_4 \sqrt{\tfrac{2}{3}} 
\left[
\partial_{[\mu} H^{15}_{\nu]}  \left( H^{9,\mu} H^{10,\nu} - H^{10,\mu} H^{9,\nu} \right)
- H^{15,\mu} \left( \partial_{[\mu} H^{9}_{\nu]}  H^{10,\nu} - \partial_{[\mu} H^{10}_{\nu]}   H^{9,\nu} \right) \right. \nonumber \\
& \left. - H^{15,\nu} \left( \partial_{[\mu} H^{10}_{\nu]}  H^{9,\mu}  - \partial_{[\mu} H^{9}_{\nu]}  H^{10,\mu} \right) 
\right] \nonumber \\ 
&= i g_4 \sqrt{\tfrac{2}{3}} 
\left[ 
\left( U^1_\mu U^{1\dag}_\nu - U^1_\nu U^{1\dag}_\mu \right) \partial^\mu H^{15,\nu} 
+ \left( \partial_\mu U^{1\dag}_\nu - \partial_\nu U^{1\dag}_\mu \right) H^{15,\mu} U^{1\nu}
\right. \nonumber \\
& \left. - \left( \partial_\mu U^1_\nu - \partial_\nu U^1_\mu \right) H^{15,\mu} U^{1\dag\nu}
\right] \, , 
\end{align}
where the last step follows from the complexification of the leptoquark basis 
(cf.~\eq{defU}). The same expression applies to the other two leptoquark components, 
after replacing $U^{1}_\mu \to U^{2,3}_\mu$. By means of the identification 
$g_4 H_\mu^{15} = g_Y \left(\sqrt{\tfrac{2}{3}} B_\mu + \frac{g_4}{g_1} Z'_\mu\right)$, 
which follows from \eq{Zptransf}, (\ref{Btransf}) and (\ref{matchinggsgY}), we finally obtain
\begin{align} 
U^\dag U B \quad &: \quad i \tfrac{2}{3} g_Y
\left[ 
\left( U_\mu U^{\dag}_\nu - U_\nu U^{\dag}_\mu \right) \partial^\mu B^\nu 
+ \left( \partial_\mu U^{\dag}_\nu - \partial_\nu U^{\dag}_\mu \right) B^\mu U^{\nu}
\right. \nonumber \\
& \left. - \left( \partial_\mu U_\nu - \partial_\nu U_\mu \right) B^\mu U^{\dag\nu}
\right] \, , \\
U^\dag U Z' \quad &: \quad i \sqrt{\tfrac{2}{3}} \frac{g_4}{g_1} g_Y
\left[ 
\left( U_\mu U^{\dag}_\nu - U_\nu U^{\dag}_\mu \right) \partial^\mu Z'^\nu 
+ \left( \partial_\mu U^{\dag}_\nu - \partial_\nu U^{\dag}_\mu \right) Z'^\mu U^{\nu}
\right. \nonumber \\
& \left. - \left( \partial_\mu U_\nu - \partial_\nu U_\mu \right) Z'^\mu U^{\dag\nu}
\right] \, .
\end{align}

\subsection{Renormalisation group equations}
\label{RGE}

The renormalisation group equations (RGEs) are defined as 
\begin{equation}
\mu \frac{dg}{d\mu} = \beta_g \, ,
\end{equation}
where $\mu$ denotes the renormalisation scale and $g$ stands for a generic coupling. 
We report here the expression of the one-loop beta functions for the gauge and Yukawa sector 
of the 4321 model (cf.~field content in \Table{fieldcontent})
for an arbitrary number $n_\Psi$ of vector like fermions 
($n_\Psi = 3$ in the case of the model studied in this paper): 
\begin{align}
(4\pi)^2\beta_{g_{1}} &= \frac{131 }{18} g_{1}^{3} \, , \nonumber \\
(4\pi)^2\beta_{g_{2}} &= \left( - \frac{19}{6} +\frac{8 n_{\Psi}}{3} \right) g_{2}^{3}  \, , \nonumber \\
(4\pi)^2\beta_{g_{3}} &= - \frac{19 }{3} g_{3}^{3} \, , \nonumber \\
(4\pi)^2\beta_{g_{4}} &= \left( - \frac{40}{3} +\frac{4 n_{\Psi}}{3} \right) g_{4}^{3} \, , 
\end{align}
\begin{align}
(4\pi)^2\beta_{Y_{u}} &= 
\frac{3}{2} Y_{u} Y_{u}^{\dag} Y_{u} 
- \frac{3}{2} Y_{d} Y_{d}^{\dag} Y_{u} 
+ 2 \lambda_{q} \lambda_{q}^{\dag} Y_{u} 
+3 \trace{ (Y_{u}Y_{u}^{\dag}  )} Y_{u}
+3 \trace{ (Y_{d}Y_{d}^{\dag}  )} Y_{u} 
+ \trace{ (Y_{e}Y_{e}^{\dag}  )} Y_{u} \nonumber \\
&- \frac{17}{12} g_{1}^{2} Y_{u} 
- \frac{9}{4} g_{2}^{2} Y_{u} - 8 g_{3}^{2} Y_{u} \, , \nonumber \\
(4\pi)^2\beta_{Y_{d}} &= 
\frac{3}{2} Y_{d} Y_{d}^{\dag} Y_{d}  
- \frac{3}{2} Y_{u} Y_{u}^{\dag} Y_{d} 
+2 \lambda_{q} \lambda_{q}^{\dag} Y_{d}
+3 \trace{ (Y_{u}Y_{u}^{\dag} )} Y_{d} 
+3 \trace{ (Y_{d}Y_{d}^{\dag} )} Y_{d}
+\trace{ (Y_{e}Y_{e}^{\dag} )} Y_{d} \nonumber \\
&- \frac{5}{12} g_{1}^{2} Y_{d} 
- \frac{9 }{4}g_{2}^{2} Y_{d} 
- 8 g_{3}^{2} Y_{d} \, , \nonumber \\
(4\pi)^2\beta_{Y_{e}} &=
\frac{3}{2} Y_{e} Y_{e}^{\dag} Y_{e} 
+2 \lambda_{\ell} \lambda_{\ell}^{\dag} Y_{e}
+3 \trace{(Y_{u}Y_{u}^{\dag}  )} Y_{e}
+3 \trace{(Y_{d}Y_{d}^{\dag})} Y_{e}
+\trace{ (Y_{e}Y_{e}^{\dag} )} Y_{e} \nonumber \\
&- \frac{15}{4} g_{1}^{2} Y_{e} - \frac{9}{4} g_{2}^{2} Y_{e} \, , \nonumber \\
(4\pi)^2\beta_{\lambda_{q}} &=
\frac{7}{2} \lambda_{q} \lambda_{q}^{\dag} \lambda_{q}
+\frac{1}{2} \lambda_{q} \lambda_{\ell}^{\dag} \lambda_{\ell} 
+\frac{15}{8} \lambda_{q} \lambda_{15}^{\dag} \lambda_{15} 
+\frac{1}{2} Y_{u} Y_{u}^{\dag} \lambda_{q}
+\frac{1}{2} Y_{d} Y_{d}^{\dag} \lambda_{q}
+2 \trace{ (\lambda_{q} \lambda_{q}^{\dag}  )} \lambda_{q} \nonumber \\
&- \frac{1}{12} g_{1}^{2} \lambda_{q} 
- \frac{9 }{2} g_{2}^{2} \lambda_{q} 
- 4 g_{3}^{2} \lambda_{q}
- \frac{45 }{8} g_{4}^{2} \lambda_{q} \, , \nonumber \\
(4\pi)^2\beta_{\lambda_{\ell}} &=
\frac{5}{2} \lambda_{\ell} \lambda_{\ell}^{\dag} \lambda_{\ell}
+ \frac{3}{2} \lambda_{\ell} \lambda_{q}^{\dag} \lambda_{q} 
+\frac{15}{8} \lambda_{\ell} \lambda_{15}^{\dag} \lambda_{15} 
+\frac{1}{2} Y_{e} Y_{e}^{\dag} \lambda_{\ell}
+2 \trace{ (\lambda_{\ell} \lambda_{\ell}^{\dag}  )} \lambda_{\ell} \nonumber \\
&- \frac{3 }{4} g_{1}^{2} \lambda_{\ell} 
- \frac{9 }{2} g_{2}^{2} \lambda_{\ell} 
- \frac{45 }{8} g_{4}^{2} \lambda_{\ell} \, , \nonumber \\
(4\pi)^2\beta_{\lambda_{15}} &=
\frac{21}{4} \lambda_{15} \lambda_{15}^{\dag} \lambda_{15}
+ \frac{3}{2} \lambda_{15} \lambda_{q}^{\dag} \lambda_{q} 
+\frac{1}{2} \lambda_{15} \lambda_{\ell}^{\dag} \lambda_{\ell} 
+4 \trace{ (\lambda_{15} \lambda_{15}^{\dag}  )} \lambda_{15} \nonumber \\
& - \frac{9 }{2} g_{2}^{2} \lambda_{15} 
- \frac{45 }{4} g_{4}^{2} \lambda_{15} \, , \nonumber \\
(4\pi)^2\beta_{M} &= 
\frac{1}{2} M \lambda_{\ell}^{\dag} \lambda_{\ell} 
+ \frac{3}{2} M \lambda_{q}^{\dag} \lambda_{q} 
+ \frac{3}{2} \lambda_{15} M^{\dag} \lambda_{15} 
+ \frac{15}{8} M \lambda_{15}^{\dag} \lambda_{15}   
+ \frac{15}{8} \lambda_{15}    \lambda_{15}^{\dag} M  \, , \nonumber \\
& -\frac{45 }{4} g_{4}^{2} M 
-\frac{9 }{2} g_{2}^{2} M   
\, .
\end{align}

One can use the RGEs above to test the radiative stability of our proposed solution by evolving the Yukawa textures in Eq.~\eqref{lamqfs} from a high scale down to the $SU(4)$-breaking scale. 
Given that the imposed $U(2)_{q+\Psi}$ symmetry in the quark sector is explicitly broken in other sectors, more specifically by $\lambda_\ell$, RGE effects are expected to introduce departures
from the original $U(2)_{q+\Psi}$ symmetry. These departures are severely constrained by $\Delta F=2$ observables and therefore they set a limit on the possible UV scale at which the Yukawa 
textures in Eq.~\eqref{lamqfs} can be generated. Interestingly, we find that the $U(2)_{q+\Psi}$ symmetry in the $Z^\prime$ and $g^\prime$ couplings is partially protected. This protection arises
from the fact that the $U(2)_{q+\Psi}$-breaking terms, i.e. those containing $\lambda_\ell$, are the same in the RGEs of $\lambda_q$, $\lambda_{15}$ and $M$, leading to partial cancellations 
of the aforementioned breaking terms for the quark mixing angles, c.f. Eq.~\eqref{eq:mixingangles}. As a result we find that the Yukawa textures in Eq.~\eqref{lamqfs} could arise from (unspecified)
UV dynamics at $\Lambda\approx10$~TeV without significantly impacting $\Delta F=2$ observables.


\subsection{$SU(4)$ generators}
\label{SU4gt}

Here we report some useful facts about the $SU(4)$ algebra. 
The generators in the fundamental representation can be written as
\beq
\begin{array}{ccc}
T^1 = \frac{1}{2}
\left(
\begin{array}{cccc}
0 & 1 & 0 & 0 \\
1 & 0 & 0 & 0 \\
0 & 0 & 0 & 0 \\
0 & 0 & 0 & 0 \\
\end{array}
\right) & 
T^2 = \frac{1}{2}
\left(
\begin{array}{cccc}
0 & -i & 0 & 0 \\
i & 0 & 0 & 0 \\
0 & 0 & 0 & 0 \\
0 & 0 & 0 & 0 \\
\end{array}
\right) &  
T^3 = \frac{1}{2}
\left(
\begin{array}{cccc}
1 & 0 & 0 & 0 \\
0 & -1 & 0 & 0 \\
0 & 0 & 0 & 0 \\
0 & 0 & 0 & 0 \\
\end{array}
\right) \\ 
& & \\
T^4 = \frac{1}{2}
\left(
\begin{array}{cccc}
0 & 0 & 1 & 0 \\
0 & 0 & 0 & 0 \\
1 & 0 & 0 & 0 \\
0 & 0 & 0 & 0 \\
\end{array}
\right) & 
T^5 = \frac{1}{2}
\left(
\begin{array}{cccc}
0 & 0 & -i & 0 \\
0 & 0 & 0 & 0 \\
i & 0 & 0 & 0 \\
0 & 0 & 0 & 0 \\
\end{array}
\right) &  
T^6 = \frac{1}{2}
\left(
\begin{array}{cccc}
0 & 0 & 0 & 0 \\
0 & 0 & 1 & 0 \\
0 & 1 & 0 & 0 \\
0 & 0 & 0 & 0 \\
\end{array}
\right) \\
& & \\
T^7 = \frac{1}{2}
\left(
\begin{array}{cccc}
0 & 0 & 0 & 0 \\
0 & 0 & -i & 0 \\
0 & i & 0 & 0 \\
0 & 0 & 0 & 0 \\
\end{array}
\right) & 
T^8 = \frac{1}{2\sqrt{3}}
\left(
\begin{array}{cccc}
1 & 0 & 0 & 0 \\
0 & 1 & 0 & 0 \\
0 & 0 & -2 & 0 \\
0 & 0 & 0 & 0 \\
\end{array}
\right) &  
T^9 = \frac{1}{2}
\left(
\begin{array}{cccc}
0 & 0 & 0 & 1 \\
0 & 0 & 0 & 0 \\
0 & 0 & 0 & 0 \\
1 & 0 & 0 & 0 \\
\end{array}
\right) \\
& & \\
T^{10} = \frac{1}{2}
\left(
\begin{array}{cccc}
0 & 0 & 0 & -i \\
0 & 0 & 0 & 0 \\
0 & 0 & 0 & 0 \\
i & 0 & 0 & 0 \\
\end{array}
\right) & 
T^{11} = \frac{1}{2}
\left(
\begin{array}{cccc}
0 & 0 & 0 & 0 \\
0 & 0 & 0 & 1 \\
0 & 0 & 0 & 0 \\
0 & 1 & 0 & 0 \\
\end{array}
\right) &  
T^{12} = \frac{1}{2}
\left(
\begin{array}{cccc}
0 & 0 & 0 & 0 \\
0 & 0 & 0 & -i \\
0 & 0 & 0 & 0 \\
0 & i & 0 & 0 \\
\end{array}
\right) \\
& & \\
T^{13} = \frac{1}{2}
\left(
\begin{array}{cccc}
0 & 0 & 0 & 0 \\
0 & 0 & 0 & 0 \\
0 & 0 & 0 & 1 \\
0 & 0 & 1 & 0 \\
\end{array}
\right) & 
T^{14} = \frac{1}{2}
\left(
\begin{array}{cccc}
0 & 0 & 0 & 0 \\
0 & 0 & 0 & 0 \\
0 & 0 & 0 & -i \\
0 & 0 & i & 0 \\
\end{array}
\right) &  
T^{15} = \frac{1}{2\sqrt{6}}
\left(
\begin{array}{cccc}
1 & 0 & 0 & 0 \\
0 & 1 & 0 & 0 \\
0 & 0 & 1 & 0 \\
0 & 0 & 0 & -3 \\
\end{array}
\right) \, , \nonumber
\end{array}
\eeq
with normalization 
\beq
\Tr \, T^\alpha T^\beta = \frac{1}{2} \delta^{\alpha\beta} \, .
\eeq
The matrices $T^\alpha$ 
satisfy the Lie algebra 
\beq
\left[ T^\alpha, T^\beta \right] = i f^{\alpha\beta\gamma} T^\gamma \, , 
\eeq 
where the completely antisymmetric structure constants can be constructed via 
the relation
\beq 
f^{\alpha\beta\gamma} = -2 i \, \Tr \left( \left[ T^\alpha, T^\beta \right] T^\gamma \right) \, .
\eeq
We have collected them for completeness in \Table{SU4f}.

\begin{table}[htbp]
\renewcommand{\arraystretch}{1.4}
\centering
\begin{tabular}{|c|c|c|c| @{}}
\hline
$\alpha$ &  $\beta$ & $\gamma$ & $f^{\alpha\beta\gamma}$ \\ 
\hline
\hline
$1$ & $2$ & $3$ & $1$ \\ 
$1$ & $4$ & $7$ & $\frac{1}{2}$ \\ 
$1$ & $5$ & $6$ & $-\frac{1}{2}$ \\ 
$1$ & $9$ & $12$ & $\frac{1}{2}$ \\ 
$1$ & $10$ & $11$ & $-\frac{1}{2}$ \\ 
$2$ & $4$ & $6$ & $\frac{1}{2}$ \\ 
$2$ & $5$ & $7$ & $\frac{1}{2}$ \\ 
$2$ & $9$ & $11$ & $\frac{1}{2}$ \\ 
$2$ & $10$ & $12$ & $\frac{1}{2}$ \\ 
$3$ & $4$ & $5$ & $\frac{1}{2}$ \\ 
$3$ & $6$ & $7$ & $-\frac{1}{2}$ \\ 
$3$ & $9$ & $10$ & $\frac{1}{2}$ \\ 
$3$ & $11$ & $12$ & $-\frac{1}{2}$ \\ 
$4$ & $5$ & $8$ & $\frac{\sqrt{3}}{2}$ \\ 
$4$ & $9$ & $14$ & $\frac{1}{2}$ \\ 
$4$ & $10$ & $13$ & $-\frac{1}{2}$ \\ 
$5$ & $9$ & $13$ & $\frac{1}{2}$ \\ 
$5$ & $10$ & $14$ & $\frac{1}{2}$ \\ 
$6$ & $7$ & $8$ & $\frac{\sqrt{3}}{2}$ \\ 
$6$ & $11$ & $14$ & $\frac{1}{2}$ \\ 
$6$ & $12$ & $13$ & $-\frac{1}{2}$ \\ 
$7$ & $11$ & $13$ & $\frac{1}{2}$ \\ 
$7$ & $12$ & $14$ & $\frac{1}{2}$ \\ 
$8$ & $9$ & $10$ & $\frac{1}{2\sqrt{3}}$ \\ 
$8$ & $11$ & $12$ & $\frac{1}{2\sqrt{3}}$ \\ 
$8$ & $13$ & $14$ & $-\frac{1}{\sqrt{3}}$ \\ 
$9$ & $10$ & $15$ & $\sqrt{\frac{2}{3}}$ \\ 
$11$ & $12$ & $15$ & $\sqrt{\frac{2}{3}}$ \\ 
$13$ & $14$ & $15$ & $\sqrt{\frac{2}{3}}$ \\
  \hline
  \end{tabular}
  \caption{\label{SU4f} 
  Non-zero $SU(4)$ structure constants. 
  }
\end{table}

\clearpage

\bibliographystyle{utphys.bst}
\bibliography{bibliography}

\providecommand{\href}[2]{#2}\begingroup\raggedright\begin{thebibliography}{100}

\bibitem{DAmbrosio:2002vsn}
G.~D'Ambrosio, G.~F. Giudice, G.~Isidori, and A.~Strumia, ``{Minimal flavor
  violation: An Effective field theory approach},''
  \href{http://dx.doi.org/10.1016/S0550-3213(02)00836-2}{{\em Nucl. Phys.}
  {\bfseries B645} (2002) 155--187},
\href{http://arxiv.org/abs/hep-ph/0207036}{{\ttfamily arXiv:hep-ph/0207036
  [hep-ph]}}.

\bibitem{Lees:2012xj}
{\bfseries BaBar} Collaboration, J.~P. Lees {\em et~al.}, ``{Evidence for an
  excess of $\bar{B} \to D^{(*)} \tau^-\bar{\nu}_\tau$ decays},''
  \href{http://dx.doi.org/10.1103/PhysRevLett.109.101802}{{\em Phys. Rev.
  Lett.} {\bfseries 109} (2012) 101802},
\href{http://arxiv.org/abs/1205.5442}{{\ttfamily arXiv:1205.5442 [hep-ex]}}.

\bibitem{Lees:2013uzd}
{\bfseries BaBar} Collaboration, J.~P. Lees {\em et~al.}, ``{Measurement of an
  Excess of $\bar{B} \to D^{(*)}\tau^- \bar{\nu}_\tau$ Decays and Implications
  for Charged Higgs Bosons},''
  \href{http://dx.doi.org/10.1103/PhysRevD.88.072012}{{\em Phys. Rev.}
  {\bfseries D88} no.~7, (2013) 072012},
\href{http://arxiv.org/abs/1303.0571}{{\ttfamily arXiv:1303.0571 [hep-ex]}}.

\bibitem{Aaij:2013qta}
{\bfseries LHCb} Collaboration, R.~Aaij {\em et~al.}, ``{Measurement of
  Form-Factor-Independent Observables in the Decay $B^{0} \to K^{*0} \mu^+
  \mu^-$},'' \href{http://dx.doi.org/10.1103/PhysRevLett.111.191801}{{\em Phys.
  Rev. Lett.} {\bfseries 111} (2013) 191801},
\href{http://arxiv.org/abs/1308.1707}{{\ttfamily arXiv:1308.1707 [hep-ex]}}.

\bibitem{Aaij:2014ora}
{\bfseries LHCb} Collaboration, R.~Aaij {\em et~al.}, ``{Test of lepton
  universality using $B^{+}\rightarrow K^{+}\ell^{+}\ell^{-}$ decays},''
  \href{http://dx.doi.org/10.1103/PhysRevLett.113.151601}{{\em Phys. Rev.
  Lett.} {\bfseries 113} (2014) 151601},
\href{http://arxiv.org/abs/1406.6482}{{\ttfamily arXiv:1406.6482 [hep-ex]}}.

\bibitem{Aaij:2015yra}
{\bfseries LHCb} Collaboration, R.~Aaij {\em et~al.}, ``{Measurement of the
  ratio of branching fractions $\mathcal{B}(\bar{B}^0 \to
  D^{*+}\tau^{-}\bar{\nu}_{\tau})/\mathcal{B}(\bar{B}^0 \to
  D^{*+}\mu^{-}\bar{\nu}_{\mu})$},''
  \href{http://dx.doi.org/10.1103/PhysRevLett.115.159901,
  10.1103/PhysRevLett.115.111803}{{\em Phys. Rev. Lett.} {\bfseries 115}
  no.~11, (2015) 111803}, \href{http://arxiv.org/abs/1506.08614}{{\ttfamily
  arXiv:1506.08614 [hep-ex]}}.
[Erratum: Phys. Rev. Lett.115,no.15,159901(2015)].

\bibitem{Aaij:2015oid}
{\bfseries LHCb} Collaboration, R.~Aaij {\em et~al.}, ``{Angular analysis of
  the $B^{0} \to K^{*0} \mu^{+} \mu^{-}$ decay using 3 fb$^{-1}$ of integrated
  luminosity},'' \href{http://dx.doi.org/10.1007/JHEP02(2016)104}{{\em JHEP}
  {\bfseries 02} (2016) 104},
\href{http://arxiv.org/abs/1512.04442}{{\ttfamily arXiv:1512.04442 [hep-ex]}}.

\bibitem{Huschle:2015rga}
{\bfseries Belle} Collaboration, M.~Huschle {\em et~al.}, ``{Measurement of the
  branching ratio of $\bar{B} \to D^{(\ast)} \tau^- \bar{\nu}_\tau$ relative to
  $\bar{B} \to D^{(\ast)} \ell^- \bar{\nu}_\ell$ decays with hadronic tagging
  at Belle},'' \href{http://dx.doi.org/10.1103/PhysRevD.92.072014}{{\em Phys.
  Rev.} {\bfseries D92} no.~7, (2015) 072014},
\href{http://arxiv.org/abs/1507.03233}{{\ttfamily arXiv:1507.03233 [hep-ex]}}.

\bibitem{Sato:2016svk}
{\bfseries Belle} Collaboration, Y.~Sato {\em et~al.}, ``{Measurement of the
  branching ratio of $\bar{B}^0 \rightarrow D^{*+} \tau^- \bar{\nu}_{\tau}$
  relative to $\bar{B}^0 \rightarrow D^{*+} \ell^- \bar{\nu}_{\ell}$ decays
  with a semileptonic tagging method},''
  \href{http://dx.doi.org/10.1103/PhysRevD.94.072007}{{\em Phys. Rev.}
  {\bfseries D94} no.~7, (2016) 072007},
\href{http://arxiv.org/abs/1607.07923}{{\ttfamily arXiv:1607.07923 [hep-ex]}}.

\bibitem{Hirose:2016wfn}
{\bfseries Belle} Collaboration, S.~Hirose {\em et~al.}, ``{Measurement of the
  $\tau$ lepton polarization and $R(D^*)$ in the decay $\bar{B} \to D^* \tau^-
  \bar{\nu}_\tau$},''
  \href{http://dx.doi.org/10.1103/PhysRevLett.118.211801}{{\em Phys. Rev.
  Lett.} {\bfseries 118} no.~21, (2017) 211801},
\href{http://arxiv.org/abs/1612.00529}{{\ttfamily arXiv:1612.00529 [hep-ex]}}.

\bibitem{Hirose:2017dxl}
{\bfseries Belle} Collaboration, S.~Hirose {\em et~al.}, ``{Measurement of the
  $\tau$ lepton polarization and $R(D^*)$ in the decay $\bar{B} \rightarrow D^*
  \tau^- \bar{\nu}_\tau$ with one-prong hadronic $\tau$ decays at Belle},''
  \href{http://dx.doi.org/10.1103/PhysRevD.97.012004}{{\em Phys. Rev.}
  {\bfseries D97} no.~1, (2018) 012004},
\href{http://arxiv.org/abs/1709.00129}{{\ttfamily arXiv:1709.00129 [hep-ex]}}.

\bibitem{Aaij:2017vbb}
{\bfseries LHCb} Collaboration, R.~Aaij {\em et~al.}, ``{Test of lepton
  universality with $B^{0} \rightarrow K^{*0}\ell^{+}\ell^{-}$ decays},''
  \href{http://dx.doi.org/10.1007/JHEP08(2017)055}{{\em JHEP} {\bfseries 08}
  (2017) 055},
\href{http://arxiv.org/abs/1705.05802}{{\ttfamily arXiv:1705.05802 [hep-ex]}}.

\bibitem{Aaij:2017deq}
{\bfseries LHCb} Collaboration, R.~Aaij {\em et~al.}, ``{Test of Lepton Flavor
  Universality by the measurement of the $B^0 \to D^{*-} \tau^+ \nu_{\tau}$
  branching fraction using three-prong $\tau$ decays},''
  \href{http://dx.doi.org/10.1103/PhysRevD.97.072013}{{\em Phys. Rev.}
  {\bfseries D97} no.~7, (2018) 072013},
\href{http://arxiv.org/abs/1711.02505}{{\ttfamily arXiv:1711.02505 [hep-ex]}}.

\bibitem{Bhattacharya:2014wla}
B.~Bhattacharya, A.~Datta, D.~London, and S.~Shivashankara, ``{Simultaneous
  Explanation of the $R_K$ and $R(D^{(*)})$ Puzzles},''
  \href{http://dx.doi.org/10.1016/j.physletb.2015.02.011}{{\em Phys. Lett.}
  {\bfseries B742} (2015) 370--374},
\href{http://arxiv.org/abs/1412.7164}{{\ttfamily arXiv:1412.7164 [hep-ph]}}.

\bibitem{Alonso:2015sja}
R.~Alonso, B.~Grinstein, and J.~Martin~Camalich, ``{Lepton universality
  violation and lepton flavor conservation in $B$-meson decays},''
  \href{http://dx.doi.org/10.1007/JHEP10(2015)184}{{\em JHEP} {\bfseries 10}
  (2015) 184},
\href{http://arxiv.org/abs/1505.05164}{{\ttfamily arXiv:1505.05164 [hep-ph]}}.

\bibitem{Greljo:2015mma}
A.~Greljo, G.~Isidori, and D.~Marzocca, ``{On the breaking of Lepton Flavor
  Universality in B decays},''
  \href{http://dx.doi.org/10.1007/JHEP07(2015)142}{{\em JHEP} {\bfseries 07}
  (2015) 142},
\href{http://arxiv.org/abs/1506.01705}{{\ttfamily arXiv:1506.01705 [hep-ph]}}.

\bibitem{Calibbi:2015kma}
L.~Calibbi, A.~Crivellin, and T.~Ota, ``{Effective Field Theory Approach to
  b?s??(?), B?K(*)?$\overline{?}$ and B?D(*)?? with Third Generation
  Couplings},'' \href{http://dx.doi.org/10.1103/PhysRevLett.115.181801}{{\em
  Phys. Rev. Lett.} {\bfseries 115} (2015) 181801},
\href{http://arxiv.org/abs/1506.02661}{{\ttfamily arXiv:1506.02661 [hep-ph]}}.

\bibitem{Bauer:2015knc}
M.~Bauer and M.~Neubert, ``{Minimal Leptoquark Explanation for the
  R$_{D^{(*)}}$ , R$_K$ , and $(g-2)_g$ Anomalies},''
  \href{http://dx.doi.org/10.1103/PhysRevLett.116.141802}{{\em Phys. Rev.
  Lett.} {\bfseries 116} no.~14, (2016) 141802},
\href{http://arxiv.org/abs/1511.01900}{{\ttfamily arXiv:1511.01900 [hep-ph]}}.

\bibitem{Fajfer:2015ycq}
S.~Fajfer and N.~Ko\v~snik, ``{Vector leptoquark resolution of $R_K$ and
  $R_{D^{(*)}}$ puzzles},''
  \href{http://dx.doi.org/10.1016/j.physletb.2016.02.018}{{\em Phys. Lett.}
  {\bfseries B755} (2016) 270--274},
\href{http://arxiv.org/abs/1511.06024}{{\ttfamily arXiv:1511.06024 [hep-ph]}}.

\bibitem{Barbieri:2015yvd}
R.~Barbieri, G.~Isidori, A.~Pattori, and F.~Senia, ``{Anomalies in $B$-decays
  and $U(2)$ flavour symmetry},''
  \href{http://dx.doi.org/10.1140/epjc/s10052-016-3905-3}{{\em Eur. Phys. J.}
  {\bfseries C76} no.~2, (2016) 67},
\href{http://arxiv.org/abs/1512.01560}{{\ttfamily arXiv:1512.01560 [hep-ph]}}.

\bibitem{Das:2016vkr}
D.~Das, C.~Hati, G.~Kumar, and N.~Mahajan, ``{Towards a unified explanation of
  $R_{D^{(\ast)}}$, $R_{K}$ and $(g-2)_{\mu}$ anomalies in a left-right model
  with leptoquarks},'' \href{http://dx.doi.org/10.1103/PhysRevD.94.055034}{{\em
  Phys. Rev.} {\bfseries D94} (2016) 055034},
\href{http://arxiv.org/abs/1605.06313}{{\ttfamily arXiv:1605.06313 [hep-ph]}}.

\bibitem{Boucenna:2016wpr}
S.~M. Boucenna, A.~Celis, J.~Fuentes-Martin, A.~Vicente, and J.~Virto,
  ``{Non-abelian gauge extensions for B-decay anomalies},''
  \href{http://dx.doi.org/10.1016/j.physletb.2016.06.067}{{\em Phys. Lett.}
  {\bfseries B760} (2016) 214--219},
\href{http://arxiv.org/abs/1604.03088}{{\ttfamily arXiv:1604.03088 [hep-ph]}}.

\bibitem{Boucenna:2016qad}
S.~M. Boucenna, A.~Celis, J.~Fuentes-Martin, A.~Vicente, and J.~Virto,
  ``{Phenomenology of an $SU(2) \times SU(2) \times U(1)$ model with
  lepton-flavour non-universality},''
  \href{http://dx.doi.org/10.1007/JHEP12(2016)059}{{\em JHEP} {\bfseries 12}
  (2016) 059},
\href{http://arxiv.org/abs/1608.01349}{{\ttfamily arXiv:1608.01349 [hep-ph]}}.

\bibitem{Becirevic:2016yqi}
D.~Becirevic, S.~Fajfer, N.~Kosnik, and O.~Sumensari, ``{Leptoquark model to
  explain the $B$-physics anomalies, $R_K$ and $R_D$},''
  \href{http://dx.doi.org/10.1103/PhysRevD.94.115021}{{\em Phys. Rev.}
  {\bfseries D94} no.~11, (2016) 115021},
\href{http://arxiv.org/abs/1608.08501}{{\ttfamily arXiv:1608.08501 [hep-ph]}}.

\bibitem{Hiller:2016kry}
G.~Hiller, D.~Loose, and K.~Schonwald, ``{Leptoquark Flavor Patterns \& B Decay
  Anomalies},'' \href{http://dx.doi.org/10.1007/JHEP12(2016)027}{{\em JHEP}
  {\bfseries 12} (2016) 027},
\href{http://arxiv.org/abs/1609.08895}{{\ttfamily arXiv:1609.08895 [hep-ph]}}.

\bibitem{Bhattacharya:2016mcc}
B.~Bhattacharya, A.~Datta, J.-P. Guevin, D.~London, and R.~Watanabe,
  ``{Simultaneous Explanation of the $R_K$ and $R_{D^{(*)}}$ Puzzles: a Model
  Analysis},'' \href{http://dx.doi.org/10.1007/JHEP01(2017)015}{{\em JHEP}
  {\bfseries 01} (2017) 015},
\href{http://arxiv.org/abs/1609.09078}{{\ttfamily arXiv:1609.09078 [hep-ph]}}.

\bibitem{Barbieri:2016las}
R.~Barbieri, C.~W. Murphy, and F.~Senia, ``{B-decay Anomalies in a Composite
  Leptoquark Model},''
  \href{http://dx.doi.org/10.1140/epjc/s10052-016-4578-7}{{\em Eur. Phys. J.}
  {\bfseries C77} no.~1, (2017) 8},
\href{http://arxiv.org/abs/1611.04930}{{\ttfamily arXiv:1611.04930 [hep-ph]}}.

\bibitem{Becirevic:2016oho}
D.~Becirevic, N.~Kosnik, O.~Sumensari, and R.~Zukanovich~Funchal, ``{Palatable
  Leptoquark Scenarios for Lepton Flavor Violation in Exclusive $b\to
  s\ell_1\ell_2$ modes},''
  \href{http://dx.doi.org/10.1007/JHEP11(2016)035}{{\em JHEP} {\bfseries 11}
  (2016) 035},
\href{http://arxiv.org/abs/1608.07583}{{\ttfamily arXiv:1608.07583 [hep-ph]}}.

\bibitem{Bordone:2017anc}
M.~Bordone, G.~Isidori, and S.~Trifinopoulos, ``{Semileptonic $B$-physics
  anomalies: A general EFT analysis within $U(2)^n$ flavor symmetry},''
  \href{http://dx.doi.org/10.1103/PhysRevD.96.015038}{{\em Phys. Rev.}
  {\bfseries D96} no.~1, (2017) 015038},
\href{http://arxiv.org/abs/1702.07238}{{\ttfamily arXiv:1702.07238 [hep-ph]}}.

\bibitem{Megias:2017ove}
E.~Megias, M.~Quiros, and L.~Salas, ``{Lepton-flavor universality violation in
  R$_{K}$ and $ {R}_{D^{{\left(\ast \right)}}} $ from warped space},''
  \href{http://dx.doi.org/10.1007/JHEP07(2017)102}{{\em JHEP} {\bfseries 07}
  (2017) 102},
\href{http://arxiv.org/abs/1703.06019}{{\ttfamily arXiv:1703.06019 [hep-ph]}}.

\bibitem{Crivellin:2017zlb}
A.~Crivellin, D.~M{\"u}ller, and T.~Ota, ``{Simultaneous explanation of
  R(D$^{(?)}$) and b?s?$^{+}$ ?$^{?}$: the last scalar leptoquarks standing},''
  \href{http://dx.doi.org/10.1007/JHEP09(2017)040}{{\em JHEP} {\bfseries 09}
  (2017) 040},
\href{http://arxiv.org/abs/1703.09226}{{\ttfamily arXiv:1703.09226 [hep-ph]}}.

\bibitem{Cai:2017wry}
Y.~Cai, J.~Gargalionis, M.~A. Schmidt, and R.~R. Volkas, ``{Reconsidering the
  One Leptoquark solution: flavor anomalies and neutrino mass},''
\href{http://arxiv.org/abs/1704.05849}{{\ttfamily arXiv:1704.05849 [hep-ph]}}.

\bibitem{Altmannshofer:2017poe}
W.~Altmannshofer, P.~Bhupal~Dev, and A.~Soni, ``{$R_{D^{(*)}}$ anomaly: A
  possible hint for natural supersymmetry with $R$-parity violation},''
  \href{http://dx.doi.org/10.1103/PhysRevD.96.095010}{{\em Phys. Rev.}
  {\bfseries D96} no.~9, (2017) 095010},
\href{http://arxiv.org/abs/1704.06659}{{\ttfamily arXiv:1704.06659 [hep-ph]}}.

\bibitem{Dorsner:2017ufx}
I.~Dorsner, S.~Fajfer, D.~A. Faroughy, and N.~Kosnik, ``{Saga of the two GUT
  leptoquarks in flavor universality and collider searches},''
\href{http://arxiv.org/abs/1706.07779}{{\ttfamily arXiv:1706.07779 [hep-ph]}}.

\bibitem{Buttazzo:2017ixm}
D.~Buttazzo, A.~Greljo, G.~Isidori, and D.~Marzocca, ``{B-physics anomalies: a
  guide to combined explanations},''
  \href{http://dx.doi.org/10.1007/JHEP11(2017)044}{{\em JHEP} {\bfseries 11}
  (2017) 044},
\href{http://arxiv.org/abs/1706.07808}{{\ttfamily arXiv:1706.07808 [hep-ph]}}.

\bibitem{Assad:2017iib}
N.~Assad, B.~Fornal, and B.~Grinstein, ``{Baryon Number and Lepton Universality
  Violation in Leptoquark and Diquark Models},''
  \href{http://dx.doi.org/10.1016/j.physletb.2017.12.042}{{\em Phys. Lett.}
  {\bfseries B777} (2018) 324--331},
\href{http://arxiv.org/abs/1708.06350}{{\ttfamily arXiv:1708.06350 [hep-ph]}}.

\bibitem{DiLuzio:2017vat}
L.~Di~Luzio, A.~Greljo, and M.~Nardecchia, ``{Gauge leptoquark as the origin of
  B-physics anomalies},''
  \href{http://dx.doi.org/10.1103/PhysRevD.96.115011}{{\em Phys. Rev.}
  {\bfseries D96} no.~11, (2017) 115011},
\href{http://arxiv.org/abs/1708.08450}{{\ttfamily arXiv:1708.08450 [hep-ph]}}.

\bibitem{Calibbi:2017qbu}
L.~Calibbi, A.~Crivellin, and T.~Li, ``{A model of vector leptoquarks in view
  of the $B$-physics anomalies},''
\href{http://arxiv.org/abs/1709.00692}{{\ttfamily arXiv:1709.00692 [hep-ph]}}.

\bibitem{Bordone:2017bld}
M.~Bordone, C.~Cornella, J.~Fuentes-Martin, and G.~Isidori, ``{A three-site
  gauge model for flavor hierarchies and flavor anomalies},''
  \href{http://dx.doi.org/10.1016/j.physletb.2018.02.011}{{\em Phys. Lett.}
  {\bfseries B779} (2018) 317--323},
\href{http://arxiv.org/abs/1712.01368}{{\ttfamily arXiv:1712.01368 [hep-ph]}}.

\bibitem{Choudhury:2017ijp}
D.~Choudhury, A.~Kundu, R.~Mandal, and R.~Sinha, ``{$R_{K^{(*)}}$ and
  $R(D^{(*)})$ anomalies resolved with lepton mixing},''
  \href{http://dx.doi.org/10.1016/j.nuclphysb.2018.06.022}{{\em Nucl. Phys.}
  {\bfseries B933} (2018) 433--453},
\href{http://arxiv.org/abs/1712.01593}{{\ttfamily arXiv:1712.01593 [hep-ph]}}.

\bibitem{Barbieri:2017tuq}
R.~Barbieri and A.~Tesi, ``{$B$-decay anomalies in Pati-Salam SU(4)},''
  \href{http://dx.doi.org/10.1140/epjc/s10052-018-5680-9}{{\em Eur. Phys. J.}
  {\bfseries C78} no.~3, (2018) 193},
\href{http://arxiv.org/abs/1712.06844}{{\ttfamily arXiv:1712.06844 [hep-ph]}}.

\bibitem{Sannino:2017utc}
F.~Sannino, P.~Stangl, D.~M. Straub, and A.~E. Thomsen, ``{Flavor Physics and
  Flavor Anomalies in Minimal Fundamental Partial Compositeness},''
  \href{http://dx.doi.org/10.1103/PhysRevD.97.115046}{{\em Phys. Rev.}
  {\bfseries D97} no.~11, (2018) 115046},
\href{http://arxiv.org/abs/1712.07646}{{\ttfamily arXiv:1712.07646 [hep-ph]}}.

\bibitem{Blanke:2018sro}
M.~Blanke and A.~Crivellin, ``{$B$ Meson Anomalies in a Pati-Salam Model within
  the Randall-Sundrum Background},''
  \href{http://dx.doi.org/10.1103/PhysRevLett.121.011801}{{\em Phys. Rev.
  Lett.} {\bfseries 121} no.~1, (2018) 011801},
\href{http://arxiv.org/abs/1801.07256}{{\ttfamily arXiv:1801.07256 [hep-ph]}}.

\bibitem{Greljo:2018tuh}
A.~Greljo and B.~A. Stefanek, ``{Third family quark?lepton unification at the
  TeV scale},'' \href{http://dx.doi.org/10.1016/j.physletb.2018.05.033}{{\em
  Phys. Lett.} {\bfseries B782} (2018) 131--138},
\href{http://arxiv.org/abs/1802.04274}{{\ttfamily arXiv:1802.04274 [hep-ph]}}.

\bibitem{Marzocca:2018wcf}
D.~Marzocca, ``{Addressing the B-physics anomalies in a fundamental Composite
  Higgs Model},'' \href{http://dx.doi.org/10.1007/JHEP07(2018)121}{{\em JHEP}
  {\bfseries 07} (2018) 121},
\href{http://arxiv.org/abs/1803.10972}{{\ttfamily arXiv:1803.10972 [hep-ph]}}.

\bibitem{Asadi:2018wea}
P.~Asadi, M.~R. Buckley, and D.~Shih, ``{It's all right(-handed neutrinos): a
  new $W'$ model for the $R_{D^{(*)}}$ anomaly},''
\href{http://arxiv.org/abs/1804.04135}{{\ttfamily arXiv:1804.04135 [hep-ph]}}.

\bibitem{Greljo:2018ogz}
A.~Greljo, D.~J. Robinson, B.~Shakya, and J.~Zupan, ``{$R(D^{(*)})$ from $W'$
  and right-handed neutrinos},''
\href{http://arxiv.org/abs/1804.04642}{{\ttfamily arXiv:1804.04642 [hep-ph]}}.

\bibitem{Robinson:2018gza}
D.~Robinson, B.~Shakya, and J.~Zupan, ``{Right-handed Neutrinos and
  $R(D^{(*)})$},''
\href{http://arxiv.org/abs/1807.04753}{{\ttfamily arXiv:1807.04753 [hep-ph]}}.

\bibitem{Azatov:2018knx}
A.~Azatov, D.~Bardhan, D.~Ghosh, F.~Sgarlata, and E.~Venturini, ``{Anatomy of
  $b \to c \tau \nu$ anomalies},''
\href{http://arxiv.org/abs/1805.03209}{{\ttfamily arXiv:1805.03209 [hep-ph]}}.

\bibitem{Bordone:2018nbg}
M.~Bordone, C.~Cornella, J.~Fuentes-Martin, and G.~Isidori, ``{Low-energy
  signatures of the $\mathrm{PS}^3$ model: from $B$-physics anomalies to
  LFV},''
\href{http://arxiv.org/abs/1805.09328}{{\ttfamily arXiv:1805.09328 [hep-ph]}}.

\bibitem{Becirevic:2018afm}
D.~Becirevic, I.~Dorsner, S.~Fajfer, D.~A. Faroughy, N.~Kosnik, and
  O.~Sumensari, ``{Scalar leptoquarks from GUT to accommodate the $B$-physics
  anomalies},''
\href{http://arxiv.org/abs/1806.05689}{{\ttfamily arXiv:1806.05689 [hep-ph]}}.

\bibitem{Kumar:2018kmr}
J.~Kumar, D.~London, and R.~Watanabe, ``{Combined Explanations of the $b \to s
  \mu^+ \mu^-$ and $b \to c \tau^- {\bar\nu}$ Anomalies: a General Model
  Analysis},''
\href{http://arxiv.org/abs/1806.07403}{{\ttfamily arXiv:1806.07403 [hep-ph]}}.

\bibitem{Trifinopoulos:2018rna}
S.~Trifinopoulos, ``{Revisiting R-parity violating interactions as an
  explanation of the B-physics anomalies},''
\href{http://arxiv.org/abs/1807.01638}{{\ttfamily arXiv:1807.01638 [hep-ph]}}.

\bibitem{Azatov:2018kzb}
A.~Azatov, D.~Barducci, D.~Ghosh, D.~Marzocca, and L.~Ubaldi, ``{Combined
  explanations of B-physics anomalies: the sterile neutrino solution},''
\href{http://arxiv.org/abs/1807.10745}{{\ttfamily arXiv:1807.10745 [hep-ph]}}.

\bibitem{Cabibbo:1963yz}
N.~Cabibbo, ``{Unitary Symmetry and Leptonic Decays},''
  \href{http://dx.doi.org/10.1103/PhysRevLett.10.531}{{\em Phys. Rev. Lett.}
  {\bfseries 10} (1963) 531--533}.
[,648(1963)].

\bibitem{Feruglio:2016gvd}
F.~Feruglio, P.~Paradisi, and A.~Pattori, ``{Revisiting Lepton Flavor
  Universality in B Decays},''
  \href{http://dx.doi.org/10.1103/PhysRevLett.118.011801}{{\em Phys. Rev.
  Lett.} {\bfseries 118} no.~1, (2017) 011801},
\href{http://arxiv.org/abs/1606.00524}{{\ttfamily arXiv:1606.00524 [hep-ph]}}.

\bibitem{Feruglio:2017rjo}
F.~Feruglio, P.~Paradisi, and A.~Pattori, ``{On the Importance of Electroweak
  Corrections for B Anomalies},''
\href{http://arxiv.org/abs/1705.00929}{{\ttfamily arXiv:1705.00929 [hep-ph]}}.

\bibitem{Faroughy:2016osc}
D.~A. Faroughy, A.~Greljo, and J.~F. Kamenik, ``{Confronting lepton flavor
  universality violation in B decays with high-$p_T$ tau lepton searches at
  LHC},'' \href{http://dx.doi.org/10.1016/j.physletb.2016.11.011}{{\em Phys.
  Lett.} {\bfseries B764} (2017) 126--134},
\href{http://arxiv.org/abs/1609.07138}{{\ttfamily arXiv:1609.07138 [hep-ph]}}.

\bibitem{Greljo:2017vvb}
A.~Greljo and D.~Marzocca, ``{High-$p_T$ dilepton tails and flavor physics},''
  \href{http://dx.doi.org/10.1140/epjc/s10052-017-5119-8}{{\em Eur. Phys. J.}
  {\bfseries C77} no.~8, (2017) 548},
\href{http://arxiv.org/abs/1704.09015}{{\ttfamily arXiv:1704.09015 [hep-ph]}}.

\bibitem{Glashow:1970gm}
S.~L. Glashow, J.~Iliopoulos, and L.~Maiani, ``{Weak Interactions with
  Lepton-Hadron Symmetry},''
\href{http://dx.doi.org/10.1103/PhysRevD.2.1285}{{\em Phys. Rev.} {\bfseries
  D2} (1970) 1285--1292}.

\bibitem{Gaillard:1974hs}
M.~K. Gaillard and B.~W. Lee, ``{Rare Decay Modes of the K-Mesons in Gauge
  Theories},''
\href{http://dx.doi.org/10.1103/PhysRevD.10.897}{{\em Phys. Rev.} {\bfseries
  D10} (1974) 897}.

\bibitem{Diaz:2017lit}
B.~Diaz, M.~Schmaltz, and Y.-M. Zhong, ``{The Leptoquark Hunter's Guide: Pair
  Production},''
\href{http://arxiv.org/abs/1706.05033}{{\ttfamily arXiv:1706.05033 [hep-ph]}}.

\bibitem{Smirnov:2018ske}
A.~D. Smirnov, ``{Vector leptoquark mass limits and branching ratios of $
  K_L^0, B^0, B_s \to l^+_i l^-_j $ decays with account of fermion mixing in
  leptoquark currents},''
  \href{http://dx.doi.org/10.1142/S0217732318500190}{{\em Mod. Phys. Lett.}
  {\bfseries A33} (2018) 1850019},
\href{http://arxiv.org/abs/1801.02895}{{\ttfamily arXiv:1801.02895 [hep-ph]}}.

\bibitem{Pati:1974yy}
J.~C. Pati and A.~Salam, ``{Lepton Number as the Fourth Color},''
  \href{http://dx.doi.org/10.1103/PhysRevD.10.275,
  10.1103/PhysRevD.11.703.2}{{\em Phys. Rev.} {\bfseries D10} (1974) 275--289}.
[Erratum: Phys. Rev.D11,703(1975)].

\bibitem{Perez:2013osa}
P.~Fileviez~Perez and M.~B. Wise, ``{Low Scale Quark-Lepton Unification},''
  \href{http://dx.doi.org/10.1103/PhysRevD.88.057703}{{\em Phys. Rev.}
  {\bfseries D88} (2013) 057703},
\href{http://arxiv.org/abs/1307.6213}{{\ttfamily arXiv:1307.6213 [hep-ph]}}.

\bibitem{Guadagnoli:2018ojc}
D.~Guadagnoli, M.~Reboud, and O.~Sumensari, ``{A gauged horizontal $SU(2)$
  symmetry and $R_{K^{(\ast)}}$},''
\href{http://arxiv.org/abs/1807.03285}{{\ttfamily arXiv:1807.03285 [hep-ph]}}.

\bibitem{Fajfer:2013wca}
S.~Fajfer, A.~Greljo, J.~F. Kamenik, and I.~Mustac, ``{Light Higgs and
  Vector-like Quarks without Prejudice},''
  \href{http://dx.doi.org/10.1007/JHEP07(2013)155}{{\em JHEP} {\bfseries 07}
  (2013) 155},
\href{http://arxiv.org/abs/1304.4219}{{\ttfamily arXiv:1304.4219 [hep-ph]}}.

\bibitem{Aaij:2017uff}
{\bfseries LHCb} Collaboration, R.~Aaij {\em et~al.}, ``{Measurement of the
  ratio of the $B^0 \to D^{*-} \tau^+ \nu_{\tau}$ and $B^0 \to D^{*-} \mu^+
  \nu_{\mu}$ branching fractions using three-prong $\tau$-lepton decays},''
  \href{http://dx.doi.org/10.1103/PhysRevLett.120.171802}{{\em Phys. Rev.
  Lett.} {\bfseries 120} no.~17, (2018) 171802},
\href{http://arxiv.org/abs/1708.08856}{{\ttfamily arXiv:1708.08856 [hep-ex]}}.

\bibitem{Amhis:2016xyh}
Y.~Amhis {\em et~al.}, ``{Averages of $b$-hadron, $c$-hadron, and $\tau$-lepton
  properties as of summer 2016},''
\href{http://arxiv.org/abs/1612.07233}{{\ttfamily arXiv:1612.07233 [hep-ex]}}.

\bibitem{Bigi:2016mdz}
D.~Bigi and P.~Gambino, ``{Revisiting $B\to D \ell \nu$},''
  \href{http://dx.doi.org/10.1103/PhysRevD.94.094008}{{\em Phys. Rev.}
  {\bfseries D94} no.~9, (2016) 094008},
\href{http://arxiv.org/abs/1606.08030}{{\ttfamily arXiv:1606.08030 [hep-ph]}}.

\bibitem{Bernlochner:2017jka}
F.~U. Bernlochner, Z.~Ligeti, M.~Papucci, and D.~J. Robinson, ``{Combined
  analysis of semileptonic $B$ decays to $D$ and $D^*$: $R(D^{(*)})$,
  $|V_{cb}|$, and new physics},''
  \href{http://dx.doi.org/10.1103/PhysRevD.95.115008,
  10.1103/PhysRevD.97.059902}{{\em Phys. Rev.} {\bfseries D95} no.~11, (2017)
  115008}, \href{http://arxiv.org/abs/1703.05330}{{\ttfamily arXiv:1703.05330
  [hep-ph]}}.
[Erratum: Phys. Rev.D97,no.5,059902(2018)].

\bibitem{Bigi:2017jbd}
D.~Bigi, P.~Gambino, and S.~Schacht, ``{$R(D^*)$, $|V_{cb}|$, and the Heavy
  Quark Symmetry relations between form factors},''
  \href{http://dx.doi.org/10.1007/JHEP11(2017)061}{{\em JHEP} {\bfseries 11}
  (2017) 061},
\href{http://arxiv.org/abs/1707.09509}{{\ttfamily arXiv:1707.09509 [hep-ph]}}.

\bibitem{Jaiswal:2017rve}
S.~Jaiswal, S.~Nandi, and S.~K. Patra, ``{Extraction of $|V_{cb}|$ from $B\to
  D^{(*)}\ell\nu_\ell$ and the Standard Model predictions of $R(D^{(*)})$},''
  \href{http://dx.doi.org/10.1007/JHEP12(2017)060}{{\em JHEP} {\bfseries 12}
  (2017) 060},
\href{http://arxiv.org/abs/1707.09977}{{\ttfamily arXiv:1707.09977 [hep-ph]}}.

\bibitem{Patrignani:2016xqp}
{\bfseries Particle Data Group} Collaboration, C.~Patrignani {\em et~al.},
  ``{Review of Particle Physics},''
\href{http://dx.doi.org/10.1088/1674-1137/40/10/100001}{{\em Chin. Phys.}
  {\bfseries C40} no.~10, (2016) 100001}.

\bibitem{UTfit2016WEB}
{\bfseries UTfit} Collaboration, ``{Results Summer 2016},'' {\em Webpage} .
  \url{http://www.utfit.org/UTfit/ResultsSummer2016SM}.

\bibitem{Crivellin:2018yvo}
A.~Crivellin, C.~Greub, D.~M{\"u}ller, and F.~Saturnino, ``{Importance of Loop
  Effects in Explaining the Accumulated Evidence for New Physics in B Decays
  with a Vector Leptoquark},''
\href{http://arxiv.org/abs/1807.02068}{{\ttfamily arXiv:1807.02068 [hep-ph]}}.

\bibitem{Wehle:2016yoi}
{\bfseries Belle} Collaboration, S.~Wehle {\em et~al.},
  ``{Lepton-Flavor-Dependent Angular Analysis of $B\to K^\ast \ell^+\ell^-$},''
  \href{http://dx.doi.org/10.1103/PhysRevLett.118.111801}{{\em Phys. Rev.
  Lett.} {\bfseries 118} no.~11, (2017) 111801},
\href{http://arxiv.org/abs/1612.05014}{{\ttfamily arXiv:1612.05014 [hep-ex]}}.

\bibitem{Capdevila:2017bsm}
B.~Capdevila, A.~Crivellin, S.~Descotes-Genon, J.~Matias, and J.~Virto,
  ``{Patterns of New Physics in $b\to s\ell^+\ell^-$ transitions in the light
  of recent data},'' \href{http://dx.doi.org/10.1007/JHEP01(2018)093}{{\em
  JHEP} {\bfseries 01} (2018) 093},
\href{http://arxiv.org/abs/1704.05340}{{\ttfamily arXiv:1704.05340 [hep-ph]}}.

\bibitem{Ciuchini:2017mik}
M.~Ciuchini, A.~M. Coutinho, M.~Fedele, E.~Franco, A.~Paul, L.~Silvestrini, and
  M.~Valli, ``{On Flavourful Easter eggs for New Physics hunger and Lepton
  Flavour Universality violation},''
  \href{http://dx.doi.org/10.1140/epjc/s10052-017-5270-2}{{\em Eur. Phys. J.}
  {\bfseries C77} no.~10, (2017) 688},
\href{http://arxiv.org/abs/1704.05447}{{\ttfamily arXiv:1704.05447 [hep-ph]}}.

\bibitem{Altmannshofer:2017yso}
W.~Altmannshofer, P.~Stangl, and D.~M. Straub, ``{Interpreting Hints for Lepton
  Flavor Universality Violation},''
\href{http://arxiv.org/abs/1704.05435}{{\ttfamily arXiv:1704.05435 [hep-ph]}}.

\bibitem{Geng:2017svp}
L.-S. Geng, B.~Grinstein, S.~Jager, J.~Martin~Camalich, X.-L. Ren, and R.-X.
  Shi, ``{Towards the discovery of new physics with lepton-universality ratios
  of $b\to s\ell\ell$ decays},''
\href{http://arxiv.org/abs/1704.05446}{{\ttfamily arXiv:1704.05446 [hep-ph]}}.

\bibitem{DAmico:2017mtc}
G.~D'Amico, M.~Nardecchia, P.~Panci, F.~Sannino, A.~Strumia, R.~Torre, and
  A.~Urbano, ``{Flavour anomalies after the $R_{K^*}$ measurement},''
\href{http://arxiv.org/abs/1704.05438}{{\ttfamily arXiv:1704.05438 [hep-ph]}}.

\bibitem{Alok:2017sui}
A.~K. Alok, B.~Bhattacharya, A.~Datta, D.~Kumar, J.~Kumar, and D.~London,
  ``{New Physics in $b \to s \mu^+ \mu^-$ after the Measurement of
  $R_{K^*}$},'' \href{http://dx.doi.org/10.1103/PhysRevD.96.095009}{{\em Phys.
  Rev.} {\bfseries D96} no.~9, (2017) 095009},
\href{http://arxiv.org/abs/1704.07397}{{\ttfamily arXiv:1704.07397 [hep-ph]}}.

\bibitem{Hiller:2017bzc}
G.~Hiller and I.~Nisandzic, ``{$R_K$ and $R_{K^{\ast}}$ beyond the standard
  model},'' \href{http://dx.doi.org/10.1103/PhysRevD.96.035003}{{\em Phys.
  Rev.} {\bfseries D96} no.~3, (2017) 035003},
\href{http://arxiv.org/abs/1704.05444}{{\ttfamily arXiv:1704.05444 [hep-ph]}}.

\bibitem{DescotesGenon:2012zf}
S.~Descotes-Genon, J.~Matias, M.~Ramon, and J.~Virto, ``{Implications from
  clean observables for the binned analysis of $B -> K*\mu^+\mu^-$ at large
  recoil},'' \href{http://dx.doi.org/10.1007/JHEP01(2013)048}{{\em JHEP}
  {\bfseries 01} (2013) 048},
\href{http://arxiv.org/abs/1207.2753}{{\ttfamily arXiv:1207.2753 [hep-ph]}}.

\bibitem{Capdevila:2017iqn}
B.~Capdevila, A.~Crivellin, S.~Descotes-Genon, L.~Hofer, and J.~Matias,
  ``{Searching for New Physics with $b\to s\tau^+\tau^-$ processes},''
  \href{http://dx.doi.org/10.1103/PhysRevLett.120.181802}{{\em Phys. Rev.
  Lett.} {\bfseries 120} no.~18, (2018) 181802},
\href{http://arxiv.org/abs/1712.01919}{{\ttfamily arXiv:1712.01919 [hep-ph]}}.

\bibitem{Miyazaki:2011xe}
{\bfseries Belle} Collaboration, Y.~Miyazaki {\em et~al.}, ``{Search for
  Lepton-Flavor-Violating tau Decays into a Lepton and a Vector Meson},''
  \href{http://dx.doi.org/10.1016/j.physletb.2011.04.011}{{\em Phys. Lett.}
  {\bfseries B699} (2011) 251--257},
\href{http://arxiv.org/abs/1101.0755}{{\ttfamily arXiv:1101.0755 [hep-ex]}}.

\bibitem{Crivellin:2015era}
A.~Crivellin, L.~Hofer, J.~Matias, U.~Nierste, S.~Pokorski, and J.~Rosiek,
  ``{Lepton-flavour violating $B$ decays in generic $Z'$ models},''
  \href{http://dx.doi.org/10.1103/PhysRevD.92.054013}{{\em Phys. Rev.}
  {\bfseries D92} no.~5, (2015) 054013},
\href{http://arxiv.org/abs/1504.07928}{{\ttfamily arXiv:1504.07928 [hep-ph]}}.

\bibitem{Becirevic:2016zri}
D.~Becirevic, O.~Sumensari, and R.~Zukanovich~Funchal, ``{Lepton flavor
  violation in exclusive $b\rightarrow s$ decays},''
  \href{http://dx.doi.org/10.1140/epjc/s10052-016-3985-0}{{\em Eur. Phys. J.}
  {\bfseries C76} no.~3, (2016) 134},
\href{http://arxiv.org/abs/1602.00881}{{\ttfamily arXiv:1602.00881 [hep-ph]}}.

\bibitem{Lees:2012zz}
{\bfseries BaBar} Collaboration, J.~P. Lees {\em et~al.}, ``{A search for the
  decay modes $B^{+-} \to h^{+-} \tau^{+-}l$},''
  \href{http://dx.doi.org/10.1103/PhysRevD.86.012004}{{\em Phys. Rev.}
  {\bfseries D86} (2012) 012004},
\href{http://arxiv.org/abs/1204.2852}{{\ttfamily arXiv:1204.2852 [hep-ex]}}.

\bibitem{Love:2008ys}
{\bfseries CLEO} Collaboration, W.~Love {\em et~al.}, ``{Search for Lepton
  Flavor Violation in Upsilon Decays},''
  \href{http://dx.doi.org/10.1103/PhysRevLett.101.201601}{{\em Phys. Rev.
  Lett.} {\bfseries 101} (2008) 201601},
\href{http://arxiv.org/abs/0807.2695}{{\ttfamily arXiv:0807.2695 [hep-ex]}}.

\bibitem{Lees:2010jk}
{\bfseries BaBar} Collaboration, J.~P. Lees {\em et~al.}, ``{Search for Charged
  Lepton Flavor Violation in Narrow Upsilon Decays},''
  \href{http://dx.doi.org/10.1103/PhysRevLett.104.151802}{{\em Phys. Rev.
  Lett.} {\bfseries 104} (2010) 151802},
\href{http://arxiv.org/abs/1001.1883}{{\ttfamily arXiv:1001.1883 [hep-ex]}}.

\bibitem{DiLuzio:2017fdq}
L.~Di~Luzio, M.~Kirk, and A.~Lenz, ``{Updated $B_s$-mixing constraints on new
  physics models for $b\to s\ell^+\ell^-$ anomalies},''
  \href{http://dx.doi.org/10.1103/PhysRevD.97.095035}{{\em Phys. Rev.}
  {\bfseries D97} no.~9, (2018) 095035},
\href{http://arxiv.org/abs/1712.06572}{{\ttfamily arXiv:1712.06572 [hep-ph]}}.

\bibitem{Branco:1999fs}
G.~C. Branco, L.~Lavoura, and J.~P. Silva, ``{CP Violation},''
{\em Int. Ser. Monogr. Phys.} {\bfseries 103} (1999) 1--536.

\bibitem{Inami:1980fz}
T.~Inami and C.~S. Lim, ``{Effects of Superheavy Quarks and Leptons in
  Low-Energy Weak Processes k(L) ---> mu anti-mu, K+ ---> pi+ Neutrino
  anti-neutrino and K0 <---> anti-K0},''
  \href{http://dx.doi.org/10.1143/PTP.65.297}{{\em Prog. Theor. Phys.}
  {\bfseries 65} (1981) 297}.
[Erratum: Prog. Theor. Phys.65,1772(1981)].

\bibitem{Artuso:2015swg}
M.~Artuso, G.~Borissov, and A.~Lenz, ``{CP violation in the $B_s^0$ system},''
  \href{http://dx.doi.org/10.1103/RevModPhys.88.045002}{{\em Rev. Mod. Phys.}
  {\bfseries 88} no.~4, (2016) 045002},
\href{http://arxiv.org/abs/1511.09466}{{\ttfamily arXiv:1511.09466 [hep-ph]}}.

\bibitem{Lenz:2011ti}
A.~Lenz and U.~Nierste, ``{Numerical Updates of Lifetimes and Mixing Parameters
  of B Mesons},'' in {\em {CKM unitarity triangle. Proceedings, 6th
  International Workshop, CKM 2010, Warwick, UK, September 6-10, 2010}}.
\newblock 2011.
\newblock \href{http://arxiv.org/abs/1102.4274}{{\ttfamily arXiv:1102.4274
  [hep-ph]}}.
\newblock
\url{https://inspirehep.net/record/890169/files/arXiv:1102.4274.pdf}.
\newblock

\bibitem{Lenz:2006hd}
A.~Lenz and U.~Nierste, ``{Theoretical update of $B_s - \bar{B}_s$ mixing},''
  \href{http://dx.doi.org/10.1088/1126-6708/2007/06/072}{{\em JHEP} {\bfseries
  06} (2007) 072},
\href{http://arxiv.org/abs/hep-ph/0612167}{{\ttfamily arXiv:hep-ph/0612167
  [hep-ph]}}.

\bibitem{Bazavov:2016nty}
{\bfseries Fermilab Lattice, MILC} Collaboration, A.~Bazavov {\em et~al.},
  ``{$B^0_{(s)}$-mixing matrix elements from lattice QCD for the Standard Model
  and beyond},'' \href{http://dx.doi.org/10.1103/PhysRevD.93.113016}{{\em Phys.
  Rev.} {\bfseries D93} no.~11, (2016) 113016},
\href{http://arxiv.org/abs/1602.03560}{{\ttfamily arXiv:1602.03560 [hep-lat]}}.

\bibitem{Bertolini:1990if}
S.~Bertolini, F.~Borzumati, A.~Masiero, and G.~Ridolfi, ``{Effects of
  supergravity induced electroweak breaking on rare $B$ decays and mixings},''
\href{http://dx.doi.org/10.1016/0550-3213(91)90320-W}{{\em Nucl. Phys.}
  {\bfseries B353} (1991) 591--649}.

\bibitem{UTfit2018}
L.~Silvestrini, ``{Talk presented at La Thuile 2018},''.
\newblock
  \url{https://agenda.infn.it/getFile.py/access?contribId=26&sessionId=11&resId=0&materialId=slides&confId=14377}.

\bibitem{Carrasco:2014uya}
N.~Carrasco {\em et~al.}, ``{$D^0$$-\bar{D}^0$ mixing in the standard model and
  beyond from $N_f$ =2 twisted mass QCD},''
  \href{http://dx.doi.org/10.1103/PhysRevD.90.014502}{{\em Phys. Rev.}
  {\bfseries D90} no.~1, (2014) 014502},
\href{http://arxiv.org/abs/1403.7302}{{\ttfamily arXiv:1403.7302 [hep-lat]}}.

\bibitem{Goertz:2015nkp}
F.~Goertz, J.~F. Kamenik, A.~Katz, and M.~Nardecchia, ``{Indirect Constraints
  on the Scalar Di-Photon Resonance at the LHC},''
  \href{http://dx.doi.org/10.1007/JHEP05(2016)187}{{\em JHEP} {\bfseries 05}
  (2016) 187},
\href{http://arxiv.org/abs/1512.08500}{{\ttfamily arXiv:1512.08500 [hep-ph]}}.

\bibitem{DiLuzio:2017chi}
L.~Di~Luzio and M.~Nardecchia, ``{What is the scale of new physics behind the
  $B$-flavour anomalies?},''
  \href{http://dx.doi.org/10.1140/epjc/s10052-017-5118-9}{{\em Eur. Phys. J.}
  {\bfseries C77} no.~8, (2017) 536},
\href{http://arxiv.org/abs/1706.01868}{{\ttfamily arXiv:1706.01868 [hep-ph]}}.

\bibitem{DiLuzio:2016sur}
L.~Di~Luzio, J.~F. Kamenik, and M.~Nardecchia, ``{Implications of perturbative
  unitarity for scalar di-boson resonance searches at LHC},''
  \href{http://dx.doi.org/10.1140/epjc/s10052-017-4594-2}{{\em Eur. Phys. J.}
  {\bfseries C77} no.~1, (2017) 30},
\href{http://arxiv.org/abs/1604.05746}{{\ttfamily arXiv:1604.05746 [hep-ph]}}.

\bibitem{Aaboud:2018uek}
{\bfseries ATLAS} Collaboration, M.~Aaboud {\em et~al.}, ``{Search for pair
  production of heavy vector-like quarks decaying into high-$p_T$ $W$ bosons
  and top quarks in the lepton-plus-jets final state in $pp$ collisions at
  $\sqrt{s}$=13 TeV with the ATLAS detector},''
\href{http://arxiv.org/abs/1806.01762}{{\ttfamily arXiv:1806.01762 [hep-ex]}}.

\bibitem{Aaboud:2018xuw}
{\bfseries ATLAS} Collaboration, M.~Aaboud {\em et~al.}, ``{Search for pair
  production of up-type vector-like quarks and for four-top-quark events in
  final states with multiple $b$-jets with the ATLAS detector},''
\href{http://arxiv.org/abs/1803.09678}{{\ttfamily arXiv:1803.09678 [hep-ex]}}.

\bibitem{vanRitbergen:1999fi}
T.~van Ritbergen and R.~G. Stuart, ``{On the precise determination of the Fermi
  coupling constant from the muon lifetime},''
  \href{http://dx.doi.org/10.1016/S0550-3213(99)00572-6}{{\em Nucl. Phys.}
  {\bfseries B564} (2000) 343--390},
\href{http://arxiv.org/abs/hep-ph/9904240}{{\ttfamily arXiv:hep-ph/9904240
  [hep-ph]}}.

\bibitem{Sirunyan:2017lzl}
{\bfseries CMS} Collaboration, A.~M. Sirunyan {\em et~al.}, ``{Search for
  vectorlike light-flavor quark partners in proton-proton collisions at $\sqrt
  s$ =8??TeV},'' \href{http://dx.doi.org/10.1103/PhysRevD.97.072008}{{\em Phys.
  Rev.} {\bfseries D97} (2018) 072008},
\href{http://arxiv.org/abs/1708.02510}{{\ttfamily arXiv:1708.02510 [hep-ex]}}.

\bibitem{Aaboud:2018eqg}
{\bfseries ATLAS} Collaboration, M.~Aaboud {\em et~al.}, ``{Measurements of
  $t\bar{t}$ differential cross-sections of highly boosted top quarks decaying
  to all-hadronic final states in $pp$ collisions at $\sqrt{s}=13\,$ TeV using
  the ATLAS detector},''
\href{http://arxiv.org/abs/1801.02052}{{\ttfamily arXiv:1801.02052 [hep-ex]}}.

\bibitem{Aaboud:2017yvp}
{\bfseries ATLAS} Collaboration, M.~Aaboud {\em et~al.}, ``{Search for new
  phenomena in dijet events using 37 fb$^{-1}$ of $pp$ collision data collected
  at $\sqrt{s}=$13 TeV with the ATLAS detector},''
  \href{http://dx.doi.org/10.1103/PhysRevD.96.052004}{{\em Phys. Rev.}
  {\bfseries D96} no.~5, (2017) 052004},
\href{http://arxiv.org/abs/1703.09127}{{\ttfamily arXiv:1703.09127 [hep-ex]}}.

\bibitem{Aaboud:2018tqo}
{\bfseries ATLAS} Collaboration, M.~Aaboud {\em et~al.}, ``{Search for
  resonances in the mass distribution of jet pairs with one or two jets
  identified as $b$-jets in proton-proton collisions at $\sqrt{s}=13$ TeV with
  the ATLAS detector},''
\href{http://arxiv.org/abs/1805.09299}{{\ttfamily arXiv:1805.09299 [hep-ex]}}.

\bibitem{Alwall:2014hca}
J.~Alwall, R.~Frederix, S.~Frixione, V.~Hirschi, F.~Maltoni, O.~Mattelaer,
  H.~S. Shao, T.~Stelzer, P.~Torrielli, and M.~Zaro, ``{The automated
  computation of tree-level and next-to-leading order differential cross
  sections, and their matching to parton shower simulations},''
  \href{http://dx.doi.org/10.1007/JHEP07(2014)079}{{\em JHEP} {\bfseries 07}
  (2014) 079},
\href{http://arxiv.org/abs/1405.0301}{{\ttfamily arXiv:1405.0301 [hep-ph]}}.

\bibitem{Alloul:2013bka}
A.~Alloul, N.~D. Christensen, C.~Degrande, C.~Duhr, and B.~Fuks, ``{FeynRules
  2.0 - A complete toolbox for tree-level phenomenology},''
  \href{http://dx.doi.org/10.1016/j.cpc.2014.04.012}{{\em Comput. Phys.
  Commun.} {\bfseries 185} (2014) 2250--2300},
\href{http://arxiv.org/abs/1310.1921}{{\ttfamily arXiv:1310.1921 [hep-ph]}}.

\bibitem{Ball:2012cx}
R.~D. Ball {\em et~al.}, ``{Parton distributions with LHC data},''
  \href{http://dx.doi.org/10.1016/j.nuclphysb.2012.10.003}{{\em Nucl. Phys.}
  {\bfseries B867} (2013) 244--289},
\href{http://arxiv.org/abs/1207.1303}{{\ttfamily arXiv:1207.1303 [hep-ph]}}.

\bibitem{Martin:2009iq}
A.~D. Martin, W.~J. Stirling, R.~S. Thorne, and G.~Watt, ``{Parton
  distributions for the LHC},''
  \href{http://dx.doi.org/10.1140/epjc/s10052-009-1072-5}{{\em Eur. Phys. J.}
  {\bfseries C63} (2009) 189--285},
\href{http://arxiv.org/abs/0901.0002}{{\ttfamily arXiv:0901.0002 [hep-ph]}}.

\bibitem{Aaboud:2017sjh}
{\bfseries ATLAS} Collaboration, M.~Aaboud {\em et~al.}, ``{Search for
  additional heavy neutral Higgs and gauge bosons in the ditau final state
  produced in 36 fb$^{?1}$ of pp collisions at $ \sqrt{s}=13 $ TeV with the
  ATLAS detector},'' \href{http://dx.doi.org/10.1007/JHEP01(2018)055}{{\em
  JHEP} {\bfseries 01} (2018) 055},
\href{http://arxiv.org/abs/1709.07242}{{\ttfamily arXiv:1709.07242 [hep-ex]}}.

\bibitem{Dorsner:2018ynv}
I.~Dorsner and A.~Greljo, ``{Leptoquark toolbox for precision collider
  studies},'' \href{http://dx.doi.org/10.1007/JHEP05(2018)126}{{\em JHEP}
  {\bfseries 05} (2018) 126},
\href{http://arxiv.org/abs/1801.07641}{{\ttfamily arXiv:1801.07641 [hep-ph]}}.

\bibitem{Blumlein:1996qp}
J.~Blumlein, E.~Boos, and A.~Kryukov, ``{Leptoquark pair production in hadronic
  interactions},'' \href{http://dx.doi.org/10.1007/s002880050538}{{\em Z.
  Phys.} {\bfseries C76} (1997) 137--153},
\href{http://arxiv.org/abs/hep-ph/9610408}{{\ttfamily arXiv:hep-ph/9610408
  [hep-ph]}}.

\bibitem{Sirunyan:2018kzh}
{\bfseries CMS} Collaboration, A.~M. Sirunyan {\em et~al.}, ``{Constraints on
  models of scalar and vector leptoquarks decaying to a quark and a neutrino at
  $\sqrt{s}=$ 13 TeV},''
\href{http://arxiv.org/abs/1805.10228}{{\ttfamily arXiv:1805.10228 [hep-ex]}}.

\bibitem{Aaboud:2017dmy}
{\bfseries ATLAS} Collaboration, M.~Aaboud {\em et~al.}, ``{Search for
  supersymmetry in final states with two same-sign or three leptons and jets
  using 36 fb$^{-1}$ of $\sqrt{s}=13$ TeV $pp$ collision data with the ATLAS
  detector},'' \href{http://dx.doi.org/10.1007/JHEP09(2017)084}{{\em JHEP}
  {\bfseries 09} (2017) 084},
\href{http://arxiv.org/abs/1706.03731}{{\ttfamily arXiv:1706.03731 [hep-ex]}}.

\bibitem{Bai:2018jsr}
Y.~Bai and B.~A. Dobrescu, ``{Collider Tests of the Renormalizable Coloron
  Model},'' \href{http://dx.doi.org/10.1007/JHEP04(2018)114}{{\em JHEP}
  {\bfseries 04} (2018) 114},
\href{http://arxiv.org/abs/1802.03005}{{\ttfamily arXiv:1802.03005 [hep-ph]}}.

\bibitem{Georgi:2016xhm}
H.~Georgi and Y.~Nakai, ``{Diphoton resonance from a new strong force},''
  \href{http://dx.doi.org/10.1103/PhysRevD.94.075005}{{\em Phys. Rev.}
  {\bfseries D94} no.~7, (2016) 075005},
\href{http://arxiv.org/abs/1606.05865}{{\ttfamily arXiv:1606.05865 [hep-ph]}}.

\bibitem{Biggio:2016wyy}
C.~Biggio, M.~Bordone, L.~Di~Luzio, and G.~Ridolfi, ``{Massive vectors and loop
  observables: the $g-2$ case},''
  \href{http://dx.doi.org/10.1007/JHEP10(2016)002}{{\em JHEP} {\bfseries 10}
  (2016) 002},
\href{http://arxiv.org/abs/1607.07621}{{\ttfamily arXiv:1607.07621 [hep-ph]}}.

\end{thebibliography}\endgroup

\end{document}